\documentclass[english,american]{article}

\usepackage[T1]{fontenc}
\usepackage[latin9]{inputenc}
\usepackage[a4paper]{geometry}
\usepackage{babel}
\usepackage{refstyle}
\usepackage{units}
\usepackage{multirow}
\usepackage{amsmath}
\usepackage{amssymb}
\usepackage{graphicx}
\usepackage{wasysym}
\usepackage[authoryear]{natbib}
\usepackage[unicode=true,pdfusetitle,
 bookmarks=true,bookmarksnumbered=false,bookmarksopen=false,
 breaklinks=false,pdfborder={0 0 1},backref=false,colorlinks=false]
 {hyperref}

\makeatletter

\AtBeginDocument{\providecommand\subsecref[1]{\ref{subsec:#1}}}
\AtBeginDocument{\providecommand\eqref[1]{\ref{eq:#1}}}
\AtBeginDocument{\providecommand\tabref[1]{\ref{tab:#1}}}
\AtBeginDocument{\providecommand\figref[1]{\ref{fig:#1}}}
\providecommand{\tabularnewline}{\\}
\RS@ifundefined{subsecref}
  {\newref{subsec}{name = \RSsectxt}}
  {}
\RS@ifundefined{thmref}
  {\def\RSthmtxt{theorem~}\newref{thm}{name = \RSthmtxt}}
  {}
\RS@ifundefined{lemref}
  {\def\RSlemtxt{lemma~}\newref{lem}{name = \RSlemtxt}}
  {}

\usepackage[CaptionAfterwards]{fltpage}

\usepackage{tabularx}
\usepackage{colortbl}
\usepackage{xcolor}

\newref{cap}{refcmd=\hyperref[#1]{Figure~\ref{#1}}}
\newref{fig}{refcmd=\hyperref[#1]{Figure~\ref{#1}}}
\newref{tab}{refcmd=\hyperref[#1]{Table~\ref{#1}}}
\newref{sec}{refcmd=\hyperref[#1]{Section~\ref{#1}}}
\newref{subsec}{refcmd=\hyperref[#1]{Section~\ref{#1}}}
\newref{eq}{refcmd=\hyperref[#1]{Equation~\ref{#1}}}

\makeatother

\begin{document}

\section*{Reconciliation of weak pairwise spike-train correlations and highly
coherent local field potentials across space}

\noindent {\large{}Johanna Senk\textsuperscript{1}, Espen Hagen\textsuperscript{1,2},
Sacha J. van Albada\textsuperscript{1}, Markus Diesmann\textsuperscript{1,3,4}}{\large\par}
\begin{flushleft}
\textsuperscript{1}Institute of Neuroscience and Medicine (INM-6)
and Institute for Advanced Simulation (IAS-6) and JARA Institute Brain
Structure-Function Relationships (INM-10), Jülich Research Centre,
Jülich, Germany
\par\end{flushleft}

\begin{flushleft}
\textsuperscript{2}Department of Physics, University of Oslo, Oslo,
Norway
\par\end{flushleft}

\begin{flushleft}
\textsuperscript{3}Department of Psychiatry, Psychotherapy and Psychosomatics,
Medical Faculty, RWTH Aachen University, Aachen, Germany
\par\end{flushleft}

\begin{flushleft}
\textsuperscript{4}Department of Physics, Faculty 1, RWTH Aachen
University, Aachen, Germany
\par\end{flushleft}

\section*{Conflict of Interest}

The authors declare no competing financial interests.

\section*{Acknowledgements}

This project received funding from the European Union Seventh Framework
Programme {[}FP7/2007-2013{]} under grant agreement No. 604102 (Human
Brain Project, HBP), the European Union\textquoteright s Horizon 2020
research and innovation programme under grant agreement No. 720270
(HBP SGA1) and No. 785907 (HBP SGA2), the Helmholtz Association through
the Helmholtz Portfolio Theme \textquotedblleft Supercomputing and
Modeling for the Human Brain\textquotedblright , and the Research
Council of Norway (NFR) through COBRA. The use of the supercomputer
JURECA in Jülich was made possible by the JARA-HPC Vergabegremium
and provided on the JARA-HPC Partition (VSR computation time grant
JINB33).

We would like to thank Hans Ekkehard Plesser for helpful suggestions
for the implementation of spatially structured networks in NEST, and
Gaute T. Einevoll for providing useful feedback on the manuscript. 

\newpage{}

\section*{Abstract}

Chronic and acute implants of multi-electrode arrays that cover several
square millimeters of neural tissue provide simultaneous access to
population signals such as extracellular potentials and the spiking
activity of one hundred or more individual neurons. While the recorded
data may uncover principles of brain function, its interpretation
calls for multiscale computational models with corresponding spatial
dimensions and signal predictions. Such models can then facilitate
the search of candidate mechanisms underlying experimentally observed
spatiotemporal activity patterns in cortex. Multi-layer spiking neuron
network models of local cortical circuits covering about $\unit[1]{mm^{2}}$
have been developed, integrating experimentally obtained neuron-type
specific connectivity data and reproducing features of observed in-vivo
spiking statistics. Using forward models, local field potentials (LFPs)
can be computed from the simulated spiking activity. To account for
the spatial scale of common neural recordings, we here extend a local
network and LFP model to an area of $\unit[4\times4]{mm^{2}}$. The
upscaling preserves the densities of neurons and local synapses, and
introduces distance-dependent connection probabilities and conduction
delays. As detailed experimental data on distance-dependent connectivity
is partially lacking, we address this uncertainty in model parameters
by testing different parameter combinations within biologically plausible
bounds. Based on model predictions of spiking activity and LFPs,
we find that the upscaling procedure preserves the overall spiking
statistics of the original model and reproduces asynchronous irregular
spiking across populations and weak pairwise spike-train correlations
experimentally observed in sensory cortex. In contrast with the weak
spike-train correlations, the correlation of LFP signals is strong
and distance-dependent, compatible with experimental observations.
Enhanced spatial coherence in the low-gamma band around $\unit[50]{Hz}$
may explain the recent experimental report of an apparent band-pass
filter effect in the spatial reach of the LFP.

\section*{Significance Statement}

Extracellular recordings with multi-electrode arrays measure both
population signals such as the local field potential (LFP) and spiking
activity of individual neurons across the cortical tissue, for instance
covering the $\unit[4\times4]{mm^{2}}$ of a Utah array. To reproduce
key features of activity data obtained from such cortical patches,
we assess spiking activity and LFPs of a multiscale neuronal network
model of this spatial extent. The circuit incorporates biological
detail such as a realistic neuron density across four cortical layers
and neuron-type-specific, layer-specific, and distance-dependent connection
rules. The model reproduces experimental observations like a frequency-dependent
LFP coherence across space despite weak pairwise spike-train correlations.

\section{Introduction}

Cortical activity on the mesoscopic scale (mesoscale), below a cortical
surface area on the order of several square millimeters to centimeters
\citep{Muller18_255}, can be recorded extracellularly with chronic
or acute implants of multi-electrode arrays \citep{Maynard97,Buzsaki12_1322,Einevoll13_770}.
The low-frequency part ($\unit[\lesssim100]{Hz}$) of the measured
extracellular potential, the local field potential (LFP), is a population
signal with contributions from up to millions of local and remote
neurons \citep{Kajikawa11_847,Linden11_859,Leski13_e1003137}. Spiking
activity of individual neurons can be obtained from the high-frequency
part ($\unit[\gtrsim100]{Hz}$) of the signal through spike sorting
\citep{Quiroga07_3583}. The number of reliably identified single
neurons is on the order of $100$ neurons for chronically implanted
Utah arrays ($10\times10$ electrodes on $\unit[4\times4]{mm^{2}}$,
Blackrock microsystems, \href{http://blackrockmicro.com}{http://blackrockmicro.com})
as in \citet{Riehle13_48}. The recordings expose LFP activity appearing
to propagate across the cortex associated with distance dependency
of statistical measures like correlations and coherences \citep{Destexhe99_4595,Smith08_12591,Wu08_487,Muller12_222,Sato12_218,Dubey16_1986,Denker18_1,Muller18_255}.
The observation of coherent LFPs across space contrasts with the often
reported low pairwise correlation in cortical spike trains obtained
in asynchronous brain states \citep[for example][]{Ecker10,Renart10_587}.

Assuming a neuron density of $\unit[10^{5}]{neurons/mm^{2}}$ across
the cortical surface \citep{Herculano-Houzel09}, the number of neurons
covered by a Utah array is more than a million. Every neuron receives
up to $10^{4}$ synapses from neighboring and distant neurons \citep{Abeles91}.
However, the local circuitry is highly specific with respect to cortical
layers and neuron types \citep{Douglas89_480,Thomson02_936,Binzegger04}.
The majority of local cortical connections are established within
a distance of $\lesssim\unit[500]{\mu m}$ from the sender/receiving
neuron \citep{Voges10_277}, with probabilities that decay with distance
according to a Gaussian or exponentially shaped profile \citep{Hellwig00_111,Boucsein11_1,Packer2011_13260,Perin11}.
Local connections are typically made by unmyelinated axons. Therefore,
typical conduction delays between pre- and postsynaptic neurons are
governed by propagation speeds estimated around $\unit[0.3]{mm/ms}$
\citep{Hirsch91_1800,Murakoshi93_211,Kang94_280}.

To date, the relationship between cortical connectivity structure
and experimentally recorded activity of spikes and LFPs on the mesoscale
remains poorly understood. Network models that encompass the relevant
anatomical and physiological detail, spatial scales, and corresponding
measurements can aid the interpretation of experimental observations
and their underlying mechanisms. We here argue for full-scale models,
in terms of realistic numbers of neurons and synapses: Downscaled
or diluted network models may not reproduce first- and second-order
statistics (rates and correlations, respectively) of full-scale networks
\citep{Albada15}. Also, \citet{Hagen16} demonstrate that biophysical
forward-model predictions of LFP signals (and by extension electroencephalographic
(EEG) and magnetoencephalographic (MEG) signals) must include the
full density of cells and connections to account for network correlations.
One such full-density model, the microcircuit model by \citet{Potjans14_785},
represents a $\unit[1]{mm^{2}}$ cortical patch of early sensory cortex
with approximately $80,000$ leaky integrate-and-fire neurons and
about $0.3$ billion synapses set up using neuron-type- and layer-specific
connection probabilities derived from anatomical and electrophysiological
data. This model produces biologically plausible firing rates across
four cortical layers with one excitatory and inhibitory population
per layer, is simple enough to allow for rigorous mathematical analysis,
is publicly available, and has by now been used also in other studies
\citep{Wagatsuma11_00031,Bos16_1,Cain16_e1005045,Hagen16,Hahne17_34,Senk17_243,Schuecker17,Schwalger17_e1005507,Schmidt18_1409,VanAlbada18_291}.

Here, we hypothesize that a version of this microcircuit model and
corresponding LFP measurements upscaled laterally to an area of at
least $\unit[4\times4]{mm^{2}}$ (similar to the Utah multi-electrode
array), while accounting for distance-dependent connection probabilities,
should not only preserve the main features of activity in the original
model, but also explain features emerging on the mesoscale such as
spatial propagation of evoked neuronal activity \citep{Bringuier99_695,Swadlow02_7766,Einevoll07_2174,Muller14_4675,Klein16_143},
and strong distance-dependent correlations and coherences in the measured
LFP \citep{Destexhe99_4595,Berens08_1,Katzner09_35,Nauhaus09_70,Kajikawa11_847,Jia11_9390,Srinath14_741,Dubey16_1986}
even for typically weak pairwise spike-train correlations in cortex
\citep[see, for example,][]{Ecker10,Renart10_587}. Furthermore, the
upscaled model should serve as a test platform for parameters that
are to date poorly constrained by available experimental data, and
expose mechanisms underlying spatiotemporal pattern formation. Indeed,
we find that the overall behavior of the original microcircuit is
preserved when upscaled, and that the resulting model reconciles the
observation of weak pairwise spike-train correlations in cortex with
spatially correlated and coherent LFPs. 

Preliminary results have been published in abstract form \citep{Senk_15,Hagen16_cns_P167}.

\section{Materials and Methods\label{sec:methods}}

\subsection{Point-neuron networks\label{subsec:point_neuron_networks}}

This section provides a compact description of the different network
models considered in this study. The full network descriptions are
given in Tables \ref{tab:model_summary} and \ref{tab:model_summary_cont}.
Each network model represents a part of early sensory cortex with
realistic densities of neurons and synapses. We first consider the
original network model proposed by \citet{Potjans14_785} which describes
a microcircuit under $\unit[1]{mm^{2}}$ cortical surface, henceforth
referred to as \textquoteleft reference model\textquoteright . We
then consider networks upscaled to greater surface areas, referred
to as \textquoteleft upscaled models\textquoteright . The eight neuron
populations within each network are organized into four cortical layers,
that is, layer $2/3$ ($\mathrm{L2/3}$), layer $4$ ($\mathrm{L4}$),
layer $5$ ($\mathrm{L5}$) and layer $6$ ($\mathrm{L6}$), respectively.
Each layer contains an excitatory ($\mathrm{E}$) and an inhibitory
($\mathrm{I}$) population of leaky integrate-and-fire (LIF) neurons,
whose sub-threshold membrane dynamics are governed by \eqref{lif_subthreshold}.
The probabilities for two neurons to be connected are layer- and neuron-type-specific
and derived from a number of anatomical and electrophysiological studies
\citep{Potjans14_785}. Postsynaptic currents have static, normally
distributed amplitudes at onset that decay exponentially (Equations
\ref{eq:lif_psc} and \ref{eq:weight_distr}). All neurons receive
stationary external input in the form of Poisson spike trains with
fixed rate parameters. In addition, one population of thalamocortical
($\mathrm{TC}$) neurons targeting $\mathrm{E}$ and $\mathrm{I}$
neurons in both $\mathrm{L4}$ and $\mathrm{L6}$ can provide transient
or stationary external input, for example to emulate stimuli of the
sensory pathway.

\subsubsection{Network model descriptions\label{subsec:network_descriptions} }

We here describe the main differences between the original network
model and upscaled models derived from it. 

\textit{Reference model:} \citet{Potjans14_785} parameterize the
original microcircuit model to cover a cortical column under a surface
area of $A^{\mathrm{r}}=\unit[1]{mm^{2}}$. The superscript $\mathrm{r}$
denotes \textquoteleft reference model\textquoteright{} here and throughout
this manuscript. The resulting network connects almost $80,000$ neurons
with approximately $0.3$ billion synapses. The calculation of connection
probabilities in the model assumes a Gaussian distance dependency
of the form (see \citet{Potjans14_785} for details) \textrm{
\begin{equation}
c^{\mathrm{r}}\left(r\right)=c_{0}\mathrm{e}^{-r^{2}/2\sigma_{0}^{2}}.\label{eq:micro_gaussian}
\end{equation}
}Here, $r$ denotes the lateral distance between the two neurons.
This distance dependency is introduced to reconcile connectivity measurements
obtained using anatomical connectivity data (retrograde/anterograde
staining, \citealp{Binzegger04}) and electrophysiological data (in
vitro, \citealp{Thomson02_936}). The computed mean values averaged
over all populations for zero-distance connection probability and
standard deviation \textrm{are $c_{0}=0.14$ and }$\sigma_{0}=\unit[0.30]{mm}$,
respectively \citep[Equations 4-8, Figure 3]{Potjans14_785}. This
spatial decay constant is large compared to the extent of a typical
cortical column, which justifies their choice of a local network connectivity
without distance dependency.\foreignlanguage{english}{ }

A neuron $j$ in a source population $X$ of size $N_{X}^{\mathrm{r}}$
connects at random to a neuron $i$ in a target population $Y$ of
size $N_{Y}^{\mathrm{r}}$ with mean connection probability \citep[Equation 1]{Potjans14_785}

\begin{equation}
C_{YX}^{\mathrm{r}}=1-\left(1-\frac{1}{N_{X}^{\mathrm{r}}N_{Y}^{\mathrm{r}}}\right)^{S_{YX}^{\mathrm{r}}},\label{eq:conn_prob_num_syn}
\end{equation}
where $S_{YX}^{\mathrm{r}}$ denotes the total number of synapses
between these populations. The connection routine draws connections
randomly between pairs of neurons $i$ and $j$ until the total number
of synapses $S_{YX}^{\mathrm{r}}$ is reached. Multiple connections
(multapses) between neuron pairs are allowed. The connection probability
$C_{YX}^{\mathrm{r}}$ is here defined as the probability that a pair
of neurons is connected via one or more synapses. Connection delays
are normally distributed according to \eqref{delay_distr_ref} with
different parameters for excitatory and inhibitory sources. The standard
deviation of delays is $50\%$ of the mean delay, and the excitatory
mean delay is twice as long as the inhibitory one.

\textit{Upscaled models:} We next consider cortical network models
based on the reference network upscaled to cover an area of $A^{\mathrm{u}}=L^{2}$.
With square layers and a chosen side length $L=\unit[4]{mm}$ this
area is similar to the area covered by the Utah array ($10\times10$
electrodes, Blackrock Microsystems). The superscript $\mathrm{u}$
denotes \textquoteleft upscaled models\textquoteright{} here and throughout
this manuscript. In the upscaled models, neuron positions are drawn
randomly within a square domain of side length $L$ with the origin
$\left(0,0\right)$ at the center. We position neurons in the $\mathrm{TC}$
layer also within the area \foreignlanguage{english}{\textrm{$A^{\mathrm{u}}$}},
which facilitates the connectivity management between $\mathrm{TC}$
neurons and cortical neurons in the model. An analogy to the early
visual pathway would be that the distance $L$ in both thalamus and
cortex corresponds to the same extent of the visual field. A source
neuron $j\in X$ at location $\left(x_{j},y_{j}\right)$ connects
to a target neuron $i\in Y$ at location $\left(x_{i},y_{i}\right)$
with a probability dependent on their distance $r_{ij}$ given in
\eqref{distance_rij}. This expression for distance accounts for periodic
boundary conditions (torus connectivity). The distance-dependent connection
probability is shaped as a two-dimensional (2D) Gaussian and cut off
at a maximal radial distance $R$ as defined in \eqref{spatial_profile}.
The zero-distance connection probability \textrm{$c_{YX}$} between
populations $X$ and $Y$ is derived in \subsecref{upscaling_procedure}.
The corresponding standard deviation $\sigma_{X}$ defines the spatial
width of the profile and depends only on the source population $X$.
Connection delays of the upscaled models are calculated using a linear
distance dependency given by \eqref{delay_distr_ups} with a constant
delay offset $d_{0}$ and a conduction speed $v$, plus a random offset
drawn from a normal distribution with zero mean and standard deviation\textrm{
$\sigma_{d}^{\mathrm{u}}$} capped at values $\pm\left(d_{0}-dt\right)$
in order to prevent delays smaller than the simulation time step $dt$.
These values are the same for all cortical populations. For the external
layer, $\mathrm{TC}$ neurons within a circle of adjustable radius
$R_{\mathrm{pulse}}^{\mathrm{TC}}$ surrounding the center emit spikes
in a synchronous and regular fashion (thalamic pulses) with time intervals
$\Delta t_{\mathrm{TC}}$.

These network model implementations rely on the neuronal network simulator
NEST (\href{http://www.nest-simulator.org}{http://www.nest-simulator.org},
\citealp{Gewaltig_07_11204}) and are set up such that the same code
is used for both the reference and upscaled models, but with different
parameters.

\subsubsection{Upscaling procedure\label{subsec:upscaling_procedure}}

We here describe the procedure used to derive parameters for the upscaled
model(s) from the original reference network model description, in
terms of neuron numbers, synapse numbers, distance-dependent connection
probabilities, in-degrees of external input, and distance-dependent
delays from available experimental data.

\textit{Neuron numbers: }The upscaled networks preserve the neuron
densities per square millimeter of the reference model. Assuming a
homogeneous neuron density across space, the size of a population
$X$ in the upscaled networks is

\begin{equation}
N_{X}^{\mathrm{u}}=N_{X}^{\mathrm{r}}\frac{A^{\mathrm{u}}}{A^{\mathrm{r}}}.
\end{equation}

\textit{Synapse numbers:} With the aim to derive zero-distance connection
probabilities \textrm{$c_{YX}$} for a Gaussian connectivity profile
\eqref{spatial_profile}, we first compute average connection probabilities
$C_{YX}^{\mathrm{u}}$ in the upscaled models similar to $C_{YX}^{\mathrm{r}}$
for the reference model (as in \eqref{conn_prob_num_syn}, but with
corresponding neuron and synapse numbers). We define this connection
probability as
\begin{equation}
C_{YX}^{\mathrm{u}}=C_{YX}^{\mathrm{ui}}\cdot\left(1-\delta C_{YX}\right).\label{eq:conn_prob_with_modification}
\end{equation}
The superscript $\mathrm{ui}$ denotes upscaled, intermediate connection
probabilities. The term $\delta C_{YX}$ is introduced to allow for
selective modifications of the connection probabilities in the final
upscaled network (for example to modify firing rate spectra, see below).
Thus, connections are unchanged for $\delta C_{YX}=0$, meaning $C_{YX}^{\mathrm{u}}=C_{YX}^{\mathrm{ui}}$,
but a small positive or negative value results in an increase or decrease
of a specific connection probability between populations $X$ and
$Y$, respectively. The connection probability\textrm{ $C_{YX}^{\mathrm{ui}}$}
depends linearly on the corresponding population-specific connection
probability of the reference model, $C_{YX}^{\mathrm{r}}$, and the
ratio of mean connection probabilities from the upscaled and reference
models \citep[Equation 6]{Schmidt18_1409}

\begin{equation}
C_{YX}^{\mathrm{ui}}=C_{YX}^{\mathrm{r}}\frac{\overline{C}^{\mathrm{u}}}{\overline{C}^{\mathrm{r}}}.\label{eq:conn_prob_ups}
\end{equation}
Like \citet{Schmidt18_1409}, we choose to use the average connection
probability of the reference model $\overline{C}^{\mathrm{r}}=0.066$
as computed in \citep[Equation 9]{Potjans14_785}. To compute the
average connection probability \textrm{$\overline{C}^{\mathrm{u}}$}
of the upscaled models, we integrate the Gaussian profile given in
\eqref{micro_gaussian} over all possible positions of a source neuron
$\left(x_{1},y_{1}\right)$ and a target neuron $\left(x_{2},y_{2}\right)$,
located on a square domain of side length $L$. Accounting for the
maximal radial distance of connections, set to $R=L/2,$ and the periodic
boundary conditions used for the upscaled model, we numerically solve

\begin{equation}
\overline{C}^{\mathrm{u}}=\frac{1}{L^{4}}\int_{-L/2}^{L/2}\int_{-L/2}^{L/2}\int_{x_{1}-L/2}^{x_{1}+L/2}\int_{y_{1}-L/2}^{y_{1}+L/2}c^{\mathrm{r}}\left(r_{21}\right)\mathrm{d}y_{2}\mathrm{d}x_{2}\mathrm{d}y_{1}\mathrm{d}x_{1}
\end{equation}
where $r_{21}=\sqrt{\left(x_{2}-x_{1}\right)^{2}+\left(y_{2}-y_{1}\right)^{2}}$
with $c^{\mathrm{r}}$ as defined in \eqref{micro_gaussian}.

The total number of synapses $S_{YX}^{\mathrm{u}}$ follows from \eqref{conn_prob_num_syn},
using connection probabilities and neuron numbers from the upscaled
models. This in turn yields the average number of incoming connections
to the target neurons, the synaptic in-degree, as $K_{YX}^{\mathrm{u}}=S_{YX}^{\mathrm{u}}/N_{Y}^{\mathrm{u}}$
. Connections in the upscaled model are drawn at random according
to the spatial profile (\eqref{spatial_profile}) and we fix only
the zero-distance connection probability $c_{YX}$ and the spatial
width $\sigma_{X}$, such that the upscaled in-degree \textrm{$K_{YX}^{\mathrm{u}}$}
is achieved. Under the assumption of a homogeneous distribution of
neurons and connections inside a disc with radius $R$ around a target
neuron, the local connection probability is then $c_{YX,R}=K_{YX}^{\mathrm{u}}/N_{X,R}^{\mathrm{u}}$,
where $N_{X,R}^{\mathrm{u}}$ denotes the number of potential source
neurons. We eliminate $N_{X,R}^{\mathrm{u}}$ from the expression
for $c_{YX,R}$ by relating neuron numbers to surface areas: $N_{X,R}^{\mathrm{u}}=N_{X}^{\mathrm{u}}\cdot A_{R}/A^{\mathrm{u}}$
with $A_{R}=\pi R^{2}$ and $A^{\mathrm{u}}=L^{2}$. To achieve the
same in-degree for the uniform connection probability $c\left(r\right)=c_{YX,R}\Theta\left(R-r\right)$
and the distance-dependent connection probability (\eqref{spatial_profile}),
the following volume integral in polar coordinates must be equal for
both choices of $c\left(r\right)$: $\int_{0}^{2\pi}\int_{0}^{\infty}\int_{0}^{c\left(r\right)}r\,\mathrm{d}z\mathrm{d}r\mathrm{d}\varphi$.
Due to isotropy, it is enough to equate $\int_{0}^{\infty}r\,c\left(r\right)\,\mathrm{d}r$
for both connection probabilities to derive the zero-distance connection
probability of the distance-dependent profile,
\begin{equation}
c_{YX}=\frac{K_{YX}^{\mathrm{u}}L^{2}}{2\pi\sigma_{X}^{2}N_{X}^{\mathrm{u}}\left[1-\exp\left(-\frac{R^{2}}{2\sigma_{X}^{2}}\right)\right]}.\label{eq:zero_distance_conn_prob}
\end{equation}
The connection routine used for the upscaled models does not fix the
total number of synapses, unlike the routine used for the reference
model. Each pair of neurons is considered only once in contrast to
the reference model which samples the neurons with replacement. If
$c_{YX}>1$, the routine is executed $N_{c}$ times with zero-distance
connection probabilities \textrm{$c_{YX}/N_{c}$ where $N_{c}=\left\lceil c_{YX}\right\rceil $.
In this case, a pair of neurons can be connected by up to $N_{c}$
synapses.}

\textit{Mean input: }To preserve the mean input to each neuron of
the reference network in the upscaled network, we adjust the in-degrees
of the external stationary Poisson input to compensate for differences
in internal in-degrees between the reference and the upscaled model
that result from the above calculation of recurrent synaptic in-degrees.
If the mean connection weight $g_{YX}\cdot J$ for internal connections,
the weight for external input $J$, the population firing rates $\nu_{X}$,
and the external Poisson rate $\nu_{\mathrm{ext}}$ are the same for
both models, the external in-degrees $K_{Y,\mathrm{ext}}^{\mathrm{u}}$
per population $Y$ of the upscaled model follow from the external
in-degrees of the reference model $K_{Y,\mathrm{ext}}^{\mathrm{r}}$
and the difference in internal in-degrees:

\begin{equation}
\begin{split}\sum_{X}K_{YX}^{\mathrm{u}}g_{YX}\nu_{X}+K_{Y,\mathrm{ext}}^{\mathrm{u}}\nu_{\mathrm{ext}}= & \sum_{X}K_{YX}^{\mathrm{r}}g_{YX}\nu_{X}+K_{Y,\mathrm{ext}}^{\mathrm{r}}\nu_{\mathrm{ext}}\\
K_{Y,\mathrm{ext}}^{\mathrm{u}}= & K_{Y,\mathrm{ext}}^{\mathrm{r}}+\sum_{X}\frac{g_{YX}\nu_{X}}{\nu_{\mathrm{ext}}}\left(K_{YX}^{\mathrm{r}}-K_{YX}^{\mathrm{u}}\right).
\end{split}
\end{equation}
This modification of external in-degrees in the upscaled network only
preserves the mean of the spiking input (which is proportional to
both in-degrees and weights), but not its variance (which is proportional
to in-degrees and to weights squared); see, for example, \citet{Brunel99,Albada15}
for details. 

\textit{Delays: }To compare the mean delays of the reference model
(\eqref{delay_distr_ref}) and mean delays resulting from linear distance
dependency in the upscaled model (\eqref{delay_distr_ups}), we compute
an effective delay for the upscaled model. The effective delay is
computed as the average delay of the distance-dependent version evaluated
on a disc of $\unit[1]{mm^{2}}$ (with radius $Q=\unit[1/\sqrt{\pi}]{mm}$),
thus equalling the extent of the reference model. Accounting for all
distances between random points on the disc, the effective delay in
polar coordinates for a disc of radius $Q$ is 

\begin{equation}
\bar{d}_{Q}\left(\sigma_{X}\right)=\frac{1}{\pi^{2}Q^{4}}\int_{0}^{Q}\int_{0}^{2\pi}\int_{0}^{Q}\int_{0}^{2\pi}\left(d_{0}+\frac{r_{21}}{v}\right)\,\frac{1}{c_{\mathrm{norm}}}\mathrm{e}^{-\frac{r_{21}^{2}}{2\sigma_{X}^{2}}}\,r_{1}r_{2}\,\mathrm{d}\varphi_{1}\mathrm{d}r_{1}\mathrm{d}\varphi_{2}\mathrm{d}r_{2}\label{eq:effective_delay}
\end{equation}
with $r_{21}=r_{1}^{2}+r_{2}^{2}-2r_{1}r_{2}\mathrm{cos\left(\varphi_{1}-\varphi_{2}\right)}$.
We here account for the Gaussian distance dependency of the spatial
profile \eqref{spatial_profile} with spatial width $\sigma_{X}$
but normalize the profile to unity for the integral over the disc
by the factor $c_{\mathrm{norm}}$, and ignore the Heaviside function
because we only consider $Q<R$. The expression simplifies \citep[Theorem 2.4]{Sheng85}
to
\begin{equation}
\begin{split}\bar{d}_{Q}\left(\sigma_{X}\right) & =\frac{\int_{0}^{2Q}\left[d_{0}+\frac{r}{v}\right]\exp\left(-\frac{r^{2}}{2\sigma_{X}^{2}}\right)r\left[4\arctan\left(\sqrt{\frac{2Q-r}{2Q+r}}\right)-\sin\left(4\arctan\left(\sqrt{\frac{2Q-r}{2Q+r}}\right)\right)\right]\mathrm{d}r}{\int_{0}^{2Q}\exp\left(-\frac{r^{2}}{2\sigma_{X}^{2}}\right)r\left[4\arctan\left(\sqrt{\frac{2Q-r}{2Q+r}}\right)-\sin\left(4\arctan\left(\sqrt{\frac{2Q-r}{2Q+r}}\right)\right)\right]\mathrm{d}r}\end{split}
,\label{eq:effective_delay_circle}
\end{equation}
which we evaluate numerically. Hence, the delay offset $d_{0}$ and
conduction speed $v$ can be set based on available experimental data,
and the mean delays in the upscaled network can be compared with the
corresponding excitatory and inhibitory mean delays of the reference
model.
\begin{table}
\begin{tabular}{|@{\hspace*{1mm}}p{0.2\linewidth}|@{\hspace*{1mm}}p{0.765\linewidth}|}
\hline 
\multicolumn{2}{|>{\color{white}\columncolor{black}}l|}{\textbf{A: Model summary}}\tabularnewline
\hline 
\textbf{Structure} & Multi-layer excitatory-inhibitory ($\mathrm{E}$-$\mathrm{I}$) network\tabularnewline
\hline 
\textbf{Populations} & $8$ cortical in $4$ layers ($\mathrm{L2/3}$, $\mathrm{L4}$, $\mathrm{L5}$,
$\mathrm{L6}$) and $1$ thalamic ($\mathrm{TC}$)\tabularnewline
\hline 
\textbf{Input } & Cortex: Independent fixed-rate Poisson spike trains to all neurons
(population-specific in-degree)\tabularnewline
\hline 
\textbf{Measurements} & Spikes, LFP, CSD, MUA\tabularnewline
\hline 
\textbf{Neuron model} & Cortex: leaky integrate-and-fire (LIF); Thalamus: point process\tabularnewline
\hline 
\textbf{Synapse model} & Exponentially shaped postsynaptic currents with normally distributed
static weights\tabularnewline
\hline 
\multicolumn{2}{|>{\color{black}\columncolor{lightgray}}c|}{\textbf{Reference model}}\tabularnewline
\hline 
\textbf{Topology} & None (no spatial information)\tabularnewline
\hline 
\textbf{Delay model } & Normally distributed delays\tabularnewline
\hline 
\textbf{Connectivity} & Random, independent, population-specific, fixed number of synapses\tabularnewline
\hline 
\multicolumn{2}{|>{\color{black}\columncolor{lightgray}}c|}{\textbf{Upscaled models}}\tabularnewline
\hline 
\textbf{Topology} & Random neuron positions on square domain of size $L\times L$; periodic
boundary conditions\tabularnewline
\hline 
\textbf{Delay model} & Distributed distance-dependent delays\tabularnewline
\hline 
\textbf{Connectivity} & Random, distance-dependent connection probability, population-specific,
number of synapses not fixed in advance\tabularnewline
\hline 
\end{tabular}

\begin{tabular}{|@{\hspace*{1mm}}p{0.2\linewidth}|@{\hspace*{1mm}}p{0.765\linewidth}|}
\hline 
\multicolumn{2}{|>{\color{white}\columncolor{black}}l|}{\textbf{B: Network models}}\tabularnewline
\hline 
\multirow{5}{*}{\textbf{Connectivity }} & Connection probabilities $C_{YX}$ from population $X$ to population
$Y$ with \newline $\{X,Y\}\in\left\{ \mathrm{L2/3,\,L4,\,L5,\,L6}\right\} \times\left\{ \mathrm{E,I}\right\} \cup\mathrm{TC}$,
$C_{YX}=0$ for $Y=\mathrm{TC}$\tabularnewline
\cline{2-2} 
 & \cellcolor{lightgray}\centering\textbf{Reference model}\tabularnewline
\cline{2-2} 
 & Fixed number of synapses $S_{YX}$ between populations $X$ and $Y$
(see \eqref{conn_prob_num_syn}), \newline binomially distributed
in-/out-degrees\tabularnewline
\cline{2-2} 
 & \cellcolor{lightgray}\centering\textbf{Upscaled models}\tabularnewline
\cline{2-2} 
 & \begin{itemize}\item Presynaptic neuron $j\in X$ at location $\left(x_{j},y_{j}\right)$
and postsynaptic neuron $i\in Y$ at $\left(x_{i},y_{i}\right)$ \item
Neuron inter-distance (periodic boundary conditions): \end{itemize}
\begin{equation}r_{ij}=\sqrt{\Delta x_{ij}^{2}+\Delta y_{ij}^{2}}\label{eq:distance_rij}\end{equation}
\begin{itemize}\item[] with $\Delta x_{ij}=\left|x_{i}-x_{j}\right|$
if \textrm{$\left|x_{i}-x_{j}\right|\leq L/2$, otherwise $\Delta x_{ij}=L-\left|x_{i}-x_{j}\right|$}\item[](same
for $\Delta y_{ij}$) \item Gaussian-shaped connection probability
with maximal distance $R$, spatial width $\sigma_{X}$ and zero-distance
connection probability $c_{YX}$ (see \eqref{zero_distance_conn_prob}):\end{itemize}
\begin{equation}c^\mathrm{u}\left(r_{ij}\right)=c_{YX}\,\mathrm{e}^{-r^{2}/2\sigma_{X}^{2}}\,\Theta\left(R-r_{ij}\right)\label{eq:spatial_profile}\end{equation}
\begin{itemize}\item[] Heaviside function $\Theta\left(t\right)=1$
for $t\geq0$, and $0$ otherwise.\end{itemize}\tabularnewline
\hline 
\end{tabular}

\caption{\textbf{Description of reference and upscaled network models following
the guidelines of }\citet{Nordlie-2009_e1000456}.\label{tab:model_summary}}
\end{table}

\begin{table}
\begin{tabular}{|@{\hspace*{1mm}}p{0.2\linewidth}@{}|@{\hspace*{1mm}}p{0.765\linewidth}|}
\hline 
\multicolumn{2}{|>{\color{white}\columncolor{black}}l|}{\textbf{C: Neuron models}}\tabularnewline
\hline 
\textbf{Cortex} & Leaky integrate-and-fire neuron (LIF) \begin{itemize}\item Dynamics
of membrane potential $V_{i}\left(t\right)$ for neuron $i$: \begin{itemize}\item
Spike emission at times $t_{s}^{i}$ with $V_{i}\left(t_{s}^{i}\right)\ge V_{\theta}$
\item Subthreshold dynamics:\end{itemize} \begin{equation}\tau_{\mathrm{m}}\dot{V}_{i}=-V_{i}+R_{\mathrm{m}}I_{i}\left(t\right)\quad\mathrm{if}\,\forall s:\,t\notin\left(t_{s}^{i},\,t_{s}^{i}+\tau_{\mathrm{ref}}\right]\quad\mathrm{with}\quad\tau_{\mathrm{m}}=R_{\mathrm{m}}C_{\mathrm{m}}\label{eq:lif_subthreshold}\end{equation}\begin{itemize}\item
Reset $+$ refractoriness: $V_{i}\left(t\right)=V_{\mathrm{reset}}\quad\mathrm{if}\,\forall s:\,t\in\left(t_{s}^{i},\,t_{s}^{i}+\tau_{\mathrm{ref}}\right]$
\end{itemize} \item Exact integration with temporal resolution $dt$
\citep{Rotter99a} \item Random, uniform distribution of membrane
potentials at $t=0$ \end{itemize}\tabularnewline
\hline 
\textbf{Thalamus} & Spontaneous activity: no thalamic input ($\nu_{\mathrm{TC}}=0$)\tabularnewline
\hline 
\multicolumn{2}{|>{\color{black}\columncolor{lightgray}}c|}{\textbf{Upscaled models}}\tabularnewline
\hline 
\textbf{Thalamus} & Thalamic pulses: coherent activation of all thalamic neurons inside
a circle with radius $R_{\mathrm{TC}}^{\mathrm{pulse}}$ centered
around $\left(0,0\right)$ at fixed time intervals $\Delta t_{\mathrm{TC}}$
\tabularnewline
\hline 
\end{tabular}

\begin{tabular}{|@{\hspace*{1mm}}p{0.2\linewidth}@{}|@{\hspace*{1mm}}p{0.765\linewidth}|}
\hline 
\multicolumn{2}{|>{\color{white}\columncolor{black}}l|}{\textbf{D: Synapse models}}\tabularnewline
\hline 
\textbf{Postsynaptic\newline currents} & \begin{itemize}\item Instantaneous onset, exponentially decaying
postsynaptic currents \item Input current of neuron $i$ from presynaptic
neuron $j$: \begin{equation}I_{i}\left(t\right)=\sum_{j}J_{ij}\sum_{s}\mathrm{e}^{-\left(t-t_{s}^{j}-d_{ij}\right)/\tau_{\mathrm{s}}}\Theta\left(t-t_{s}^{j}-d_{ij}\right)\label{eq:lif_psc}\end{equation}
\end{itemize}\tabularnewline
\hline 
\textbf{Weights} & \begin{itemize} \item Normal distribution with static weights, clipped
to preserve sign: \begin{equation}J_{ij}\sim\mathcal{N}\left\{ \mu=g_{YX}\cdot J,\,\sigma^{2}=\sigma_{\mathrm{J},YX}^2\right\}   \label{eq:weight_distr}\end{equation}
\item Probability density of normal distribution:\begin{equation}f\left(x|\mu,\sigma^{2}\right)=\frac{1}{\sqrt{2\pi\sigma^{2}}}\mathrm{e}^{-\frac{\left(x-\mu\right)^{2}}{2\sigma^{2}}}\label{eq:density_normal_distr}\end{equation}
\end{itemize}\tabularnewline
\hline 
\multicolumn{2}{|>{\color{black}\columncolor{lightgray}}c|}{\textbf{Reference model}}\tabularnewline
\hline 
\textbf{Delays} & Normal distribution, left-clipped at $dt$: \begin{equation}d_{ij}=d^\mathrm{r}_{ij}\sim\mathcal{N}\left\{ \mu=\bar{d}_X,\,\sigma^{2}=\left(\sigma_{\mathrm{d},X}^\mathrm{r}\right)^{2}\right\}\label{eq:delay_distr_ref}\end{equation}\tabularnewline
\hline 
\multicolumn{2}{|>{\color{black}\columncolor{lightgray}}c|}{\textbf{Upscaled models}}\tabularnewline
\hline 
\textbf{Delays} & Linear distance dependency with delay offset $d_{0}$ and conduction
speed $v$. Normally distributed additive noise, left-clipped at $-\left(d_{0}-dt\right)$
and right-clipped at $d_{0}-dt$: \begin{equation}d_{ij}=d^\mathrm{u}_{ij}\sim d_{0}+\frac{r_{ij}}{v} + \mathcal{N}\left\{ \mu=0,\,\sigma^{2}=\left(\sigma_\mathrm{d}^\mathrm{u}\right)^{2}\right\}\label{eq:delay_distr_ups}\end{equation}\tabularnewline
\hline 
\end{tabular}

\caption{\textbf{Description of reference and upscaled network models (continuation
of \tabref{model_summary})}. \label{tab:model_summary_cont}}
\end{table}

\begin{table}
\begin{tabular}{|@{\hspace*{1mm}}p{0.1\linewidth}@{}|@{\hspace*{1mm}}p{0.2\linewidth}|@{\hspace*{1mm}}p{0.665\linewidth}|}
\hline 
\multicolumn{3}{|>{\color{white}\columncolor{black}}l|}{\textbf{A: Global simulation parameters}}\tabularnewline
\hline 
\textbf{Symbol} & \textbf{Value} & \textbf{Description}\tabularnewline
\hline 
$T_{\mathrm{sim}}$ & $\unit[5,000]{ms}$ & Simulation duration\tabularnewline
$dt$ & $\unit[0.1]{ms}$ & Temporal resolution\tabularnewline
$T_{\mathrm{trans}}$ & $\unit[500]{ms}$ & Startup transient\tabularnewline
\hline 
\end{tabular}

\begin{tabular}{|@{\hspace*{1mm}}p{0.1\linewidth}@{}|@{\hspace*{1mm}}p{0.2\linewidth}|@{\hspace*{1mm}}p{0.665\linewidth}|}
\hline 
\multicolumn{3}{|>{\color{white}\columncolor{black}}l|}{\textbf{B: Preprocessing}}\tabularnewline
\hline 
\textbf{Symbol} & \textbf{Value} & \textbf{Description}\tabularnewline
\hline 
$\Delta t$ & $\unit[0.5]{ms}$ & Temporal bin size\tabularnewline
$\Delta l$ & $\unit[0.1]{mm}$ & Spatial bin size\tabularnewline
\hline 
\end{tabular}

\begin{tabular}{|@{\hspace*{1mm}}p{0.1\linewidth}@{}|@{\hspace*{1mm}}p{0.2\linewidth}|@{\hspace*{1mm}}p{0.665\linewidth}|}
\hline 
\multicolumn{3}{|>{\color{white}\columncolor{black}}l|}{\textbf{C: Global network parameters}}\tabularnewline
\hline 
\multicolumn{3}{|>{\color{black}\columncolor{lightgray}}c|}{\textbf{Connection parameters and external input}}\tabularnewline
\hline 
\textbf{Symbol} & \textbf{Value} & \textbf{Description}\tabularnewline
\hline 
$J$ & $\unit[87.81]{pA}$ & Reference synaptic strength. All synapse weights are measured in units
of $J$.\tabularnewline
$g_{YX}$ &  & Relative synaptic strengths:\tabularnewline
 & $1$ & $X\in\left\{ \text{\ensuremath{\mathrm{L2/3E,\,L4E,\,L5E,\,L6E,\,TC}}}\right\} $\tabularnewline
 & $-4$ & $\ensuremath{X\in\{\mathrm{L2/3I,\,L4I,\,L5I,\,L6I}\}}$, except for:\tabularnewline
 & $2$ & $\left(X,Y\right)=\left(\mathrm{L4E,\,L2/3E}\right)$\tabularnewline
$\sigma_{\mathrm{J},YX}$ & $0.1\cdot g_{YX}\cdot J$ & Standard deviation of weight distribution\tabularnewline
$\nu_{\mathrm{ext}}$ & $\unit[8]{s^{-1}}$ & Rate of external input with Poisson inter-spike interval statistics\tabularnewline
\hline 
\multicolumn{3}{|>{\color{black}\columncolor{lightgray}}c|}{\textbf{LIF neuron model}}\tabularnewline
\hline 
\textbf{Symbol} & \textbf{Value} & \textbf{Description}\tabularnewline
\hline 
$C_{\mathrm{m}}$ & $\unit[250]{pF}$ & Membrane capacitance \tabularnewline
$\tau_{\mathrm{m}}$ & $\unit[10]{ms}$ & Membrane time constant \tabularnewline
$E_{\mathrm{L}}$ & $\unit[-65]{mV}$ & Resistive leak reversal potential \tabularnewline
$V_{\theta}$ & $\unit[-50]{mV}$ & Spike detection threshold \tabularnewline
$V_{\mathrm{reset}}$ & $\unit[-65]{mV}$ & Spike reset potential \tabularnewline
$\tau_{\mathrm{ref}}$ & $\unit[2]{ms}$ & Absolute refractory period after spikes\tabularnewline
$\tau_{\mathrm{s}}$ & $\unit[0.5]{ms}$ & Postsynaptic current time constant \tabularnewline
\hline 
\end{tabular}

\caption{\textbf{Global simulation, preprocessing, and network parameters used
for both reference and upscaled network models.} \label{tab:parameters_ref_and_ups}}
\end{table}

\begin{table}
\begin{tabular}{|@{\hspace*{1mm}}p{0.01\linewidth}@{}@{\hspace*{1mm}}p{0.07\linewidth}|@{\hspace*{1mm}}p{0.07\linewidth}|@{\hspace*{1mm}}p{0.07\linewidth}|@{\hspace*{1mm}}p{0.07\linewidth}|@{\hspace*{1mm}}p{0.07\linewidth}|@{\hspace*{1mm}}p{0.07\linewidth}|@{\hspace*{1mm}}p{0.07\linewidth}|@{\hspace*{1mm}}p{0.07\linewidth}|@{\hspace*{1mm}}p{0.07\linewidth}|@{\hspace*{1mm}}p{0.1\linewidth}|}
\hline 
\multicolumn{11}{|>{\color{white}\columncolor{black}}l|}{\textbf{Additional network parameters for reference model}}\tabularnewline
\hline 
\multicolumn{11}{|>{\color{black}\columncolor{lightgray}}c|}{\textbf{Populations and external input}}\tabularnewline
\hline 
\multicolumn{1}{|l|}{\textbf{Symbol}} & \multicolumn{1}{l}{\textbf{Value}} & \multicolumn{1}{l}{} & \multicolumn{1}{l}{} & \multicolumn{1}{l}{} & \multicolumn{1}{l}{} & \multicolumn{1}{l}{} & \multicolumn{1}{l}{} & \multicolumn{1}{l}{} &  & \textbf{Description}\tabularnewline
\hline 
\multicolumn{1}{|l|}{$X$} & $\mathrm{L2/3E}$ & $\mathrm{L2/3I}$ & $\mathrm{L4E}$ & $\mathrm{L4I}$ & $\mathrm{L5E}$ & $\mathrm{L5I}$ & $\mathrm{L6E}$ & $\mathrm{L6I}$ & $\mathrm{TC}$ & Name\tabularnewline
\hline 
\multicolumn{1}{|l|}{$N_{X}^{\mathrm{r}}$} & $20,683$ & $5,834$ & $21,915$ & $5,479$ & $4,850$ & $1,065$ & $14,395$ & $2,948$ & $902$ & Size\tabularnewline
\hline 
\multicolumn{1}{|l|}{$K_{X,\mathrm{ext}}^{\mathrm{r}}$} & $1,600$ & $1,500$ & $2,100$ & $1,900$ & $2,000$ & $1,900$ & $2,900$ & $2,100$ & - & External\newline in-degree\tabularnewline
\hline 
\multicolumn{11}{|>{\color{black}\columncolor{lightgray}}c|}{\textbf{Connection probabilities}}\tabularnewline
\hline 
$C_{YX}^{\mathrm{r}}$ & \multicolumn{10}{c|}{from~$X$}\tabularnewline
 &  & $\mathrm{L2/3E}$ & $\mathrm{L2/3I}$ & $\mathrm{L4E}$ & $\mathrm{L4I}$ & $\mathrm{L5E}$ & $\mathrm{L5I}$ & $\mathrm{L6E}$ & $\mathrm{L6I}$ & $\mathrm{TC}$\tabularnewline
\cline{2-11} 
 & $\mathrm{L2/3E}$ & $0.1009$ & $0.1689$ & $0.0437$ & $0.0818$ & $0.0323$ & $0.0$ & $0.0076$ & $0.0$ & $0.0$\tabularnewline
\cline{2-11} 
 & $\mathrm{L2/3I}$ & $0.1346$ & $0.1371$ & $0.0316$ & $0.0515$ & $0.0755$ & $0.0$ & $0.0042$ & $0.0$ & $0.0$\tabularnewline
\cline{2-11} 
 & $\mathrm{L4E}$ & $0.0077$ & $0.0059$ & $0.0497$ & $0.1350$ & $0.0067$ & $0.0003$ & $0.0453$ & $0.0$ & $0.0983$\tabularnewline
\cline{2-11} 
to~$Y$ & $\mathrm{L4I}$ & $0.0691$ & $0.0029$ & $0.0794$ & $0.1597$ & $0.0033$ & $0.0$ & $0.1057$ & $0.0$ & $0.0619$\tabularnewline
\cline{2-11} 
 & $\mathrm{L5E}$ & $0.1004$ & $0.0622$ & $0.0505$ & $0.0057$ & $0.0831$ & $0.3726$ & $0.0204$ & $0.0$ & $0.0$\tabularnewline
\cline{2-11} 
 & $\mathrm{L5I}$ & $0.0548$ & $0.0269$ & $0.0257$ & $0.0022$ & $0.0600$ & $0.3158$ & $0.0086$ & $0.0$ & $0.0$\tabularnewline
\cline{2-11} 
 & $\mathrm{L6E}$ & $0.0156$ & $0.0066$ & $0.0211$ & $0.0166$ & $0.0572$ & $0.0197$ & $0.0396$ & $0.2252$ & $0.0512$\tabularnewline
\cline{2-11} 
 & $\mathrm{L6I}$ & $0.0364$ & $0.0010$ & $0.0034$ & $0.0005$ & $0.0277$ & $0.0080$ & $0.0658$ & $0.1443$ & $0.0196$\tabularnewline
\hline 
\end{tabular}

\begin{tabular}{|@{\hspace*{1mm}}p{0.1\linewidth}@{}|@{\hspace*{1mm}}p{0.2\linewidth}|@{\hspace*{1mm}}p{0.665\linewidth}|}
\hline 
\multicolumn{3}{|>{\color{black}\columncolor{lightgray}}c|}{\textbf{Connection Parameters}}\tabularnewline
\hline 
\textbf{Symbol} & \textbf{Value} & \textbf{Description}\tabularnewline
\hline 
$\overline{d}_{\mathrm{E}}$ & $\unit[1.5]{mm}$ & Mean excitatory delay\tabularnewline
$\overline{d}_{\mathrm{\ensuremath{I}}}$ & $\unit[0.75]{mm}$ & Mean inhibitory delay\tabularnewline
$\sigma_{\mathrm{d},X}^{\mathrm{r}}$ & $0.5\cdot\overline{d}_{\mathrm{X}}$ & Standard deviation of delay distribution\tabularnewline
\hline 
\end{tabular}

\caption{\textbf{Additional network parameters for the reference model.} \label{tab:parameters_ref}}
\end{table}

\begin{table}
\begin{tabular}{|@{\hspace*{1mm}}p{0.01\linewidth}@{}@{\hspace*{1mm}}p{0.07\linewidth}|@{\hspace*{1mm}}p{0.07\linewidth}|@{\hspace*{1mm}}p{0.07\linewidth}|@{\hspace*{1mm}}p{0.07\linewidth}|@{\hspace*{1mm}}p{0.07\linewidth}|@{\hspace*{1mm}}p{0.07\linewidth}|@{\hspace*{1mm}}p{0.07\linewidth}|@{\hspace*{1mm}}p{0.07\linewidth}|@{\hspace*{1mm}}p{0.07\linewidth}|@{\hspace*{1mm}}p{0.1\linewidth}|}
\hline 
\multicolumn{11}{|>{\color{white}\columncolor{black}}l|}{\textbf{Additional network parameters for the final upscaled model}}\tabularnewline
\hline 
\multicolumn{11}{|>{\color{black}\columncolor{lightgray}}c|}{\textbf{Populations and external input}}\tabularnewline
\hline 
\multicolumn{1}{|l|}{\textbf{Symbol}} & \multicolumn{1}{l}{\textbf{Value}} & \multicolumn{1}{l}{} & \multicolumn{1}{l}{} & \multicolumn{1}{l}{} & \multicolumn{1}{l}{} & \multicolumn{1}{l}{} & \multicolumn{1}{l}{} & \multicolumn{1}{l}{} &  & \textbf{Description}\tabularnewline
\hline 
\multicolumn{1}{|l|}{$X$} & $\mathrm{L2/3E}$ & $\mathrm{L2/3I}$ & $\mathrm{L4E}$ & $\mathrm{L4I}$ & $\mathrm{L5E}$ & $\mathrm{L5I}$ & $\mathrm{L6E}$ & $\mathrm{L6I}$ & $\mathrm{TC}$ & Name\tabularnewline
\hline 
\multicolumn{1}{|l|}{$N_{X}^{\mathrm{u}}$} & $330,928$ & $93,344$ & $350,640$ & $87,664$ & $77,600$ & $17,040$ & $230,320$ & $47,168$ & $14,432$ & Size\tabularnewline
\hline 
\multicolumn{1}{|l|}{$K_{X,\mathrm{ext}}^{\mathrm{u}}$} & $1,702$ & $1,621$ & $1,864$ & $2,443$ & $1,939$ & $1,724$ & $3,051$ & $2,246$ & - & External\newline in-degree\tabularnewline
\hline 
\multicolumn{11}{|>{\color{black}\columncolor{lightgray}}c|}{\textbf{Connection probabilities}}\tabularnewline
\hline 
$C_{YX}^{\mathrm{u}}$ & \multicolumn{10}{c|}{from~$X$}\tabularnewline
 &  & $\mathrm{L2/3E}$ & $\mathrm{L2/3I}$ & $\mathrm{L4E}$ & $\mathrm{L4I}$ & $\mathrm{L5E}$ & $\mathrm{L5I}$ & $\mathrm{L6E}$ & $\mathrm{L6I}$ & $\mathrm{TC}$\tabularnewline
\cline{2-11} 
 & $\mathrm{L2/3E}$ & $0.007540$ & $0.012622$ & $0.003266$ & $0.006113$ & $0.002414$ & $0.0$ & $0.000568$ & $0.0$ & $0.0$\tabularnewline
\cline{2-11} 
 & $\mathrm{L2/3I}$ & $0.010059$ & $0.010245$ & $0.002361$ & $0.003849$ & $0.005642$ & $0.0$ & $0.000314$ & $0.0$ & $0.0$\tabularnewline
\cline{2-11} 
 & $\mathrm{L4E}$ & $0.000575$ & $0.000441$ & $0.003714$ & $0.008575$ & $0.000501$ & $0.000022$ & $0.003385$ & $0.0$ & $0.007346$\tabularnewline
\cline{2-11} 
to~$Y$ & $\mathrm{L4I}$ & $0.005164$ & $0.000217$ & $0.005934$ & $0.013725$ & $0.000247$ & $0.0$ & $0.007899$ & $0.0$ & $0.004626$\tabularnewline
\cline{2-11} 
 & $\mathrm{L5E}$ & $0.007503$ & $0.004648$ & $0.003774$ & $0.000426$ & $0.006210$ & $0.029237$ & $0.001524$ & $0.0$ & $0.0$\tabularnewline
\cline{2-11} 
 & $\mathrm{L5I}$ & $0.004095$ & $0.002010$ & $0.001921$ & $0.000164$ & $0.003587$ & $0.021240$ & $0.000643$ & $0.0$ & $0.0$\tabularnewline
\cline{2-11} 
 & $\mathrm{L6E}$ & $0.001166$ & $0.000493$ & $0.001577$ & $0.001241$ & $0.004275$ & $0.001472$ & $0.002959$ & $0.016829$ & $0.003826$\tabularnewline
\cline{2-11} 
 & $\mathrm{L6I}$ & $0.002720$ & $0.000075$ & $0.000254$ & $0.000037$ & $0.002070$ & $0.000598$ & $0.004917$ & $0.010784$ & $0.001465$\tabularnewline
\hline 
\end{tabular}

\begin{tabular}{|@{\hspace*{1mm}}p{0.1\linewidth}@{}|@{\hspace*{1mm}}p{0.2\linewidth}|@{\hspace*{1mm}}p{0.665\linewidth}|}
\hline 
\multicolumn{3}{|>{\color{black}\columncolor{lightgray}}c|}{\textbf{Connection probability modifications}}\tabularnewline
\hline 
\textbf{Symbol} & \textbf{Value} & \textbf{Description}\tabularnewline
\hline 
$\delta C_{YX}$ & $0$ & $\left\{ X,Y\right\} \in\left\{ \text{\ensuremath{\mathrm{L2/3E,\,L2/3I,\,L4E,\,L4I,\,L5E,\,L5I,\,L6E,\,L6I,\,TC}}}\right\} $,
except for:\tabularnewline
 & $-0.15$ & $\left(X,Y\right)=\left(\mathrm{L4I,\,L4E}\right)$\tabularnewline
 & $0.15$ & $\left(X,Y\right)=\left(\mathrm{L4I,\,L4I}\right)$\tabularnewline
 & $-0.2$ & $\left(X,Y\right)=\left(\mathrm{L5E,\,L5I}\right)$\tabularnewline
 & $0.05$ & $\left(X,Y\right)=\left(\mathrm{L5I,\,L5E}\right)$\tabularnewline
 & $-0.1$ & $\left(X,Y\right)=\left(\mathrm{L5I,\,L5I}\right)$\tabularnewline
\hline 
\end{tabular}

\begin{tabular}{|@{\hspace*{1mm}}p{0.1\linewidth}@{}|@{\hspace*{1mm}}p{0.2\linewidth}|@{\hspace*{1mm}}p{0.665\linewidth}|}
\hline 
\multicolumn{3}{|>{\color{black}\columncolor{lightgray}}c|}{\textbf{Connection Parameters}}\tabularnewline
\hline 
\textbf{Symbol} & \textbf{Value} & \textbf{Description}\tabularnewline
\hline 
$d_{0}$ & $\unit[0.5]{ms}$ & Delay offset\tabularnewline
$v$ & $\unit[0.3]{mm/ms}$ & Conduction speed\tabularnewline
$\sigma_{\mathrm{d}}^{\mathrm{u}}$ & $\unit[0.1]{ms}$ & Width of jitter distribution for delay\tabularnewline
$\sigma_{\mathrm{E}}$ & $\unit[0.35]{mm}$ & Excitatory spatial width\tabularnewline
$\sigma_{\mathrm{I}}$ & $\unit[0.1]{mm}$ & Inhibitory spatial width\tabularnewline
\hline 
\end{tabular}

\begin{tabular}{|@{\hspace*{1mm}}p{0.1\linewidth}@{}|@{\hspace*{1mm}}p{0.2\linewidth}|@{\hspace*{1mm}}p{0.665\linewidth}|}
\hline 
\multicolumn{3}{|>{\color{black}\columncolor{lightgray}}c|}{\textbf{Thalamus}}\tabularnewline
\hline 
\textbf{Symbol} & \textbf{Value} & \textbf{Description}\tabularnewline
\hline 
$R_{\mathrm{TC}}^{\mathrm{pulse}}$ & $\unit[0.3]{mm}$ & $\mathrm{TC}$ neuron activation radius of disc around $\left(0,0\right)$,
all $\mathrm{TC}$ neurons in the disc are active during pulses\tabularnewline
$\sigma_{\mathrm{TC}}$ & $\unit[0.3]{mm}$ & Spatial width of $\mathrm{TC}$ neuron connections\tabularnewline
$\Delta t_{\mathrm{TC}}$ & $\unit[100]{ms}$ & Interval between thalamic pulses\tabularnewline
\hline 
\end{tabular}

\caption{\textbf{Additional network parameters for the final upscaled model.}
\label{tab:parameters_ups}}
\end{table}

\subsection{Forward modeling of extracellular potentials\label{subsec:forward_modeling}}

In the present study we use a now well-established method to compute
extracellular potentials from neuronal activity. The method relies
on multicompartment neuron modeling to compute transmembrane currents
\citep[see, for example,][]{DeSchutter09_260} and volume conduction
theory \citep{Nunez06_2ed,Einevoll13_37} which relates current sources
and electric potentials in space. Assuming a volume conductor model
that is linear (frequency-independent), homogeneous (the same in all
locations), isotropic (the same in all directions), and ohmic (currents
depend linearly on the electric field $\mathbf{E}$), as represented
by the scalar electric conductivity $\sigma_{\mathrm{e}}$, the electric
potential in location $\mathbf{r}\equiv(x,y,z)$ of a time-varying
point current with magnitude $I(t)$ in location $\mathbf{r'}$ is
given by
\begin{equation}
\phi(\textbf{r},t)=\frac{I(t)}{4\pi\sigma_{\text{e}}|\textbf{r}-\textbf{r'}|}.\label{eq:extracellular1-1}
\end{equation}
The potential is assumed to be measured relative to an ideal reference
at infinite distance from the source. Consider a set of transmembrane
currents of $n_{\mathrm{comp}}$ individual cylindrical compartments
indexed by $n$ in an $N-$sized population of cells indexed by $j$
with time-varying magnitude $I_{jn}^{\mathrm{m}}(t)$ embedded in
a volume conductor representing the surrounding neural tissue. The
extracellular electric potential is then calculated as the linear
sum
\begin{equation}
\phi(\textbf{r},t)=\sum_{j=1}^{N}\sum_{n=1}^{n_{\text{comp}}}\frac{I_{jn}^{\mathrm{m}}(t)}{4\pi\sigma_{\text{e}}}\int\frac{1}{|\textbf{r}-\textbf{r}_{jn}|}\,\mathrm{d}\textbf{r}_{jn}.\label{eq:extracellular2-1}
\end{equation}
The integral term here enters as we utilize the \textit{line-source}
approximation \citep{Holt99_169} which amounts to assuming a homogeneous
transmembrane current density per unit length and integrating \eqref{extracellular1-1}
along the center axis of each cylindrical compartment. The thick soma
compartments (with $n=1$) with magnitude $I_{j}^{\mathrm{m,soma}}(t)$,
however, are approximated as spherical current sources, which amounts
to combining Equations \ref{eq:extracellular1-1} and \ref{eq:extracellular2-1}
as \citet{Linden14}
\begin{align}
\phi(\textbf{r},t) & =\sum_{j=1}^{N}\frac{1}{4\pi\sigma_{\text{e}}}\left(\frac{I_{j}^{\mathrm{m,soma}}(t)}{|\textbf{r}-\textbf{r}_{j}^{\text{soma}}|}+\sum_{n=2}^{n_{\text{comp}}}\int\frac{I_{jn}^{\mathrm{m}}(t)}{|\textbf{r}-\textbf{r}_{jn}|}\,\mathrm{d}\textbf{r}_{jn}\right)\nonumber \\
 & =\sum_{j=1}^{N}\frac{1}{4\pi\sigma_{\text{e}}}\left(\frac{I_{j}^{\mathrm{m,soma}}(t)}{|\textbf{r}-\textbf{r}_{j}^{\text{soma}}|}+\sum_{n=2}^{n_{\text{comp}}}\frac{I_{jn}^{\mathrm{m}}(t)}{\Delta s_{jn}}\ln\left|\frac{\sqrt{h_{jn}^{2}+r_{\perp jn}^{2}}-h_{jn}}{\sqrt{l_{jn}^{2}+r_{\perp jn}^{2}}-l_{jn}}\right|\right).\label{eq:extracellular3-1}
\end{align}
Here, lengths of compartments $n$ of cells $j$ are denoted by \textrm{$\Delta s_{jn}$,
}perpendicular distances from the electrode point contact to the axis
of the line compartments by $r_{\perp jn}$, and longitudinal distances
measured from the start of the compartment by $h_{jn}$. The distances
$l_{jn}=\Delta s_{jn}+h_{jn}$ are measured longitudinally from the
end of the compartment. As the above denominators can be arbitrarily
small and cause singularities in the computed extracellular potential,
we set the minimum separation $|\textbf{r}-\textbf{r}_{j}^{\text{soma}}|$
or $r_{\perp jn}$ equal to the radius of the corresponding compartment. 

 The above equations assume point electrode contacts, while real
electrode contacts have finite extents. We employ the \textit{disc-electrode}
approximation \citep{Camunas-Mesa2013,Linden14,Ness2015}
\begin{equation}
\phi_{\text{disc}}(\mathbf{u},t)=\frac{1}{A_{S}}\iint_{S}\phi(\mathbf{u},t)\,\mathrm{d}^{2}r\approx\frac{1}{m}\sum_{h=1}^{m}\phi({\bf u}_{h},t)\label{eq:extracellular4-1}
\end{equation}
to approximate the averaged potential across the uninsulated contact
surface \citep{Robinson1968,Nelson08_141,Nelson10_2315,Ness2015}.
We average the potential (\eqref{extracellular3-1}) in $m=50$ randomized
locations $\mathbf{u}_{h}$ on each circular and flat contact surface
$S$ with surface area $A_{S}$ and radius $\unit[5]{\mu m}$. The
surface normal vector on the disc representing each contact is the
unit vector along the vertical $z-$axis. All forward-model calculations
are performed with the simulation tool LFPy (\href{https://lfpy.readthedocs.io}{https://lfpy.readthedocs.io},
\citealp{Linden14,Hagen18_biorxiv_281717}), which uses the NEURON
simulation software (\href{https://neuron.yale.edu}{https://neuron.yale.edu},
\citealp{Carnevale06,Hines09_1}) to calculate transmembrane currents.

\subsubsection{Modifications to the hybrid scheme\label{subsec:The-hybrid-scheme}}

Extracellular potentials from the point-neuron network models are
here calculated using a slightly modified version of the biophysics-based
hybrid scheme introduced by \citet{Hagen16}. The scheme combines
forward modeling of extracellular potentials, or more specifically
its low-frequency part termed the local field potential (LFP), from
spatially extended multicompartment neuron models described above
instead of point neurons. Point neurons cannot generate an extracellular
potential, as the sum of all in- and outgoing currents vanishes in
a point, in contrast to multicompartment neuron models, which can
account for in- and outgoing currents distributed in space. We refer
the reader to the Methods of \citet{Hagen16} for an in-depth technical
description of the implementation for randomly connected point-neuron
network models. Here, we only summarize its main steps and list the
main changes which allow accounting for extracellular potentials of
networks with distance-dependent connectivity and periodic boundary
conditions. This hybrid modeling scheme for extracellular potentials
combines the simplicity and efficiency of point-neuron network models
with multicompartment neuron models for LFP generation accounting
for the biophysical origin of extracellular potentials. As in \citet{Hagen16},
we assume that cortical network dynamics are well captured by the
point-neuron network, and implement the hybrid scheme as follows:
\begin{itemize}
\item Spike trains of individual point neurons are mapped to synapse activation
times on corresponding postsynaptic multicompartment neurons while
overall connection parameters are preserved, that is, the distribution
of delays, the mean postsynaptic currents, and the mean number of
incoming connections onto individual cells (in-degree). 
\item Each multicompartment neuron has its equivalent in the point-neuron
network and receives input spikes from presynaptic point neurons with
the same distribution as in the point-neuron network (the mean in-degree
of neurons in the network and the cell-type and layer specificity
of connections is preserved, as in \citet{Hagen16}. 
\item The multicompartment neurons are mutually unconnected, and synaptic
activations are translated into a distribution of transmembrane currents
that contributes to the total LFP.
\item Activity in multicompartment neuron models (and the corresponding
LFP) does not interact with other multicompartment neurons or the
activity in the point-neuron network model, that is, there are no
ephaptic interactions. 
\end{itemize}
The first version of the hybrid scheme implemented in hybridLFPy (\href{https://INM-6.github.com/hybridLFPy}{https://INM-6.github.com/hybridLFPy})
is developed for random networks such as the layered cortical microcircuit
model of \citet{Potjans14_785} that is our reference network. In
contrast to this reference model that contains no spatial information,
the upscaled models described in \subsecref{point_neuron_networks}
assign spatial coordinates to the neurons within each layer but ignore
information about cortical depth, and draw connections between neurons
with probabilities depending on lateral distance. Modifications to
the hybrid scheme to account for upscaled networks thus include:
\begin{itemize}
\item We use the lateral locations of the point neurons also for the multicompartment
neuron models, and assign population-dependent somatic depths as in
\citet{Hagen16}.
\item We record the spiking activity from all neurons in the point-neuron
network and associate each spike train to the corresponding neuron
ID.
\item Presynaptic neuron IDs are drawn for each multicompartment neuron
using the same distance-dependent probability rule as is used when
constructing the point-neuron network (the connectivity is thus statistically
reproduced). The same distance-dependent delay rule is also implemented
in the hybrid scheme, and can be set separately for each pair of populations.
\item We compute the extracellular potential at $100$ contact sites arranged
on a square regular grid with each contact separated by $\unit[400]{\mu m}$,
similar to the layout of the Blackrock \textquoteleft Utah\textquoteright{}
multi-electrode array. The local field potential is computed at the
center of layer $2/3$ ($\mathrm{L}2/3$). 
\item LFPy, which implements the above forward model and is used internally
in the hybrid scheme, accounts for periodic boundary conditions.
\end{itemize}

\subsubsection{Modifications to LFPy to account for periodic boundary conditions}

As the upscaling procedure of the $1\text{\ensuremath{\mathrm{mm}}}^{2}$
reference point-neuron network model incorporates periodic boundary
conditions, we modify the forward-model calculations in LFPy (\href{https://LFPy.readthedocs.io}{https://LFPy.readthedocs.io},
\citealp{Linden14,Hagen18_biorxiv_281717}) to also account for such
boundaries. The basic premise for this modification is that transmembrane
currents of a neuron positioned near the network layer boundary should
result in a fluctuation of the extracellular potential also due to
sources across the boundary. This is analogous to input from network
connections across the boundaries resulting from the distance-dependent
connectivity rule. Thus, for a current source located in location
$\mathbf{r}_{jn}=(x_{jn},y_{jn},z_{jn})$ the extracellular potential
in location $\mathbf{r}$ is computed as the sum
\begin{equation}
\phi(\mathbf{r},t)=\sum_{p=-M}^{M}\sum_{q=-M}^{M}\phi_{pq}(\mathbf{r},t),
\end{equation}
where $\phi_{pq}(\mathbf{r},t)$ corresponds to the extracellular
potential with horizontally shifted source coordinates $(x_{jn}+pL,y_{jn}+qL,z_{jn})$,
$L$ the network layer side length and $M=2$ a chosen integer setting
the number of \textquoteleft mirror\textquoteright{} sources to either
side.

\subsection{Statistical analysis}

As simulation output, we consider the spiking activity of the point-neuron
networks (\subsecref{network_descriptions}), and corresponding multi-unit
activity (MUA), LFP (\subsecref{forward_modeling}) and current-source
density (CSD) estimates. We use simulated output data only after an
initial time period of $T_{\mathrm{trans}}$ to avoid startup transients,
and compute all measures for the whole time interval of the following
simulation duration $T_{\mathrm{sim}}$. Parameters are given in Tables
\ref{tab:parameters_ref_and_ups}, \ref{tab:parameters_ref} and \ref{tab:parameters_ups}.

\subsubsection{Temporal binning of spike trains\label{subsec:temporal_binning}}

Spike times $t_{i}^{\mathrm{s}}$ of the point-neuron networks simulated
using temporal resolution $dt$ are assigned to bins with width $\Delta t$.
Temporally binned spike trains are used to compute pairwise spike-train
correlations and population-rate power spectral densities, and to
illustrate population-averaged rate histograms. The bin width $\Delta t$
is an integer multiple of the simulation resolution $dt$. The simulation
duration $T_{\mathrm{sim}}$ is an integer multiple of the bin width
such that the number of bins is $K=T/\Delta t$. Time bins have indices
\textrm{$k\in\{0,1,...,K-1\}$, spanning time points in }$t\in[k\Delta t,(k+1)\Delta t)$.

\subsubsection{Spatiotemporal binning of spike trains\label{subsec:spatiotemporal-binning}}

In order to compute the propagation speed of evoked activity in the
network, we perform a spatiotemporal binning operation of spiking
activity in the network. As introduced in \subsecref{network_descriptions},
neuron positions $\left(x_{i},y_{i}\right)$ of the point-neuron network
are randomly drawn with $\left\{ x_{i},y_{i}\right\} \in$$\left[-L/2,L/2\right)$
. We subdivide the spatial domain of each layer into square bins of
side length $\Delta l$ such that the integer numbers of bins along
the $x-$ and $y-$axis are $L_{\left\{ x,y\right\} }=L/\Delta l$.
The bin indices are $l_{\left\{ x,y\right\} }\in\{0,1,...,L_{\left\{ x,y\right\} }-1\}$,
spanning $\left\{ x,y\right\} \in[l_{\left\{ x,y\right\} }\Delta l-L/2,(l_{\left\{ x,y\right\} }+1)\Delta l-L/2)$.
Temporal bins of width $\Delta t$ are defined as above. We compute
for each population a spatially and temporally binned instantaneous
spike-count rate in units of $\mathrm{s^{-1}}$ as the number of spike
events from all neurons inside the spatial bin divided by $\Delta t$.

\subsubsection{Current-source density (CSD) analysis\label{subsec:current_source_density}}

We estimate the current-source density (CSD) using the kernel CSD
(kCSD) method introduced by \citealt{Potworowski12_541}. The CSD
is an estimate of the volume density of transmembrane currents nearby
each LFP measurement site (in units of current per volume). Based
on the Poisson equation in electrostatics,

\begin{equation}
\nabla(\sigma\nabla)\phi=-C,
\end{equation}
which relates the electric potential $\phi\equiv\phi(\mathbf{r})$,
conductivity $\sigma_{\mathrm{e}}\equiv\sigma_{\mathrm{e}}(\mathbf{r})$
(which is here assumed to be scalar as above), and current density
$C\equiv C(\mathbf{r})$, one can make the assumption that the measured
LFP at each electrode results from a sum of $M$ current sources distributed
across space. Similar to \citealt{Leski11_401,Potworowski12_541},
we consider the underlying CSD as a product
\begin{equation}
\widetilde{f}(x,y,z)=\widetilde{f}(x,y)H(z),
\end{equation}
where the term $\widetilde{f}(x,y)$ describes a spatial profile in
the horizontal $xy-$plane and $H(z)$ the step function along the
vertical $z-$axis,
\begin{equation}
H(z)=\begin{cases}
1 & -h\leq z\leq h,\\
0 & \text{otherwise}.
\end{cases}
\end{equation}
The variable $h$ denotes the half-thickness of the current-generating
region. Under the assumption of a linear (frequency-independent) and
homogeneous (equal in all locations) conductivity, it follows that
the electric potential in a location $(x,y,0)$ is
\begin{equation}
f(x,y,0)=\frac{1}{2\pi\sigma}\int\text{arcsinh}\bigg(\frac{2h}{\sqrt{(x-x')^{2}+(y-y')^{2}}}\bigg)\widetilde{f}(x,y)\,\mathrm{d}y'\mathrm{d}x'.
\end{equation}
We here choose to define $\widetilde{f}(x,y)$ in terms of 2D Gaussians
of the form
\begin{equation}
\widetilde{b}_{i}(x,y)=\exp\bigg(-\frac{(x-x_{i})^{2}+(y-y_{i})^{2}}{2\sigma_{R}^{2}}\bigg),
\end{equation}
resulting in
\begin{equation}
b_{i}(x,y)=\frac{1}{2\pi\sigma_{\mathrm{e}}}\int\text{arcsinh}\bigg(\frac{2h}{\sqrt{(x-x')^{2}+(y-y')^{2}}}\bigg)\widetilde{b}_{i}(x,y)\,\mathrm{d}y'\mathrm{d}x'.
\end{equation}
Introducing 
\begin{equation}
\phi(x,y)=\mathcal{A}\,C(x,y)=\sum_{j=1}^{M}a_{j}b_{j}(x,y),
\end{equation}
where $\mathcal{A}:\mathcal{\widetilde{F}\rightarrow F}$ is a linear
operator connecting electric potentials and the underlying sources,
the CSD is estimated as
\begin{equation}
C^{*}(x,y)=\mathbf{\widetilde{K}}^{T}(x,y)\cdot\mathbf{K}^{-1}\cdot\mathbf{V},
\end{equation}
which minimizes the norm $||\phi||^{2}=\sum_{i=1}^{M}|a_{i}|^{2}$.
Here $\mathbf{V}=[\phi_{1},\phi_{2},\dots,\phi_{N}]^{T}$ is the observed
LFP across channels, $\mathbf{\widetilde{K}}^{T}(x,y)=[\widetilde{K}_{1}(x_{1},y_{1},x,y),\widetilde{K}_{2}(x_{2},y_{2},x,y),\dots,\widetilde{K}_{N}(x_{N},y_{N},x,y)]$
and 
\begin{equation}
\mathbf{K}=\bigg[\begin{array}{ccc}
K(x_{1},y_{1},x_{1},y_{1}) & \cdots & K(x_{1},y_{1},x_{N},y_{N})\\
\vdots & \ddots & \vdots\\
K(x_{N},y_{N},x_{1},y_{1}) & \cdots & K(x_{N},y_{N},x_{N},y_{N})
\end{array}\bigg],
\end{equation}
defined in terms of the kernel functions $K(x,y,x',y')=\sum_{i=1}^{M}b_{i}(x,y)b_{i}(x',y')$
and cross-kernel functions $\widetilde{K}(x,y,x',y')=\sum_{i=1}^{M}b_{i}(x,y)\widetilde{b}_{i}(x',y')$.
See \citealt{Potworowski12_541} for details on the procedure. We
use the implementation of the 2D kCSD method available in Elephant
(Electrophysiology Analysis Toolkit, \href{https://github.com/neuralensemble/elephant}{https://github.com/neuralensemble/elephant}),
with default parameters $\sigma_{\mathrm{e}}=\unit[0.3]{S/m}$, $M=1000$,
$h=\unit[1]{mm},$ $\sigma_{R}^{2}=\unit[0.23]{mm}^{2}$, and return
the estimate at the space spanned by the LFP electrodes with resolution
$\text{\ensuremath{\unit[0.4]{mm}}}$.

\subsubsection{Calculation of MUA signal\label{subsec:multi_unit_activity}}

For each electrode contact point located in $\mathrm{L2/3}$, we compute
a signal representative of the so-called multi-unit activity (MUA)
signal that can be obtained from recordings of extracellular potentials
by high-pass filtering the signal ($\gtrsim500\mathrm{Hz}$), followed
by signal rectification, temporal smoothing, and downsampling \citep[see, for example,][]{Einevoll07_2174}.
In a biophysical modeling study \citep{Pettersen08_291} it is shown
that this signal is approximately linearly related to the firing rate
of the local population of neurons in the vicinity of the measurement
device. Neuron coordinates $\left(x_{i},y_{i}\right)$ of the upscaled
point-neuron network are randomly drawn on the interval $\left\{ x_{i},y_{i}\right\} \in$$\left[-L/2,L/2\right)$.
We subdivide the layers into square bins of side length $\Delta l_{\mathrm{MUA}}=\unit[0.4]{mm}$
resulting in $10$ bins along the $x-$ and $y-$axis, respectively.
Each electrode contact point is located at the center of the respective
bin. We also define temporal bins of width $\Delta t$. We then compute
for each population a spatially and temporally binned spike-count
rate in units of $\mathrm{s^{-1}}$ by summing the number of spike
events from all neurons inside the spatial bin and divide by the width
of the temporal bin $\Delta t$. We then define the MUA signal as
the sum of the per-bin contributions of the populations $\mathrm{L2/3E}$
and $\mathrm{L2/3I}$. 

\subsubsection{Visual analysis}

The \textit{spike raster} diagrams or dot displays show information
on spiking activity. Each dot marks a spike event, and the dot position
along the horizontal axis denotes the time of the event. Spike data
of different neuron populations are stacked and the number of neurons
shown is proportional to the population size. Within each population,
neurons are sorted according to their lateral $x-$position and arranged
accordingly on the vertical axis of the dot display. 

We compute \textit{population-averaged rate histograms} by deriving
the per-neuron spike rates in time bins $\Delta t$ and units of $\mathrm{s^{-1}}$,
averaged over all neurons per population within the center disc of
$\unit[1]{mm^{2}}$. The corresponding histogram shows the rates in
a time interval of $\pm\unit[25]{ms}$ around the occurrence of a
thalamic pulse. Such a display is comparable to the Peri-Stimulus
Time Histogram (PSTH, \citealp{Perkel67a}) that typically shows the
spike count summed over different neurons or trials versus binned
time.

Image plots with color bars can have a linear or a logarithmic scaling\textit{
}as specified in the respective captions. Since values of the distance-dependent
cross-correlation functions can be positive or negative, we plot these
with linear scaling up to a threshold, beyond which the scaling is
logarithmic.

\subsubsection{Statistical measures\label{subsec:statistical_measures}}

\textit{Per-neuron spike rates} $\nu$ are defined as the number of
spikes per neuron during each simulation divided by the simulation
duration $T_{\mathrm{sim}}$. Distributions of per-neuron spike rates
are computed from all spike trains of each population separately for
an interval from $0$ to $\unit[30]{s^{-1}}$ using bins of width
$\unit[1]{s^{-1}}$. Histograms are normalized such that the cumulative
sum over the histogram equals unity. We define the mean rate per population
$\overline{\nu}$ as the arithmetic mean of all per-neuron spike rates
of each population.

The \textit{coefficient of local variation} $LV$ is a measure of
spike-train irregularity computed from a sequence of length $n$ of
consecutive inter-spike intervals $T_{i}$ \citep[Equation 2.2]{Shinomoto03_2823},
defined as

\begin{equation}
LV=\frac{1}{n-1}\sum_{i=1}^{n-1}\frac{3\left(T_{i}-T_{i+1}\right)^{2}}{T_{i}+T_{i+1}}.\label{eq:lv}
\end{equation}
Like the conventional coefficient of variation $CV$ \citep[Equation 2.1]{Shinomoto03_2823},
a sequence of intervals generated by a stationary Poisson process
results in a value of unity, but the $LV$ statistic is less affected
by rate fluctuations compared to the $CV$; thus, a non-stationary
Poisson process should result in $LV\approx1$. We compute the $LV$
from the inter-spike intervals of the spike trains of all neurons
within each population. Distributions of $LV$s are computed using
bins of width $0.1$, and histograms are normalized such that the
cumulative sum over the histogram equals unity. We define the mean
$LV$ per population $\overline{LV}$ as the arithmetic mean of all
$LV$s of each population.

The \textit{Pearson (product-moment) correlation coefficient} $CC$
is a measure of synchrony that is defined for two signals $u$ und
$v$ as

\begin{equation}
CC_{uv}=\frac{\mathrm{cov}(u,v)}{\sqrt{\mathrm{cov}(u,u)\,\mathrm{cov}(v,v)}},\label{eq:cc}
\end{equation}
with the covariance denoted by $\mathrm{cov}$. The calculation is
implemented using \texttt{numpy.corrcoef}. To compute distributions
of correlation coefficients from spike trains, we randomly select
$1000$ neurons per population and assign their spike times to temporal
bins with width $\Delta t_{CC}=\unit[5]{ms}$ (see \subsecref{temporal_binning}).
Then, we compute pairwise $CC$s for the spike counts $u=n_{i}$ and
$v=n_{j}$ of selected neurons $i$ from a population $X$ and neurons
$j$ from a population $Y$ (ignoring autocorrelations). Within each
population, meaning $X=Y$, the $CC$ is denoted by $E-E$ for an
excitatory population or $I-I$ for an inhibitory population. Correlations
between neurons from the excitatory and the inhibitory population
in each layer are denoted by $E-I$. $CC$ histograms have bins of
width $0.003$, are restricted to a range with a minimum and maximum
$CC$ of $\pm0.08$, respectively, and are normalized such that the
integral over the histogram equals unity. We also compute correlation
coefficients for assessing the distance dependency of spikes, LFP,
CSD, and MUA signals. In these cases, $u$ and $v$ are LFP, CSD,
or MUA time series in different spatial locations. For spikes, we
sample $40$ excitatory and $10$ inhibitory spike trains, bin them
as above, compute their correlation coefficients (ignoring autocorrelations),
and plot them according to distance between the pairs of neurons.

\textit{Coherences} are computed as
\begin{equation}
\gamma_{uv}(f)=\frac{|\mathcal{S}_{uv}(f)|}{\sqrt{\mathcal{S}_{uu}(f)\mathcal{S}_{vv}(f)}},
\end{equation}
where $\mathcal{S}_{uv}(f)$ is the cross-spectral density between
$u$ and $v$, and $\mathcal{S}_{uu}(f)$ and $\mathcal{S}_{vv}(f)$
are the power spectral densities (PSDs) of each signal. The \textit{cross-spectral
density} and \textit{power spectra} are computed using Welch's average
periodogram method \citep{Welch1967} as implemented by \texttt{matplotlib.mlab}'s
\texttt{csd} and \texttt{psd} functions, respectively, with number
of data points used in each block for the fast Fourier transform (FFT),
that is, segment length $N_{\mathrm{FFT}}=256$, overlap between segments
$N_{\mathrm{overlap}}=192$ and signal sampling frequency $F_{\mathrm{s}}=\unit[2]{kHz}$.
To compute the \textit{population-rate power spectral density}, we
use the spike trains of all neurons per population (in \figref{statistics_ref_interm_ups}N
and H only within the center disc of $\unit[1]{mm^{2}}$), resampled
into bins of size $\Delta t$, and with the arithmetic mean of the
binned spike trains subtracted.

The effect of thalamic pulses is analyzed by means of \textit{distance-dependent
cross-correlation functions }$CC^{\nu}\left(\tau,r\right)$ evaluated
for time lags $\tau$. We discretize the network of size $L\times L$
into an even number of square bins of side length $\Delta l$ . The
spike trains from all neurons within each spatial bin are resampled
into time bins of size $\Delta t$ and averaged across neurons to
obtain spatially and temporally resolved per-neuron spike rates. We
select spatial bins on the diagonals of the network such that each
distance to the center with coordinates $\left(0,0\right)$ is represented
by four bins. For $14$ distances from consecutive spatial bins along
the diagonal, we compute the temporal correlation function between
the rates in the respective spatial bins with a binary vector containing
ones at spike times of the thalamic pulses and zeros elsewhere, and
then average over the four spatial bins at equal distance. The sequences
are normalized by subtracting their mean and dividing by their standard
deviation. Correlations between the sequences $u$ and \textbf{$v$
}with time steps $k$ and the length of the sequences $K$ are then
computed as

\begin{equation}
CC_{u,v}\left(\tau\right)=\frac{1}{K}\sum_{k=1}^{K}u_{k+\tau}v_{k}
\end{equation}
for $\tau\in\left[-25,25\right]\,\mathrm{ms}$ in steps of $\Delta t$.
Finally, we subtract the baseline correlation value, obtained by averaging
over all negative time lags (before thalamic activation at $\tau=0$),
and get $CC^{\nu}\left(\tau,r\right)$. 

To estimate the \textit{propagation speed} $v_{\mathrm{prop}}$ from
the cross-correlation functions, we find for each distance the time
lag corresponding to the largest $CC^{\nu}$. Values of $CC^{\nu}$
smaller than $\unit[10]{\%}$ of the maximum of all $CC^{\nu}$ per
population across distances and time lags are excluded. We further
exclude distances smaller than the thalamic radius $R_{\mathrm{TC}}^{\mathrm{pulse}}$
plus the spatial width of thalamic connections $\sigma_{\mathrm{TC}}$
because a large part of neurons within this radius are simultaneously
receiving spikes directly from thalamus upon thalamic pulses. A linear
fit for the distance as function of time lag, $r_{p}\left(\tau\right)=r_{p,0}+v_{p}\cdot\tau$,
yields the speed $v_{p}$ and its fitting error, the standard deviation
$\sigma_{\mathrm{v,p}}$. We compute the speed for different populations
$p$ and obtain the propagation speed as weighted mean with its uncertainty:
\begin{equation}
v_{\mathrm{prop}}=\frac{\sum_{p}v_{p}/\sigma_{\mathrm{v,p}}^{2}}{\sum_{p}1/\sigma_{\mathrm{v,p}}^{2}},\qquad\sigma_{\mathrm{v,prop}}=\sqrt{\frac{1}{\sum_{p}1/\sigma_{\mathrm{v,p}}^{2}}}.
\end{equation}

\subsubsection{Curve fitting\label{subsec:curve_fitting}}

For certain measures, such as pairwise correlation coefficients computed
for different distances between LFP electrode locations, we fit exponential
functions of the form
\begin{equation}
y(r)=a\cdot\mathrm{e}^{-r/b}+c,
\end{equation}
where $\beta=(a,b,c)$ are constant parameters that minimize the sum
$\sum_{i=1}^{m}\left|y_{i}(r_{i})-y(r_{i},\beta))\right|^{2}$ for
the $m$ data points $y_{i}$ computed for distance $r_{i}$. The
parameter fitting is implemented using the non-linear least squares
function \texttt{curve\_fit} provided by the \texttt{scipy.optimize}
module, with initial guess $\beta=(0.1,0.1,0.1)$. Goodness of fit
is quantified by the coefficient of determination, defined as
\begin{equation}
\mathrm{R}^{2}=1-\frac{\sum_{i=1}^{m}\left(y_{i}(r_{i})-y(r_{i},\beta)\right){}^{2}}{\sum_{i=1}^{m}\left(y_{i}(r_{i})-\overline{y}\right){}^{2}},
\end{equation}
where $\overline{y}$ is the mean of the observed data.

\subsection{Software accessibility\label{subsec:implementation}}

We here summarize the details of software and hardware used to generate
the results presented throughout this study. Point-neuron network
simulations are implemented using the SLI interface of NEST v2.12.0
\citep{Nest2120}, and Python v2.7.11. We use the same network implementation
for reference and all upscaled models and switch between them by adjusting
parameters. Parameter scans rely on the parameters module of NeuroTools
(\href{http://neuralensemble.org/NeuroTools/}{http://neuralensemble.org/NeuroTools/}).
LFP signals are computed using NEURON v7.5 and LFPy from \href{http://lfpy.github.io/}{http://lfpy.github.io/}
(branch \textquoteleft som\_as\_point\_periodic\textquoteright{} at
SHA:4cab667), hybridLFPy (\href{https://github.com/INM-6/hybridLFPy}{https://github.com/INM-6/hybridLFPy},
branch \textquoteleft LFPy\_dev\textquoteright{} at SHA:0f1bfb2).
Analysis and plotting rely on Python with numpy v1.10.4, SciPy v0.17.0,
and matplotlib v2.1.2. All simulations and analyses are conducted
on the JURECA supercomputer (\href{http://www.fz-juelich.de/ias/jsc/EN/Expertise/Supercomputers/JURECA/JURECA_node.html}{http://www.fz-juelich.de/ias/jsc/EN/Expertise/Supercomputers/JURECA/JURECA\_node.html})
based on Intel Xeon E5-2680 v3 Haswell CPUs running the CentOS 7 Linux
distribution. Simulations are run using $1152$ and $2304$ physical
cores for the network and LFP simulations, respectively. All source
codes to reproduce these results and figures will be made publicly
available upon final publication of this manuscript.

\section{Results\label{sec:results}}

\subsection{Upscaling of a cortical microcircuit model using lateral distance-dependent
connectivity\label{subsec:upscaling_results}}

\begin{figure}
\begin{centering}
\includegraphics[width=1\textwidth]{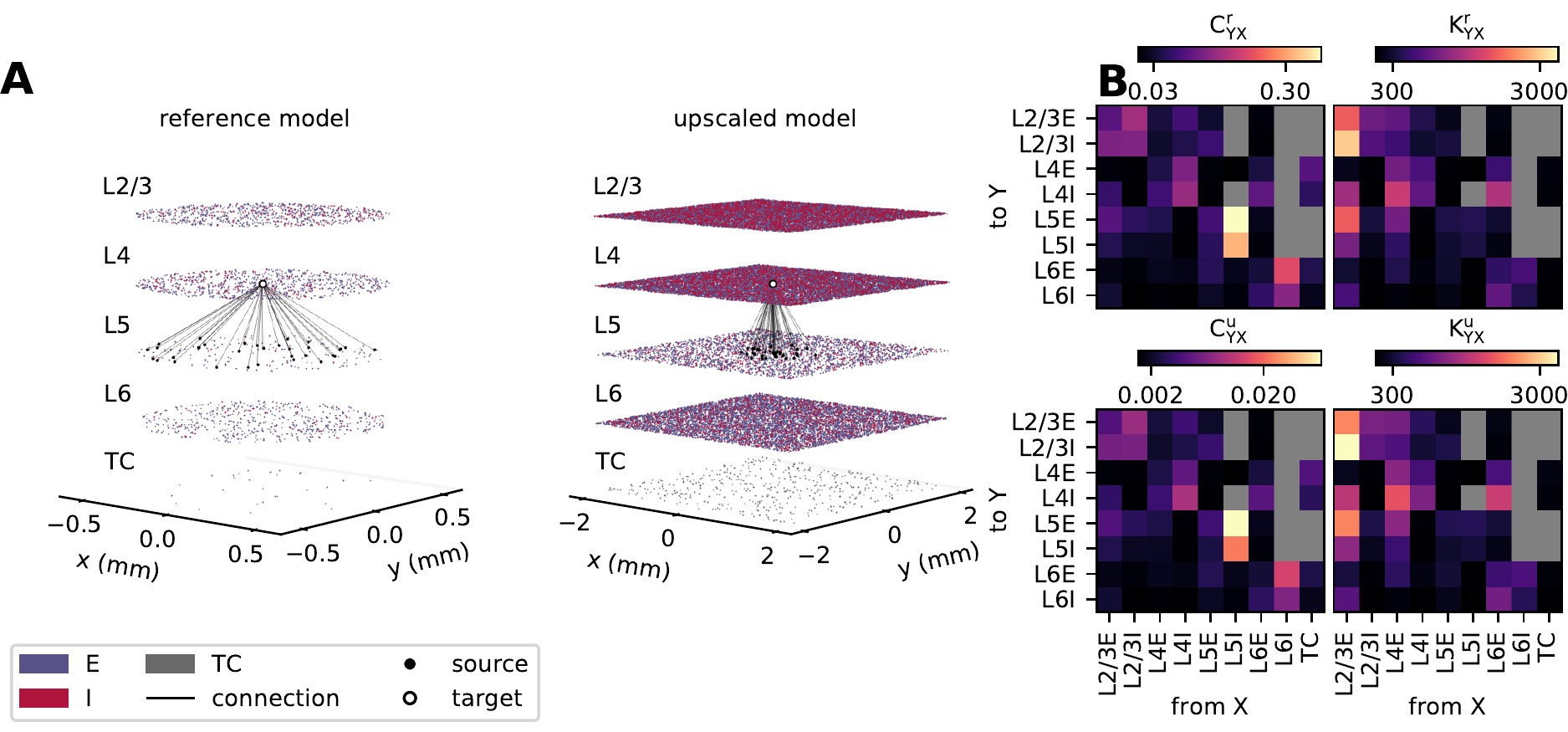}
\par\end{centering}
\begin{centering}
\par\end{centering}
\centering{}\caption{\textbf{Layered cortical point-neuron network models.} \textbf{A}~Illustrations
of the network geometry of the reference model (left, $\unit[1]{mm^{2}}$
cortical microcircuit, introduced by \citealp{Potjans14_785}) and
an upscaled model (right, $\unit[4\times4]{mm^{2}}$ cortical layers).
Both models consist of four cortical layers ($\mathrm{L2/3}$, $\mathrm{L4}$,
$\mathrm{L5}$, $\mathrm{L6}$) with an excitatory ($\mathrm{E}$)
and an inhibitory ($\mathrm{I}$) population each, and an external
thalamic population ($\mathrm{TC}$). Colored dots represent individual
neurons at their $\left(x,y\right)-$coordinates; excitatory neurons
in blue, inhibitory neurons in red, and thalamic neurons in gray.
The number of neurons shown per population is reduced by a factor
$32$ compared to the actual neuron number in each network to not
saturate the illustrated layers. Black lines illustrate convergent
connections from sources in $\mathrm{L5E}$ (black dots) to a target
neuron in $\mathrm{L4E}$ (white dot). In-degrees correspond to the
actual average in-degrees in both models rounded to the nearest integer:
$33$ in the reference model and $39$ in the upscaled model. Sources
are drawn at random in the reference model, but with lateral distance
dependency (Gaussian-shaped profile) in the upscaled model. \textbf{B}~Network
connectivity of the reference model (top panels) and the upscaled
model (bottom panels). Upscaled connection probabilities are computed
as in \eqref{conn_prob_with_modification}. Left panels show color-coded
connection probabilities $C_{YX}^{\mathrm{r}}$ and $C_{YX}^{\mathrm{u}}$
(different color code) with the values given in Tables \ref{tab:parameters_ref}
and \ref{tab:parameters_ups}, and right panels show derived in-degrees
$K_{YX}^{\mathrm{r}}$ and $K_{YX}^{\mathrm{u}}$ (same color code).
Color maps have linear scaling with zero-values masked in gray. \label{fig:model_sketch} }
\end{figure}

Starting with a model of the cortical microcircuit (the reference
model, see \citealp{Potjans14_785}), we construct full-scale multi-layer
neuronal network models with distance-dependent connectivity via the
upscaling procedure described in \subsecref{upscaling_procedure}.
The full network descriptions are provided in \subsecref{network_descriptions}
and in Tables \ref{tab:model_summary} and \ref{tab:model_summary_cont}.
Here, we point out similarities and differences between the reference
model and an upscaled model with parameters set to the values given
in Tables \ref{tab:parameters_ref_and_ups}-\ref{tab:parameters_ups}.
We refer to this parameterization as the \textquoteleft base parameters\textquoteright .
\figref{model_sketch}A  illustrates the reference model next to
the laterally upscaled version. The reference model comprises almost
$80,000$ neurons under $\unit[1]{mm^{2}}$ of cortical surface area,
while the upscaled model consists of approximately $1.2$ million
neurons and covers an area of $\unit[4\times4]{mm^{2}}$, similar
to the the area covered by the Utah multi-electrode array. To illustrate
their connectivities, the figure shows in both network sketches incoming
connections from population $\mathrm{L5E}$ to an example target neuron
in population $\mathrm{L4E}$. In the reference model without spatial
structure, source neurons are picked randomly from the source population.
In the upscaled model, source neurons are picked around the target
neuron in layer $4$ according to distance-dependent probabilities
with Gaussian profiles of outgoing connections from layer $5$ excitatory
neurons. The width of the profile is $\unit[0.3]{mm}$ which is
the average value $\sigma_{0}$ from the connectivity data underlying
the reference model, see \eqref{micro_gaussian}. A major fraction
of source neurons falls into the center $\unit[1]{mm^{2}}$, justifying
the assumption of random connectivity in the reference model.

The population-specific connection probabilities in the reference
model $C_{YX}^{\mathrm{r}}$, shown in \figref{model_sketch}B, are
equal to those in \citet[Table 5]{Potjans14_785}. The upscaling procedure
yields connection probabilities $C_{YX}^{\mathrm{u}}$ that are decreased
by approximately one order of magnitude in comparison to the reference
model.  The derived in-degrees $K_{YX}^{\mathrm{u}}$, however,
are slightly larger than $K_{YX}^{\mathrm{r}}$ for all population
pairs. This is expected since the upscaling procedure adds connections
at distances not accounted for within the limited extent of the reference
model.

For the final upscaled model, we increase the excitatory and decrease
the inhibitory spatial widths of the connection probability profiles
(\eqref{spatial_profile}) compared to the average value $\sigma_{0}$
of the reference model to $\sigma_{\mathrm{E}}=\unit[0.35]{mm}$ and
$\sigma_{\mathrm{I}}=\unit[0.1]{mm}$, respectively. Accumulating
experimental data indicate Gaussian or exponentially decaying connection
probabilities with distance for both excitatory and inhibitory local
connections;see, for example, the review by \citet{Boucsein11_1},
or \citet{Hellwig00_111} for pyramidal cells in layers $2$ and $3$
of rat visual cortex, \citet{Budd01} for clutch cells in layer $4$
of cat visual cortex, \citet{Perin11} for pyramidal cells in layer
$5$ of rat somatosensory cortex, \citet{Levy2012_5609} for pyramidal
cells and (non-)fast-spiking inhibitory cells in deep layer $2/3$
and layer $4$ of mouse auditory cortex, \citet{Schnepel15_3818}
for excitatory input to pyramidal neurons in layer $\mathrm{5B}$
of rat somatosensory cortex, \citet{Jiang2015} for pyramidal cells
and different interneurons in layers $1$, $2/3$, and $5$ of mouse
visual cortex, \citet{Packer2011_13260} for parvalbumin-positive
cells connected to pyramidal cells in multiple layers of mouse neocortex,
and \citet{Reimann17} for morphologically classified cell types in
an anatomical reconstruction and simulation of a rat hindlimb somatosensory
cortex column \citep{Markram2015_456}. Such profiles result largely
from the axo-dendritic overlap of the neuronal morphologies \citep{Amirikian05,Brown09_1133,Hill12_E2885}.
Broader excitation than inhibition is in line with the experimental
data since excitatory neurons, in particular pyramidal types, develop
axons with larger horizontal reach compared to most inhibitory interneuron
types \citep{Budd01,Binzegger04,Buzas06_861,Binzegger07_12242,Stepanyants2008_13,Stepanyants09_3555,Ohana12_e40601}.
Certain interneuron types may, however, have elaborate axons that
span and form synapses across different layers within the cortical
column \citep[see, for example,][Figure 2]{Markram2015_456}. Others
may also form longer-range lateral connections \citep{McDonald93}.

The chosen value for the conduction speed $v=\unit[0.3]{mm/ms}$
is in the range of speeds reported for action potential propagation
along unmyelinated nerve fibers in cortex. Conduction speeds can be
measured, for example, in brain slices using electrical stimulation
combined with electrophysiological recordings: $\unit[0.2-0.35]{mm/ms}$
in guinea pig hippocampus \citep{Andersen78_11}, $1/\left(\unit[3.5]{ms/mm}\right)\approx\unit[0.29]{mm/ms}$
at $34-35{^\circ}\mathrm{C}$ in cat visual cortex \citep{Hirsch91_1800},
$\unit[0.3]{mm/ms}$ at $35{^\circ}\mathrm{C}$ in rat hippocampus
\citep{Berg-Johnsen92_319}, $\unit[0.15-0.55]{mm/ms}$ at $31\pm0.5{^\circ}\mathrm{C}$
in rat visual cortex \citep{Murakoshi93_211}, $\unit[0.28-0.48]{mm/ms}$
(mean $\pm$ standard deviation, $\unit[0.37\pm0.37]{mm/ms}$) at
$35{^\circ}\mathrm{C}$ in cat motor cortex \citep{Kang94_280}, $\unit[0.28\pm0.19]{mm/ms}$
at $34{^\circ}\mathrm{C}$ in rat visual cortex \citep{Lohmann94},
$\unit[0.06-0.2]{mm/ms}$ at $34-35{^\circ}\mathrm{C}$ in rat somatosensory
cortex \citep{Salin96_1589}, $\unit[0.508]{mm/ms}$ at $32-35{^\circ}\mathrm{C}$
in rat somatosensory cortex \citep[back-propagating action-potentials in dendrites]{Larkum01_447},
and $\unit[0.34-0.44]{mm/ms}$ at $34\pm1{^\circ}\mathrm{C}$ in rat
somatosensory cortex. Some of these values are likely underestimated
because the separation of conduction speed from both the synaptic
delay and spike initiation time is difficult \citep{Hirsch91_1800}.
The bath temperature is provided if specified by the study because
the conduction speed and the timing of synaptic processing depend
strongly on environmental temperature \citep{Katz65_656,Berg-Johnsen92_319,Sabatini96_170,Hardingham98_249}.
We are here primarily interested in physiologically relevant body
temperatures. Connections in the upscaled models have a delay offset
$d_{0}=\unit[0.5]{ms}$ comparable to the experimental estimates
$\unit[0.5-1]{ms}$ \citep{Murakoshi93_211}, $\unit[0.6-0.8]{ms}$
\citep{Hirsch91_1800} and $\unit[0.6]{ms}$ \citep{Kang94_280}.
To account for this variability in experimental data the delays have
an additive normally distributed random component, see \eqref{delay_distr_ups}.
From a theoretical perspective, a wide delay distribution expands
the region of stability in the phase space of stationary network activity
\citep[Section 5.2]{Brunel00_183}.

Although delay offset and conduction speed have the same parameter
values for excitatory and inhibitory connections in the upscaled model,
the effective delays (\eqref{effective_delay_circle}) within a given
surface area differ due to the different space constants of the connectivity.
Computing the mean delay for connections within a circle of $\unit[1]{mm^{2}}$
with the respective spatial widths according to \eqref{effective_delay_circle}
results in a shorter mean delay for inhibitory connections. The effective
excitatory and inhibitory delays up to single decimal precision are
$\unit[1.6]{ms}$ and $\unit[0.9]{ms}$, respectively. Hence, a shorter
inhibitory delay in a network model without distance dependence like
the reference model is justified by a narrower inhibitory connectivity
of the corresponding model with spatial structure.

\begin{figure}
\begin{centering}
\includegraphics[width=1\textwidth]{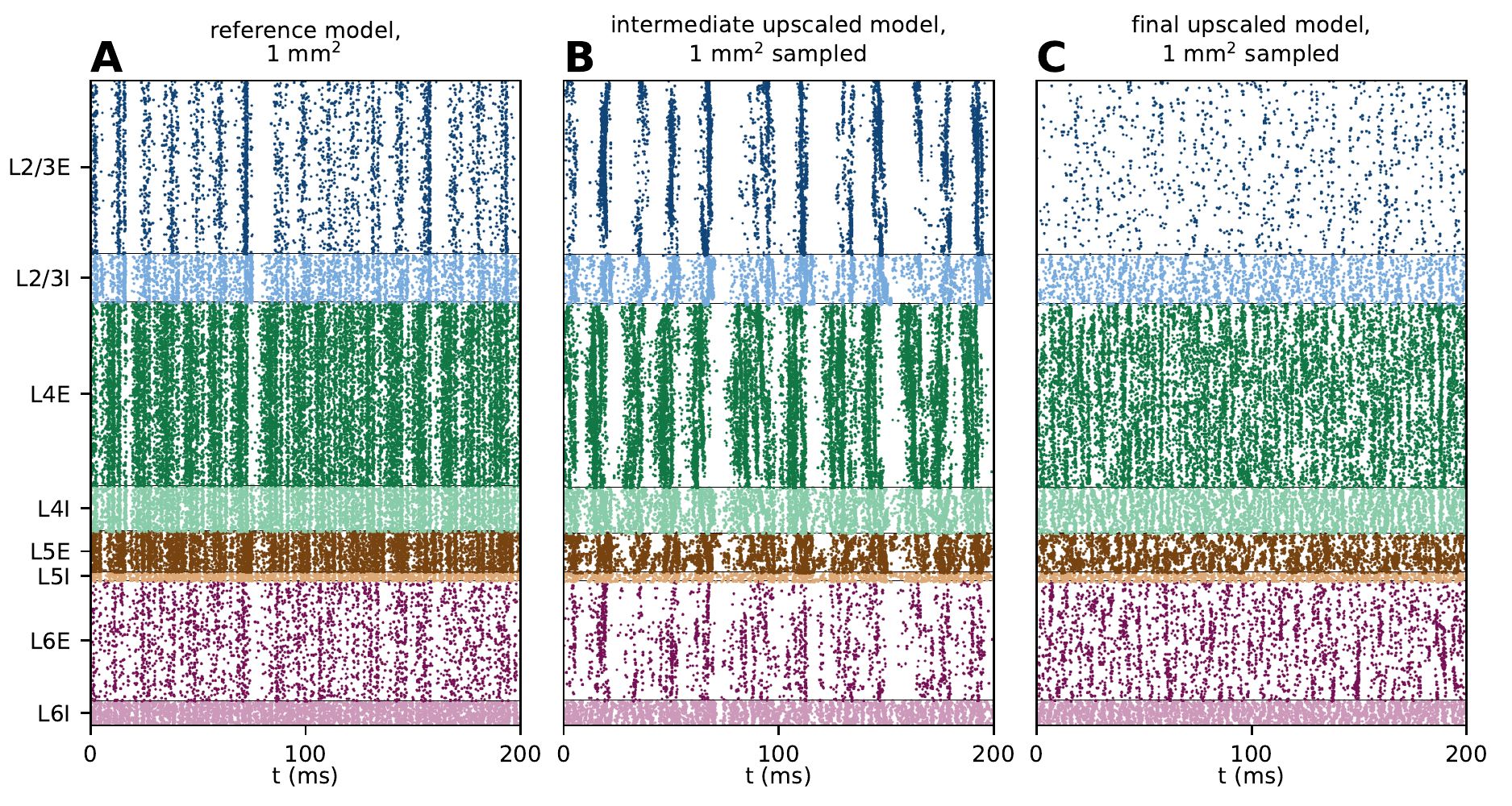}
\par\end{centering}
\begin{centering}
\par\end{centering}
\caption{\textbf{Spiking activity of the reference, intermediate upscaled,
and final upscaled models. A}~Spike raster showing the spike times
(horizontal) of all neurons of the reference model network (microcircuit
below $\unit[1]{mm^{2}}$ of cortical surface, no spatial connectivity
structure) vertically organized according to layer (axes labeling
and colors) and neuron type (lighter for inhibitory). \textbf{B}~Spike
raster of a model network upscaled to $\unit[4\times4]{mm^{2}}$ with
distance-dependent connectivity. The intermediate connection probabilities
$C_{YX}^{\mathrm{ui}}$ resulting from the upscaling procedure are
not modified ($\delta C_{YX}=0$). Spike times of all neurons located
inside a disc of $\unit[1]{mm^{2}}$ shown (neurons are always sorted
vertically according to their $x-$position). \textbf{C}~Same as
panel B, but with modified connection probabilities $C_{YX}^{\mathrm{u}}$
according to $\delta C_{YX}$ given in \tabref{parameters_ups}. The
parameters of the final upscaled model are referred to as \textquoteleft base
parameters\textquoteright{} and given in \tabref{parameters_ups}.\label{fig:rasters_ref_interm_ups}}
\end{figure}

\begin{center}
\par\end{center}

\begin{FPfigure} \centering

\includegraphics[width=1\textwidth]{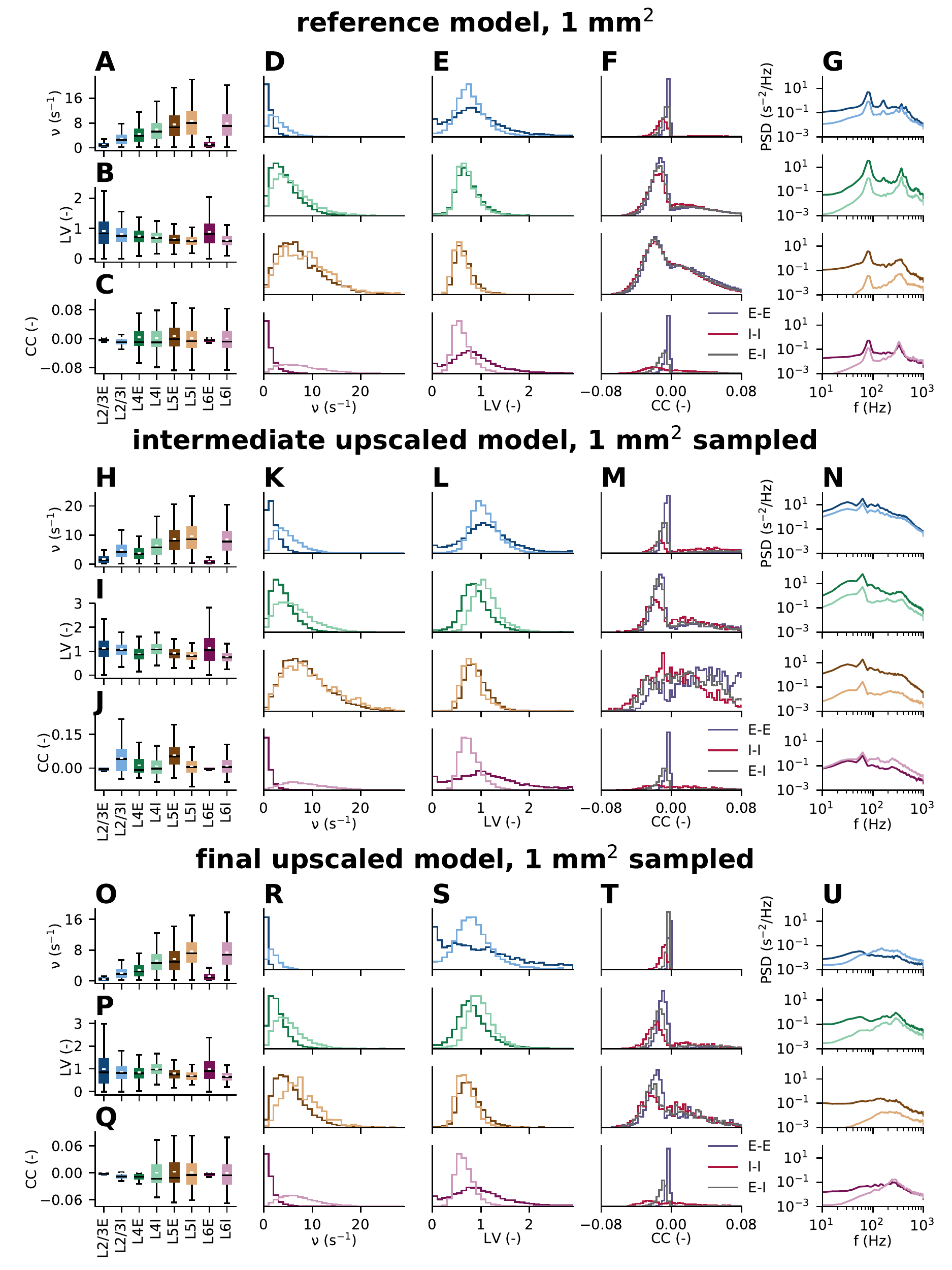}

\caption{ 

\textbf{Statistics of spiking activity of the reference, intermediate
upscaled, and final upscaled models. A\textendash G}~Statistics of
spiking activity of reference model shown in \figref{rasters_ref_interm_ups}A.
\textbf{A}~Heterogeneity of spike rates $\nu$ for each population
(horizontal black lines: median, short white lines: mean, boxes in
population-specific colors: lower and upper quartiles of the data,
whiskers extend to most extreme observations within $1.5\times\text{IQR}$
beyond the $\mathrm{IQR}$ (interquartile range) without outliers,
see documentation of \texttt{matplotlib.pyplot.boxplot}). \textbf{B~}Coefficients
of local variation $LV$, see \eqref{lv}.\textbf{ C}~Pearson correlation
coefficients $CC$, see \eqref{cc}. \textbf{D}~Distributions of
spike rates $\nu$.\textbf{ E~}Distributions of coefficients of local
variation $LV$.\textbf{ F}~Distributions of Pearson correlation
coefficients $CC$. \textbf{G}~Population-rate power spectral densities
$PSD$.\textbf{ H\textendash N~}Same as panels A\textendash G for
spiking activity of intermediate model shown in \figref{rasters_ref_interm_ups}B.
\textbf{O\textendash U}~Same as panels A\textendash G for spiking
activity of upscaled model shown in \figref{rasters_ref_interm_ups}C.

} \label{fig:statistics_ref_interm_ups} \end{FPfigure}

The spike raster in \figref{rasters_ref_interm_ups}A shows that
the reference model produces asynchronous irregular spiking with low
firing rates \citep{Softky93,Brunel99,Brunel00_183} across all populations.
Network oscillations appear as weakly pronounced vertical stripes.
The firing rates are on average higher for inhibitory populations
than for excitatory populations within the same layer, see \figref{statistics_ref_interm_ups}A,
and the mean illustrated in each box-chart is larger than the median.
The latter corresponds to the long-tailed distributions of spike rates
in \figref{statistics_ref_interm_ups}D with most neurons firing at
lower rates, while few neurons have high ($\unit[>20]{s^{-1}}$) rates.
This type of non-symmetric distribution of firing rates in the model
resembles approximately lognormally distributed firing rates observed
experimentally (reviewed in \citealp{buzsaki14_264}). The mean values
of the coefficients of local variation (\figref{statistics_ref_interm_ups}B,E)
are slightly below unity, indicating more regular spike trains than
events produced by a Poisson point process ($LV=1$). The distributions
are broad, that is, a fraction of neurons in each population has spike-train
statistics with $LV>1$. The mean $LV$ values are comparable to values
observed in visual cortex across different species \citep[Figure 5B]{Mochizuki16_5736}.
The box charts in \figref{statistics_ref_interm_ups}A,B are similar
to \citep[Figure 6]{Potjans14_785} showing firing rates and the conventional
coefficient of variation \citep[Equation 2.1]{Shinomoto03_2823}.
The Pearson correlation coefficients (\figref{statistics_ref_interm_ups}C,F)
are distributed and have a mean close to zero. Weak pairwise spike-train
correlations (with mean values $<0.1$ using $\unit[50]{ms}$ windows)
are reported, for example, by \citet{Ecker10} who record from nearby
neurons in primary visual cortex of awake monkey under different stimulation
conditions, and by \citet{Renart10_587} in somatosensory and auditory
cortex of anesthetized rats. The latter study finds that the mean
correlations are not distance-dependent, but their standard deviations
decay with distance (their Figure S11). The authors also include
a theoretical analysis of this phenomenon for networks of infinite
size and find that excitatory and inhibitory synaptic currents are
anticorrelated, thereby leading to a suppression of shared-input correlations,
and, hence, weak overall correlations in the asynchronous state. \citet{Tetzlaff12_e1002596}
and \citet{Helias14} identify the mechanism underlying the suppression
of shared-input correlations for the realistic case of finite-sized
networks, which differs from the mechanism in the infinite-size limit.
They show that the decorrelation is due to dominant negative feedback,
which leads to small correlations in both excitatory-inhibitory and
purely inhibitory networks.. However, depending on factors such
as brain state and distance, stronger correlations are also detected
in some cases \citep{Smith08_12591,Kriener09_177,Peyrache12_1731,Smith2012,Doiron16_383,Rosenbaum16_107}.
The population-rate power spectral densities in \figref{statistics_ref_interm_ups}G
show that the power tends to be higher in the activity of excitatory
compared to inhibitory populations due to the overall larger density
of excitatory neurons, except for layer $6$, where the inhibitory
rate is very high compared to the excitatory rate. Across layers the
power is highest in layer $4$, explained by the comparatively high
spike rates and high cell densities. The power spectra reveal two
dominant oscillation frequencies of the network in the low and high
gamma ranges ($\unit[\sim80]{Hz}$ and $\unit[\sim320]{Hz}$). Recent
theoretical work by \citet{Bos16_1} provides insight into the main
pathways between the recurrently connected populations involved in
generating these high-frequency oscillations. The low-gamma peak is
predominantly generated by a sub-circuit of layer $2/3$ and layer
$4$ populations of excitatory and inhibitory neurons (pyramidal-interneuron
gamma or ``PING'' mechanism \citep{Leung1982,Boergers03,Boergers05_557},
while the high-gamma peak results from interneuron-interneuron interactions
(interneuron-interneuron gamma or ``ING'' mechanism, see \citealp{Whittington1995,Wang1996,Chow1998,Whittington2000})
within each layer. See \citet{Buzsaki12_203} for a review on the
various mechanisms underlying gamma oscillations.

Before we discuss the final upscaled model in comparison to the reference
model, we first introduce an intermediate model in order to differentiate
between effects of pure upscaling and effects of modified connection
probabilities on network activity. This intermediate model is upscaled
as described in \subsecref{upscaling_procedure} resulting in connection
probabilities $C_{YX}^{\mathrm{ui}}$ derived directly from $C_{YX}^{\mathrm{r}}$
(from \eqref{conn_prob_ups}). No connection probabilities are otherwise
perturbed ($\delta C_{YX}=0$ for all $X$ and $Y$). All model parameters
are as specified in Tables \ref{tab:parameters_ref_and_ups}-\ref{tab:parameters_ups}
apart from the connection probabilities and the in-degrees of external
input, which are derived as specified in \subsecref{upscaling_procedure}.
This intermediate model covers an area of $\unit[4\times4]{mm^{2}}$,
but we here choose to analyze only the spiking activity of neurons
inside a disc of $\unit[1]{mm^{2}}$ at the center to obtain a representative
sample for comparison with the reference model in terms of neuron
numbers and spatial scale. The spike raster of the intermediate model
(\figref{rasters_ref_interm_ups}B) exhibits by visual inspection
spatially inhomogeneous activity and network synchrony that are more
pronounced than observed in the reference model. Compared to the reference
model, spike-train correlations in this intermediate model are increased
by approximately an order of magnitude (\figref{statistics_ref_interm_ups}J,M),
the coefficients of local variation are slightly increased (\figref{statistics_ref_interm_ups}I),
and finally the overall power in the rate spectra is increased across
all frequencies (\figref{statistics_ref_interm_ups}N). The spectra
also exhibit reduced low- and high-gamma peaks, and the activity is
generally more broadband.

The high global synchrony observed in the spiking of the intermediate
upscaled model is most likely exaggerated. There is accumulating evidence
that the typical operating regime of sensory cortices is asynchronous
and irregular in particular when no particular stimulus is present.
Measures of LFP signals, which are assumed to mainly reflect synaptic
activity, in for example visual cortex also do not show pronounced
peaks in their spectra in the absence of stimuli \citep[see, for example,][]{Berens08_1,Jia11_9390,Ray2011,Jia13_762,vanKerkoerle14}.
We therefore modify the network to suppress the amplitudes of the
two dominant oscillations in the low- and high-gamma range, and reduce
their frequencies to better resemble the low and high-gamma peaks
more commonly reported in the literature. For the final upscaled model,
we adapt connection probabilities by applying the modifications \textrm{$\delta C_{YX}$}
given in \tabref{parameters_ups}. The connection probabilities in
the reference model are estimated across different areas and species
and are merely suggestive of typical cortical connectivity\textemdash we
therefore consider small modifications to these values to be within
the bounds of uncertainties of these probabilities. Our choices on
which connections to perturb rely on the framework developed by \citet{Bos16_1}
who provide a \textquoteleft sensitivity measure\textquoteright{}
that relates population rate spectra to the connectivity of the underlying
neuron network in a systematic manner. With the example of our reference
model, they expose which individual connections are crucial for peak
amplitudes and frequencies of emerging oscillations, and demonstrate
how modifications of these connections affect the power spectra. By
applying this sensitivity measure to the intermediate upscaled network,
we find that its rate spectra are primarily shaped by the same specific
connections as in the reference network. To stabilize the circuit,
\citet{Bos16_1} reduce the number of connections from $\mathrm{L4I}$
to $\mathrm{L4E}$ of the reference model for their analysis. With
the same aim, we here reduce the connection probability from $\mathrm{L4I}$
to $\mathrm{L4E}$ and also increase that from $\mathrm{L4I}$ to
$\mathrm{L4I}$. Both of these modifications reduce amplitude and
frequency of the low-gamma peak \citep[Figure 8A for L4I-L4I]{Bos16_1}.
In addition, we increase the number of connections slightly from $\mathrm{L5I}$
to $\mathrm{L5E}$ and reduce the number of connections from $\mathrm{L5I}$
to $\mathrm{L5I}$ to further decrease the amplitude of this peak.
A decrease of the number of connections from $\mathrm{L5E}$ to $\mathrm{L5I}$
amplifies low-frequency oscillations \citep[Figure 8B]{Bos16_1}.
The resulting spike raster of the final upscaled model, similarly
sampled in the center $\unit[1]{mm^{2}}$, exhibits temporally and
spatially more homogeneous activity (\figref{rasters_ref_interm_ups}C)
compared to the reference and intermediate networks. The mean spike-train
correlations (\figref{statistics_ref_interm_ups}Q) are even lower
than in the reference model. The power spectra have overall reduced
power and its peaks are attenuated (\figref{statistics_ref_interm_ups}U).
Most visible in populations $\mathrm{L2/3E}$ and $\mathrm{L4E}$,
a broad low-gamma peak spans roughly $\unit[40-60]{Hz}$. Across all
interneuron populations, a broad high-gamma peak above $\unit[100]{Hz}$
is present. The per-population spike rates of the reference model
are now largely retained in the upscaled model (\figref{statistics_ref_interm_ups}O),
as the upscaling procedure preserves the mean input of the neurons
(see \subsecref{upscaling_procedure}). The coefficients of local
variation (\figref{statistics_ref_interm_ups}P) are similar to those
of the reference model, although the $LV$ of $\mathrm{L4I}$ is increased,
which we also observe in the intermediate model (\figref{statistics_ref_interm_ups}H).

\subsection{Spiking activity of the point-neuron networks\label{subsec:spike_data}}

We have so far established an upscaling procedure of the reference
network from an area of $\unit[1]{mm^{2}}$ to an area of $\unit[16]{mm^{2}}$,
which includes small perturbations to connection probabilities between
key pre- and post-synaptic populations. The final upscaled network
exhibits a stable network state that ($1$) is asynchronous and irregular
across populations, ($2$) preserves the population rates, ($3$)
preserves the distribution of firing rates, ($4$) preserves the variability
of spike trains, ($5$) has very low average pairwise spike-train
correlations, and ($6$) has rate spectra without pronounced peaks
. We next investigate the spontaneous behavior around this network
parameterization (\textquoteleft base parameters\textquoteright )
by varying external input rates, inhibitory feedback weights, spatial
connection widths, and the delay offset, which are all hard to constrain
with available experimental data. We also study evoked thalamocortical
activity in different network states in order to quantify the lateral
propagation speed of the evoked network response, motivated by reports
of propagating cortical activity. 

\subsubsection{Sensitivity to parameter perturbation during spontaneous activity\label{subsec:spontaneous_activity}}

\begin{center}
\par\end{center}

\begin{FPfigure} \centering

\includegraphics[width=1\columnwidth]{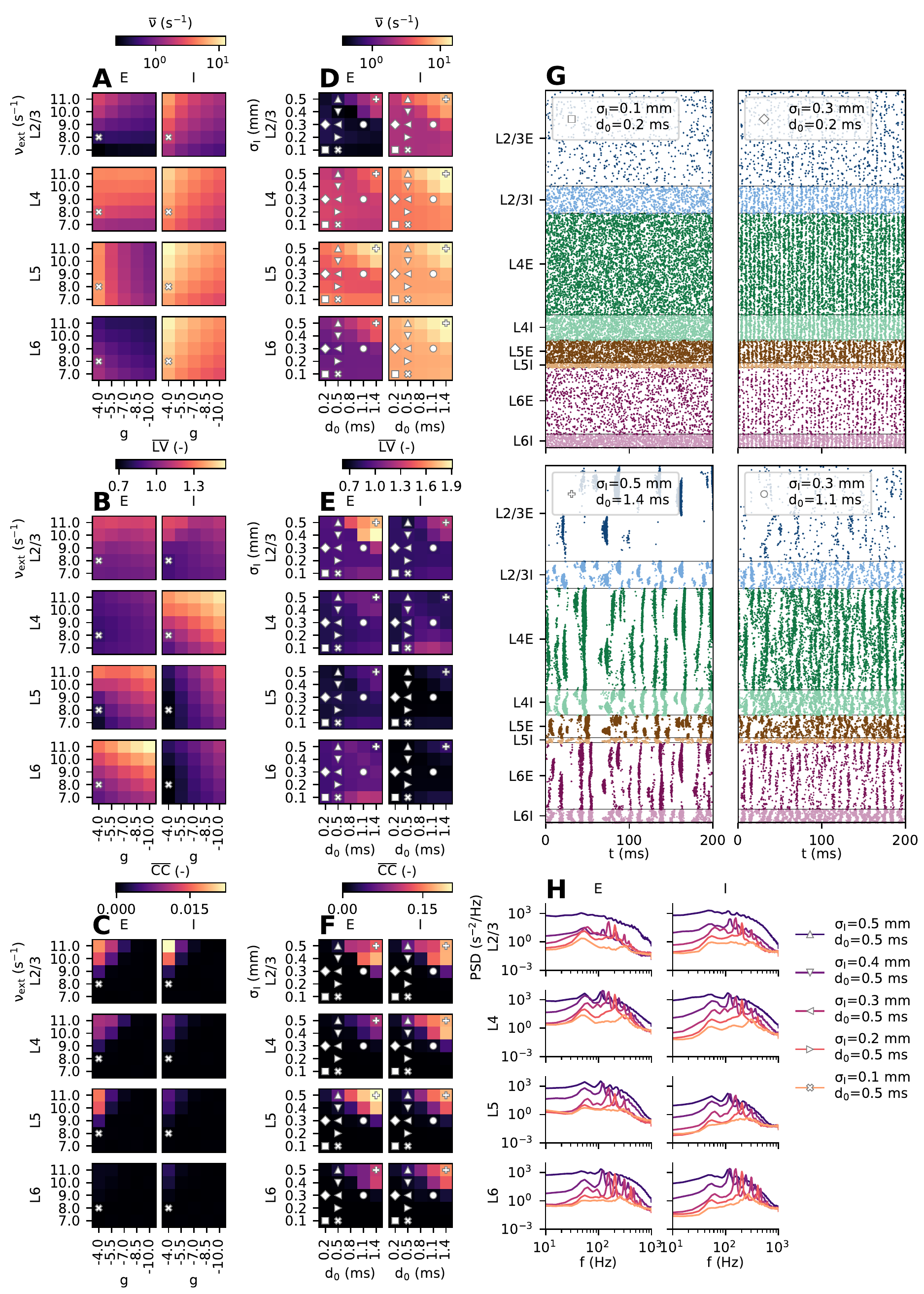}

\caption{

\textbf{Parameter sensitivity in the upscaled model.} \textbf{A\textendash C~}Dependency
on external rate $\nu_{\mathrm{ext}}$ and relative weight of inhibition
$g$ ($=g_{YX}$ with any inhibitory presynaptic population $X$).
\textbf{A}~Mean per-neuron spike rates $\overline{\nu}$ for each
population (color map with logarithmic scaling). The cross marker
denotes the default \textquoteleft base parameters\textquoteright{}
in this and subsequent panels. \textbf{B~}Mean coefficients of local
variation $\overline{LV}$ for each population, see \eqref{lv} (color
map with linear scaling). \textbf{C}~Mean Pearson correlation coefficients
$\overline{CC}$ between pairs of spike trains for each population,
see \figref{statistics_ref_interm_ups} (color map with linear scaling).
\textbf{D\textendash F~}Same as panels A\textendash C, but for dependency
on inhibitory spatial width $\sigma_{\mathrm{I}}$ and delay offset
$d_{0}$. Additional markers refer to parameter combinations used
in panels G and H.\textbf{ G}~Spike rasters of selected parameter
combinations (showing $\unit[3]{\%}$ of all neurons sampled from
the full network of size $\unit[4\times4]{mm^{2}}$, neurons are sorted
as in \figref{rasters_ref_interm_ups}). The symbols in each raster
plot legend mark the corresponding locations in the parameter space
spanned by $d_{0}$ and $\sigma_{\mathrm{I}}$ (panels D\textendash F).
\textbf{H~}Population-rate power spectral densities ($PSD$) of selected
parameter combinations. The markers correspond to the chosen parameter
combinations in panels D\textendash F.

} \label{fig:spontaneous_activity} \end{FPfigure}

We here explore the state space of the upscaled network model by running
parameter scans of both global network parameters (external input
rate and inhibitory weights) and parameters governing distance-dependent
connectivity (width of inhibition and delay offset). Theoretical work
exposes a crucial sensitivity to network parameters studying the existence
and stability of diverse dynamical states \citep{Brunel00_183,Roxin05,Senk18_arxiv_06046v1}.
However, experimental data from the literature are often sparse and
disparate and the mapping of measured quantities to specific model
parameters is not straightforward. Therefore, an exploration of the
parameter space is necessary in order to characterize the range of
possible model behaviors given the experimental constraints on the
parameter values and also to obtain an intuition of the model behavior.

We first choose to vary the rate of the external Poisson input $\nu_{\mathrm{ext}}$
and the relative inhibitory weight $g$. As shown in a simpler, analytically
tractable case \citep{Brunel00_183}, spatially unstructured networks
of randomly and sparsely connected excitatory and inhibitory leaky
integrate-and-fire neurons can transition between distinct activity
states with respect to the regularity of individual neuron firing
and the synchrony of population activity upon changing these two parameters.
Jumps in $LV$ (or the conventional coefficient of variation $CV$,
see \citealt[Equation 2.1]{Shinomoto03_2823}) and $CC$ during parameter
scans of comparable two-population networks typically indicate transitions
between states. It is, however, not a priori clear whether or not
this analytical insight obtained with a smaller random network generalizes
to spatially extended networks incorporating multiple layers and realistic
density of neurons and connections such as our upscaled network model.
\citet[Figure 2]{Mehring03_395}, \citet[Figure 4]{Voges10_51} and
\citet[Figure 2]{Voges12} study the same parameter space with spatially
organized network models; however, only in single-layer and diluted
networks. 

While the mean population rates $\overline{\nu}$ in a two-population
network typically increase when increasing $\nu_{\mathrm{ext}}$ or
decreasing $g$ (see \citet[Figure 2D]{Mehring03_395}, \citet[Figure 4]{Voges10_51}
and \citet[Figure 2]{Voges12} for examples), \figref{spontaneous_activity}A
shows that a similar trend does not appear for all populations of
our multi-layer upscaled model. Within the parameter range tested,
the mean rate of $\mathrm{L4E}$ is nearly unaffected upon varying
$g$, and varying $\nu_{\mathrm{ext}}$ has little effect on the rate
of $\mathrm{L5E}$. For $\mathrm{L6E}$, the trend is even reversed.
Different responses in different populations is explained by the population-specific
network connectivity and competing inhibition and excitation between
the different populations. Both recurrent (excitatory and inhibitory)
and external (only excitatory) in-degrees and corresponding presynaptic
rates result in population-specific means and variances of synaptic
inputs. Spike-train irregularity, here quantified by the mean coefficient
of local variation $\overline{LV}$ in \figref{spontaneous_activity}B,
also shows different trends per population. For all populations, the
$\overline{LV}$ increases when increasing $\nu_{\mathrm{ext}}$.
Increasing $g$ results in an increased $\overline{LV}$ only in layers
4 to 6, while the effect on $\mathrm{L2/3}$ does not show a clear
dependency on either parameter in the tested parameter range. The
$\overline{LV}$ remains below 1 across the whole parameter space
for populations $\mathrm{L4E}$ and $\mathrm{L6I}$, while the highest
values (above $1.3$) are observed in $\mathrm{L4I}$ and $\mathrm{L6E}$.
Mean pairwise spike train correlations $\overline{CC}$ in \figref{spontaneous_activity}C,
increase for all populations by increasing $\nu_{\mathrm{ext}}$ and
decreasing $g$. 

Next, we vary the spatial width $\sigma_{\mathrm{I}}$ of inhibitory
connections and the delay offset $d_{0}$, to assess the sensitivity
of the upscaled network dynamics to variations in their chosen values.
Although inhibitory spatial widths in terms of lateral axonal branching
patterns are generally assumed to be shorter than excitatory widths
\citep{Stepanyants09_3555}, estimates for the local excitatory and
inhibitory decay of connection probabilities are broadly distributed
and differ between brain areas, pre- and post-synaptic neuron types,
and species \citep{Hellwig00_111,Budd01,Boucsein11_1,Katzel2011_100,Perin11,Hill12_E2885,Levy2012_5609,Jiang2015,Schnepel15_3818,Reimann17}.
The reduction of multiple cell types and classes into only one excitatory
and one inhibitory neuron type per layer in the reference model \citep{Potjans14_785}
implicitly collapses the diversity of neuron morphologies \citep{Amirikian05,Brown09_1133,Hill12_E2885}
which have different spatial connectivity characteristics. Just as
for the spatial widths of connections, experimental evidence on distance-dependent
delay parameters is also sparse. As reviewed in \subsecref{upscaling_results},
the estimates for the conduction speed in unmyelinated nerve fibers
as well as for delay offsets are also widely distributed. In addition,
experimentally obtained spiking statistics exhibit a high variability,
even within the same brain area \citep{Mochizuki16_5736}. While available
experimental data on the typical widths of connections of different
types and corresponding conduction delays is inherently uncertain,
theoretical neural-field model studies frequently investigate the
strong influence of these parameters on the stability of the system
\citep{Ermentrout98b,Coombes05,Roxin05,Bressloff12}. In our upscaled
model, broader inhibition and larger delays increase the mean per-neuron
spike rates and the correlations in all populations, shown in \figref{spontaneous_activity}D
and F. The effect of changing the parameters $d_{0}$ and $\sigma_{\mathrm{I}}$
on $\overline{LV}$ in \figref{spontaneous_activity}E is again population-specific.
The highest $\overline{LV}$ values (above $1.6$) are obtained for
long delays ($d_{0}>1$), and broader inhibition than excitation in
$\mathrm{L2/3E}$; the $\overline{LV}$ remains low ($\apprle0.7$)
in $\mathrm{L5I}$ and $\mathrm{L6I}$ across the whole parameter
space. \figref{spontaneous_activity}G shows spike rasters of four
distinct network states emerging from this parameter space. Short-range
inhibition and short delays yield a spatially and temporally homogeneous
state (square marker). Increasing the width of inhibition to an intermediate
value results in fast global oscillations (diamond marker). For broader
inhibition than excitation, we observe localized activity spreading
outwards (plus marker). Finally, we show show an intermediate state
(circular marker). These results are in line with predictions from
neural-field studies, which indicate that long-range inhibition promotes
localized states such as spatially periodic patterns. By contrast,
long-range excitation promotes temporally periodic states that can
also combine with spatial patterns; see \citet[Chapter 8]{Ermentrout98b}
and \citet{Senk18_arxiv_06046v1}. For a network of spiking neurons,
\citet{Rosenbaum14} show that a balanced state of excitation and
inhibition requires broader excitation than inhibition. They demonstrate
that the balanced state loses stability if excitation is too narrow
compared to inhibition, leading to the emergence of spatial activity
patterns.

Finally, \figref{spontaneous_activity}H shows population rate spectra
($PSD$) varying with the delay offset of the base parameters, $d_{0}=\unit[0.5]{ms}$,
and different values for widths of inhibitory connections. While spatially
inhomogeneous activity with localized patterns (large $\sigma_{\mathrm{I}}$)
are manifested as comparatively flat spectra with high power across
all frequencies, reducing the spatial width also reduces the overall
power, while peaks at the dominant oscillation frequencies emerge.
Decreasing $\sigma_{\mathrm{I}}$ not only reduces amplitudes of the
power spectra, the frequency of the high-gamma peak is also gradually
shifted to higher values. Both observations can be related to a reduction
of the mean inhibitory delay averaged over all connections in the
network due to the shorter-range connectivity. The faster inhibitory
feedback results in a stronger decorrelation effect that reduces global
oscillations \citep{Tetzlaff12_e1002596,Helias14}. The upward shift
of the high-gamma frequency is explained by a shorter time period
for the ING mechanism \citep{Bos16_1}. 

\subsubsection{Sensitivity to perturbed parameters during evoked activity\label{subsec:evoked_activity}}

\begin{center}
\par\end{center}

\begin{FPfigure} \centering

\includegraphics[width=1\textwidth]{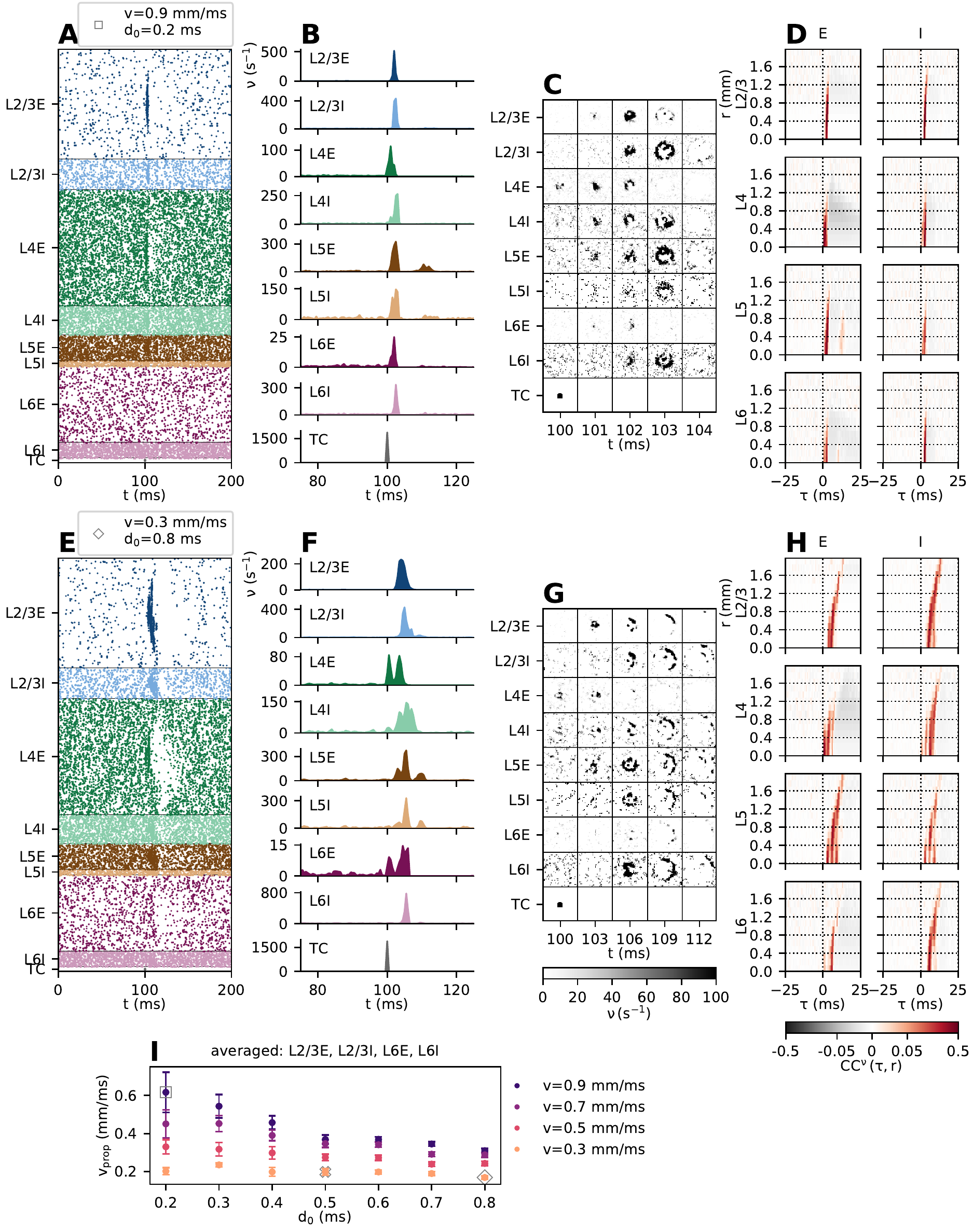}

\caption{

\textbf{Activity evoked by thalamic pulses. A}~Spike raster (showing
$\unit[3]{\%}$ of all neurons in $\unit[4\times4]{mm^{2}}$, neurons
are sorted as in \figref{rasters_ref_interm_ups}). A single thalamic
pulse occurs at $t=\unit[100]{ms}$. \textbf{B}~Population-averaged
rate histogram for neurons within the center disc of $\unit[1]{mm^{2}}$
with bin size $\Delta t$ for a time interval around the thalamic
pulse shown in panel A. \textbf{C}~Series of snapshots of spatiotemporally
binned activity per population over the whole $\unit[4\times4]{mm^{2}}$
network. \textbf{D}~Distance-dependent cross-correlation functions
between thalamic activation and spatially binned spiking activity
$CC^{\nu}\left(\tau,r\right)$ where $r$ is the distance to the center
of the network and $\tau$ is the time lag. Color maps have a symmetric
logarithmic scaling (linear up to threshold of $\pm0.05$ indicated
by ticks in the color bar). Panels A\textendash D are obtained in
a network with conduction speed $v$ and delay offset $d_{0}$ as
indicated in the legend of panel A. \textbf{E\textendash H} Same as
panels A\textendash D but with parameters as indicated in the legend
of panel E. \textbf{I}~Propagation speed $v_{\mathrm{prop}}$ estimated
for parameter combinations of conduction speeds and delay offsets
and averaged across populations named above the panel; error bars
denote standard deviation $\sigma_{v,\mathrm{prop}}$. The same markers
correspond to the same parameter combinations throughout this figure.
Base parameters are marked with a cross.

} \label{fig:evoked_activity} \end{FPfigure}

We have so far only considered networks receiving external inputs
with stationary rates. Cortical areas are, however, recurrently connected
to other parts of cortex and subcortical structures, and receive inputs
with large rate fluctuations. We here mimic a stimulation experiment,
by activating all thalamic neurons inside a disc of radius $R_{\mathrm{TC}}^{\mathrm{pulse}}$
around $\left(x,y\right)=\left(0,0\right)$ once every time interval
of $\Delta t_{\mathrm{TC}}$ (see \tabref{parameters_ups} for values).
The activation could for example represent a visual stimulation experiment
where activity in lateral geniculate nucleus (LGN, or visual thalamus)
thalamocortical ($\mathrm{TC}$) projection neurons is evoked by a
brief flash stimulus to a part of the visual field \citep{Bringuier99_695,Muller14_4675},
air puffs or mechanical whisker deflections to stimulate whisker barrel
cortex \citep{Swadlow02_7766,Einevoll07_2174}, or direct electric
or optogenetic stimulation of the thalamocortical pathway \citep{Klein16_143}.
In its population-specific responses to thalamic pulses, the reference
model of \citet[page 802]{Potjans14_785} exhibits a ``handshake
principle'', in which the receiving layer inhibits the sending layer
as if to signal that it has received the message, so that the sending
layer can stop transmitting. We test whether this effect and its strength
are preserved in the upscaled model. Furthermore, we derive the propagation
speed of evoked spiking activity spreading outward from the center
of stimulation. Finally, we test the robustness of the propagation
speed to parameter perturbations by varying the conduction speed and
the delay offset.

Panels A\textendash D and E-H in \figref{evoked_activity} show results
for two different choices of conduction speed $v$ and the delay offset
$d_{0}$. At times prior to a thalamic pulse at $t=\unit[100]{ms}$,
the spiking activities in \figref{evoked_activity}A and E are comparable,
and both asynchronous and irregular, despite the different parameterization.
However, the effect of the pulse on the network activity is more pronounced
in panel E than in panel A according to visual inspection; the initial
response lasts longer and the subsequent activity vanishes for tens
of milliseconds in different populations. In \citet[Figure 3]{Hao16_104}
a similar suppression period of tens of milliseconds is observed following
a single-pulse electrical micro-stimulation in monkey motor cortex,
often followed by a rebound of excitation. In panel E, the effective
delay is larger due to the choice of a larger $d_{0}$ and a smaller
$v$. In the population-averaged rate histograms of activity within
$\unit[1]{mm^{2}}$ in \figref{evoked_activity}B and F, corresponding
to the spike rasters in panels A and E, respectively, we highlight
the transient network responses by zooming into a smaller time window
around the pulse. The strong initial response visible in populations
$\mathrm{L4E}$ and $\mathrm{L6E}$ is expected since the thalamocortical
input targets layers $4$ and $6$ directly (see \tabref{parameters_ups}).
This evoked activity affects the other network populations via recurrent
network connections across and within layers. The larger effective
delay (panel H) here increases the response latency of the populations,
and increases the duration of the responses while their maximum rates
in some populations are reduced. The duration of the activation is
overall similar to evoked multi-unit activity (MUA) following whisker
stimulation as reported by \citet{Einevoll07_2174}. The multiple
peaks in the rate histograms in panel D, most prominent in populations
$\mathrm{L4E}$, $\mathrm{L5E},$ $\mathrm{L5I}$ and $\mathrm{L6E}$,
are due to recurrent excitation and inhibition within and across layers.
The overall increased delays expectedly break balance, that is, the
high temporal correlation of excitatory and inhibitory spiking activity
\citep[see, for example,][]{Renart10_587}. These results are comparable
with \citet[Figure 10]{Potjans14_785} and \citet[Figure 7]{Hagen16},
and we therefore conclude that the upscaling procedure does not fundamentally
affect the response of the network to transient external input.

While the population-averaged rate histograms in \figref{evoked_activity}B
and F expose the temporal effect of the perturbation of network activity,
we next focus on the corresponding spatiotemporal responses. \figref{evoked_activity}C
and G show series of snapshots of spatiotemporally binned activity
of each population in the full network of size $\unit[4\times4]{mm^{2}}$
(similar to \citealp[Figure 2]{Mehring03_395,Yger2011_229}). The
temporal bin size is $\Delta t$ as in the rate histograms, but we
show snapshots only for selected time points as indicated below the
frames. The thalamic pulse is visible only at $t=\unit[100]{ms}$
in the center of the network. The cortical populations respond with
a ring-like outward spread of activity which can be described as a
traveling wave in contrast to a stationary bump \citep{Muller18_255}.
The wave travels at a lower speed in the network with larger effective
delay (compare selected time points in \figref{evoked_activity}C
and G). In order to derive the radial propagation speed of activity
evoked by thalamic pulses, we compute the distance-dependent cross-correlation
functions (see \subsecref{statistical_measures}) shown in \figref{evoked_activity}D
and H. The maximum value of $CC^{\nu}\left(\tau,r\right)$ shifts
faster to larger time lags $\tau$ with increasing distances $r$
in panel H compared to panel D, which indicates a lower propagation
speed. \figref{evoked_activity}I summarizes the propagation speed
estimates $v_{\mathrm{prop}}$ as a function of $v$ and $d_{0}$.
The estimated propagation speeds increase with increasing conduction
speed $v$ and decreasing delay offset $d_{0}$. Estimating the propagation
speed in this way from spatially resolved spike trains can help to
infer underlying network parameters from experimental data. It is
to date difficult to observe wave-like activity on the spiking level
\citep{Takahashi15}. However, model predictions for spiking propagation
speeds can be compared with population measures, keeping in mind potential
differences between spiking activity and population measures such
as the LFP. Both types of signals can reflect propagation along long-range
horizontal connections which also includes synaptic processing times,
but they are also affected by intrinsic dendritic filtering \citep{Grinvald94_2545,Nauhaus09_70,Takahashi15,Zanos15_615}.
\citet{Muller18_255} remark that macroscopic waves traveling across
the whole brain typically exhibit propagation speeds of $\unit[1-10]{mm/ms}$
similar to axonal conduction speeds of myelinated white matter fibers
in cortex, while mesoscopic waves (as considered here) show propagation
speeds of $\unit[0.1-0.8]{mm/ms}$ similar to axonal conduction speeds
of unmyelinated long-range horizontal fibers within the superficial
layers of cortex. For example, LFP \textquoteleft waves\textquoteright{}
in visual cortex travel with such speeds. \citet{Nauhaus09_70} study
the propagation of spike-triggered LFPs both in spontaneous activity
and with visual stimulation and and derive speeds (mean $\pm$ standard
deviation) of $\unit[0.31\pm0.23]{mm/ms}$ in cat and $\unit[0.24\pm0.2]{mm/ms}$
in monkey (both anesthetized). \citet{Nauhaus12_3088} reanalyze the
data from \citet{Nauhaus09_70} and further report a speed of $\unit[0.18]{mm/ms}$
in cat and $\unit[0.29]{mm/ms}$ in monkey for the impulse response
of ongoing activity; for data from awake monkey \citep{Ray2011} they
compute a speed of $\unit[0.13]{mm/ms}$. \citet{Zanos15_615} measure
a speed of $\unit[0.31\pm0.08]{mm/ms}$ triggered by saccades in monkey
visual cortex. Propagation speeds obtained via voltage-sensitive dye
imaging in visual cortex are comparable as well: an average speed
of $\unit[0.28]{mm/ms}$ with a $75\%$ confidence interval of $0.19$
to $\unit[0.55]{mm/ms}$ in cat \citep{Benucci07_103}, $\unit[0.1-0.25]{mm/ms}$
in monkey \citep{Grinvald94_2545}, and a range of $\unit[0.25-1.35]{mm/ms}$
with median $\pm$ standard deviation of $\unit[0.57\pm0.18]{mm/ms}$
in monkey \citep{Muller14_4675}. Estimates from monkey motor cortex
are in the same range \citep{Rubino-2006_1549,Takahashi15,Denker18_1}.
For the biologically plausible ranges of delay offsets and conduction
speeds tested in the model, $d_{0}\in\left[0.2,0.8\right]\,\mathrm{ms}$
and $v\in\left[0.3,0.9\right]\,\mathrm{mm/ms}$, the resulting propagation
speeds are mainly between $0.2$ and $\unit[0.6]{mm/ms}$. These derived
propagation speeds are smaller than the corresponding conduction speeds
because propagation through the network includes neuronal integration
and the delay offsets. The values in the model cover the range of
experimentally measured propagation speeds.

\subsection{LFP predictions\label{subsec:lfp_predictions}}

\begin{figure}
\includegraphics[width=1\columnwidth]{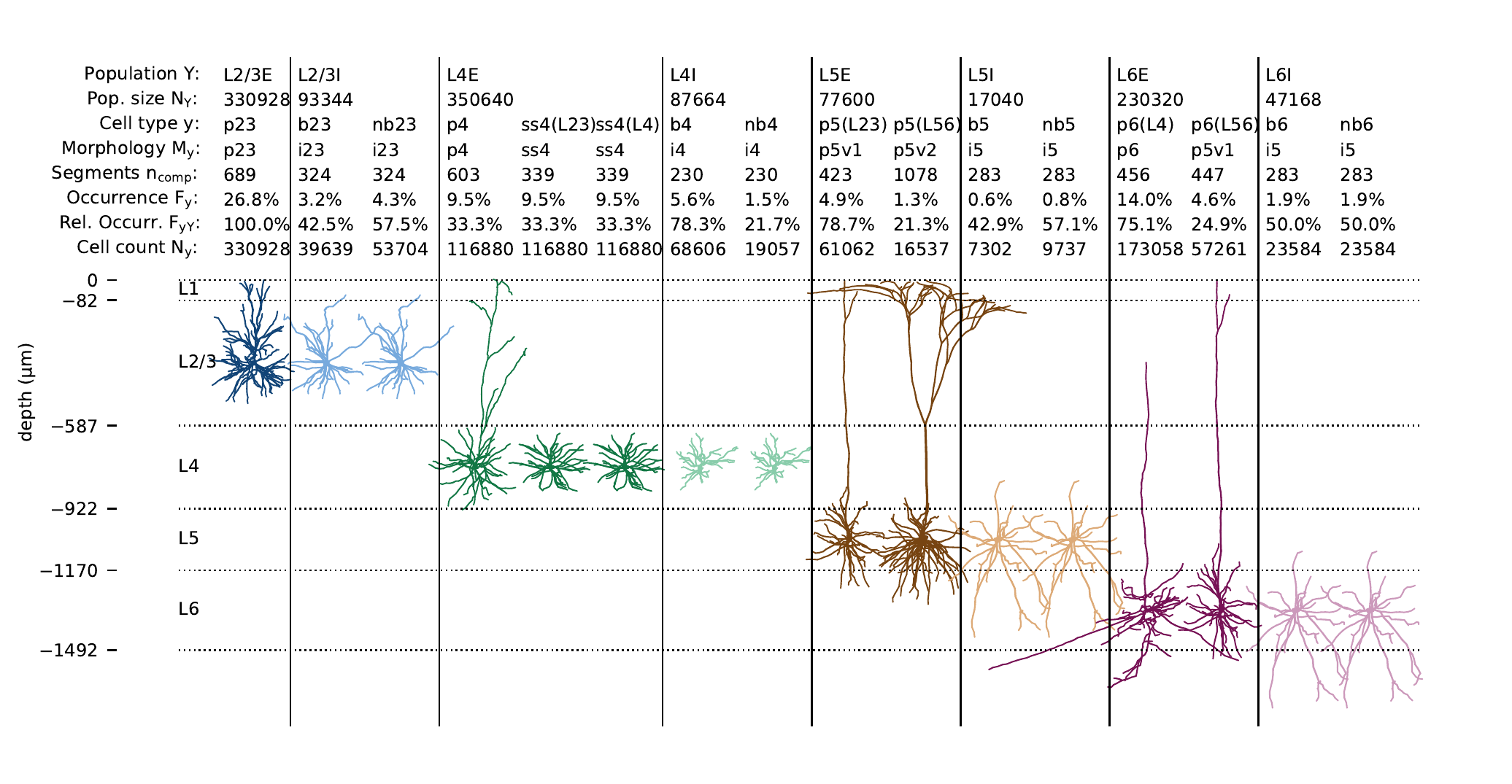}

\caption{\textbf{Cell types and morphologies of the multicompartment-neuron
populations.} The $8$ cortical populations $Y$ of size $N_{Y}$
in the $\unit[4\times4]{mm^{2}}$ network model are represented by
$16$ subpopulations of cell type $y$ with detailed morphologies
$M_{y}$ \citep{Binzegger04,Izhikevich08_3593}. Neuron reconstructions
are obtained from cat visual cortex and cat somatosensory cortex (source:
NeuroMorpho.org by \citet{Kisvarday92,Mainen96,Contreras97_335,Ascoli07_9247,Stepanyants2008_13},
see \citealp[Table 7]{Hagen16}). Each morphology $M_{y}$ is here
shown in relation to the layer boundaries (horizontal lines). Colors
distinguish between network populations as in \figref{rasters_ref_interm_ups}.
The number of compartments $n_{\text{comp}}$, frequencies of occurrence
$F_{y}$, relative occurrence $F_{yY}$ and cell count $N_{y}$ are
given for each cell type $y\in Y$. \label{fig:morphologies}}
\end{figure}

We here summarize our findings for the predicted LFP signal across
cortical space, with recording geometry similar to a $\unit[4\times4]{mm^{2}}$
Utah multi-electrode array. As in \citet{Hagen16}, the eight cortical
network populations spanning layers $2/3$, $4$, $5$ and $6$ are
expanded into $16$ different cell types in order to account for differences
in layer specificity of synaptic connections among cell types in a
single layer when predicting the LFP. While we here refrain from discussing
the detailed derivation of these layer specificities (see \citealp{Hagen16})
from available anatomical data \citep[i.e.,][]{Binzegger04}, in \figref{morphologies}
we show the reconstructed morphology used for each cell type $y$
in population $Y$, with compartment counts and occurrences summarized
in the table contained within the figure. The cortical layer boundaries
and depths are also illustrated, and each morphology is positioned
such that the soma is at the center of the corresponding layer. Different
cell types belonging to the same population within a layer may have
different geometries supporting different layer specificities of synaptic
connections. This is the case for example for the p4 pyramidal cell
type versus the ss4 spiny stellate cell types that both belong to
population $\mathrm{L4E}$ of the point-neuron network. Previous modeling
studies demonstrate the major effect of the geometry of the morphology
on the measured extracellular potential due to intrinsic dendritic
filtering of synaptic input \citep[e.g.,][]{Linden10,Linden11_859,Leski13_e1003137}.

\begin{figure}

\includegraphics[width=1\textwidth]{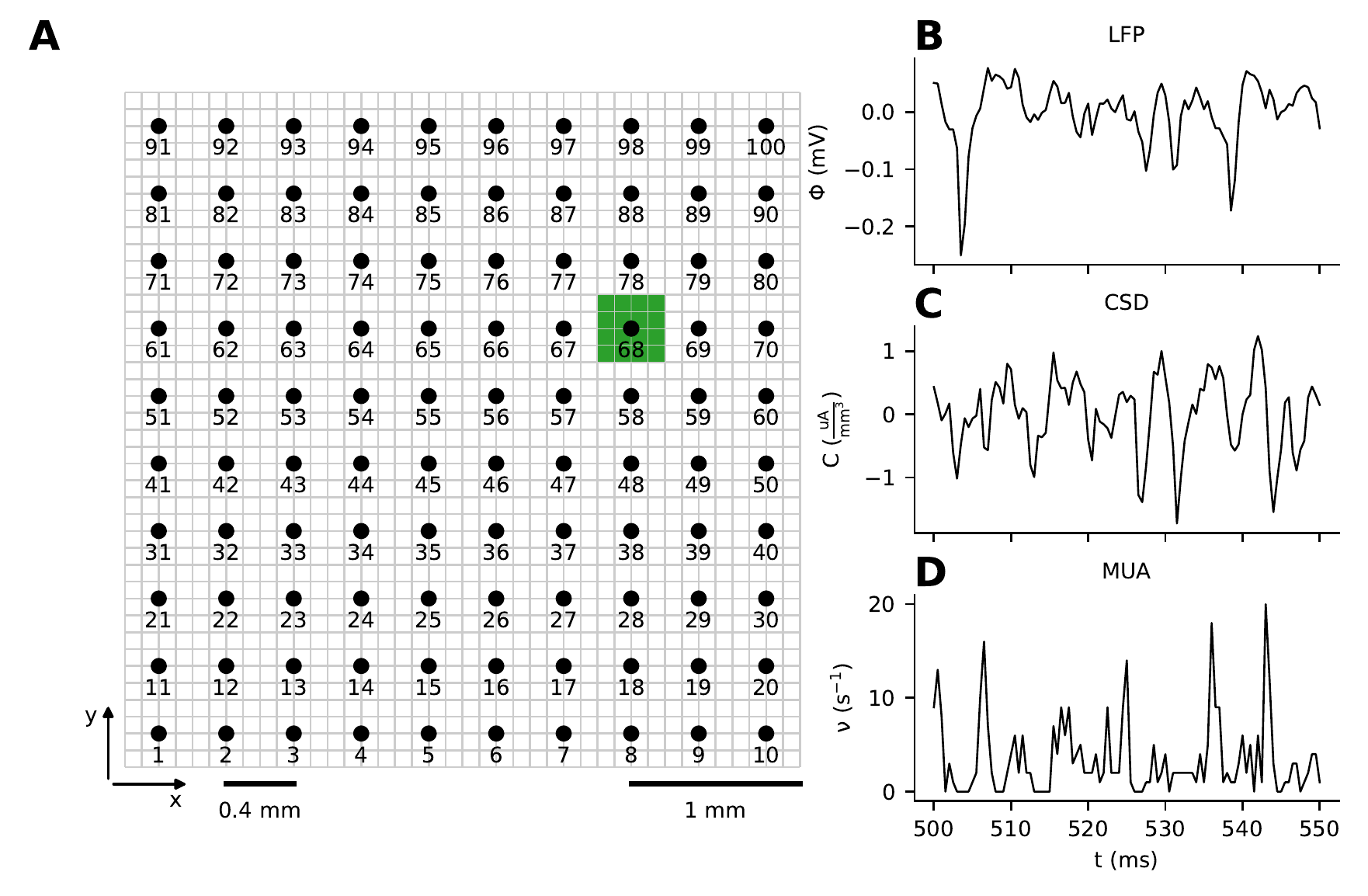}

\caption{\textbf{Illustration of multi-electrode array geometry for LFP, CSD,
and MUA predictions.} \textbf{A~}Extracellular potentials are computed
in $10\times10$ electrode locations denoted by circular markers at
the depth corresponding to the center of layer $\mathrm{2/3}$. The
electrode inter-contact distance is $\unit[400]{\mu m}$. The number
under each circular marker denotes the channel number. \textbf{B\textendash D~}Example
LFP, CSD, and MUA from one arbitrarily chosen contact (here channel
number $68$). The CSD is estimated from the LFP using an inverse
method, and the MUA is calculated as the sum of excitatory and inhibitory
spike events from layer $\mathrm{2/3}$ neurons in spatiotemporal
bins of duration $\unit[0.5]{ms}$ and width $\unit[400]{\mu m}$
around each contact. \label{fig:illustration_array_geometry}}
\end{figure}

The geometry of the recording locations corresponding to the \textrm{$\unit[4\times4]{mm^{2}}$}
Utah multi-electrode array is illustrated in \figref{illustration_array_geometry}A.
The $100$ contact locations denoted by circular markers are positioned
on a $10\times10$ grid with $\unit[400]{\mu m}$ separation between
contact sites. LFPs are computed at the center of layer $\mathrm{2/3}$
(at $z=\unit[-334]{\mu m}$). An example LFP signal segment from one
chosen channel (channel $68$) is shown in panel B, corresponding
to the spontaneous activity in our laminar, upscaled point neuron
network with \textquoteleft base parameters\textquoteright{} introduced
above (in \figref{rasters_ref_interm_ups}C and corresponding text).
The signal fluctuates with amplitudes similar to experimentally observed
spontaneous potentials ($\unit[0.1-1]{mV}$, \citet{Maier10,Hagen15_182,Reyes-Puerta2016}),
with occasional larger transients. Further, we estimate from the LFP
the underlying current source density (CSD) across space using the
so-called kernel CSD method in two dimensions \citep[2DkCSD,][]{Potworowski12_541}.
The CSD signal is expected to suppress correlations in the LFP resulting
from volume conduction, and is therefore less correlated across space
as it is taken to reflect the gross in\textendash{} and outgoing transmembrane
currents in vicinity to the recording device \citep{Nicholson75_356,Mitzdorf85_37,Pettersen06_116,Pettersen08_291,Potworowski12_541}.
The LFP and corresponding CSD in general reflect correlations in synaptic
input nearby the measurement site and therefore contain contributions
from both local and remote neuronal activities. In contrast, the high-frequency
($\unit[\gtrsim100]{Hz}$) part of experimentally obtained extracellular
potentials contains information on spiking activity of local neurons.
Activity of high-amplitude single neurons may be separated from the
background based on classification of their extracellular action-potential
waveforms \citep[through \textquoteleft spike sorting\textquoteright ,][]{Quiroga07_3583}.
Even if no units are clearly discernible in the high-frequency part
of the signal, a previous biophysical forward-modeling study using
biophysically detailed neuron models \citep{Pettersen08_291} shows
that the envelope of the rectified high-pass filtered ($\unit[750]{Hz}$
cutoff frequency) signal correlates with the spike rate in the local
population of neurons. In this study, this rectified signal is referred
to as the multi-unit activity (MUA), which we approximate by summing
up all spiking activities of layer $2/3$ neurons in $\unit[400\times400]{\mu m^{2}}$
spatial bins around each contact. The presently used LFP predictions
rely on passive neuron models which do not generate spikes; spiking
only occurs in the network. The contribution from excitatory and inhibitory
spikes are weighted identically. One example MUA trace obtained at
the same location as the LFP and CSD is shown in \figref{illustration_array_geometry}D.
A notable observation is that the MUA signal and its relations to
the corresponding LFP and CSD signals are non-trivial.

\subsubsection{Distance-dependent correlations of spike trains and LFPs\label{subsec:distance_dep_correlations_spike_LFP}}

We next investigate the temporal correlation and coherence with distance
for these measures of activity. The observation of weak pairwise spike-train
correlations in cortical neuronal networks \citep[for example,][]{Ecker10}
is seemingly at odds with the typical observation of highly correlated
LFPs across cortical space \citep[for example,][]{Nauhaus09_70}.
We have so far established that the mean pairwise spike-train correlations
within populations in our upscaled layered network are typically near
zero (\subsecref{upscaling_results}), and that the perturbation of
key network parameters such as the external rate and delays affect
the mean correlation (or \textquoteleft synchrony\textquoteright )
in the network (\subsecref{spike_data}), as well as other measures
like regularity (as measured by their mean coefficients of local variation
$\overline{LV}$). It is, however, not clear how this weakly correlated
network activity translates into population signals such as the LFP.
Previous modeling studies of mechanisms of the spatial reach of the
LFP highlight the crucial role of correlation in synaptic inputs to
the LFP-generating neurons \citep{Linden11_859,Leski13_e1003137}.
In contrast to these studies, which use input spike trains with Poisson
inter-spike statistics, we here account for ongoing network interactions,
and realistic numbers of neurons and connections under \textrm{$\unit[4\times4]{mm^{2}}$}
of cortical surface using the methods to compute LFPs introduced by
\citet{Hagen16}. We thus extend our analysis to distance-dependent
correlations in LFP, CSD, MUA, and pairs of spike trains.

\begin{figure}
\includegraphics[width=1\textwidth]{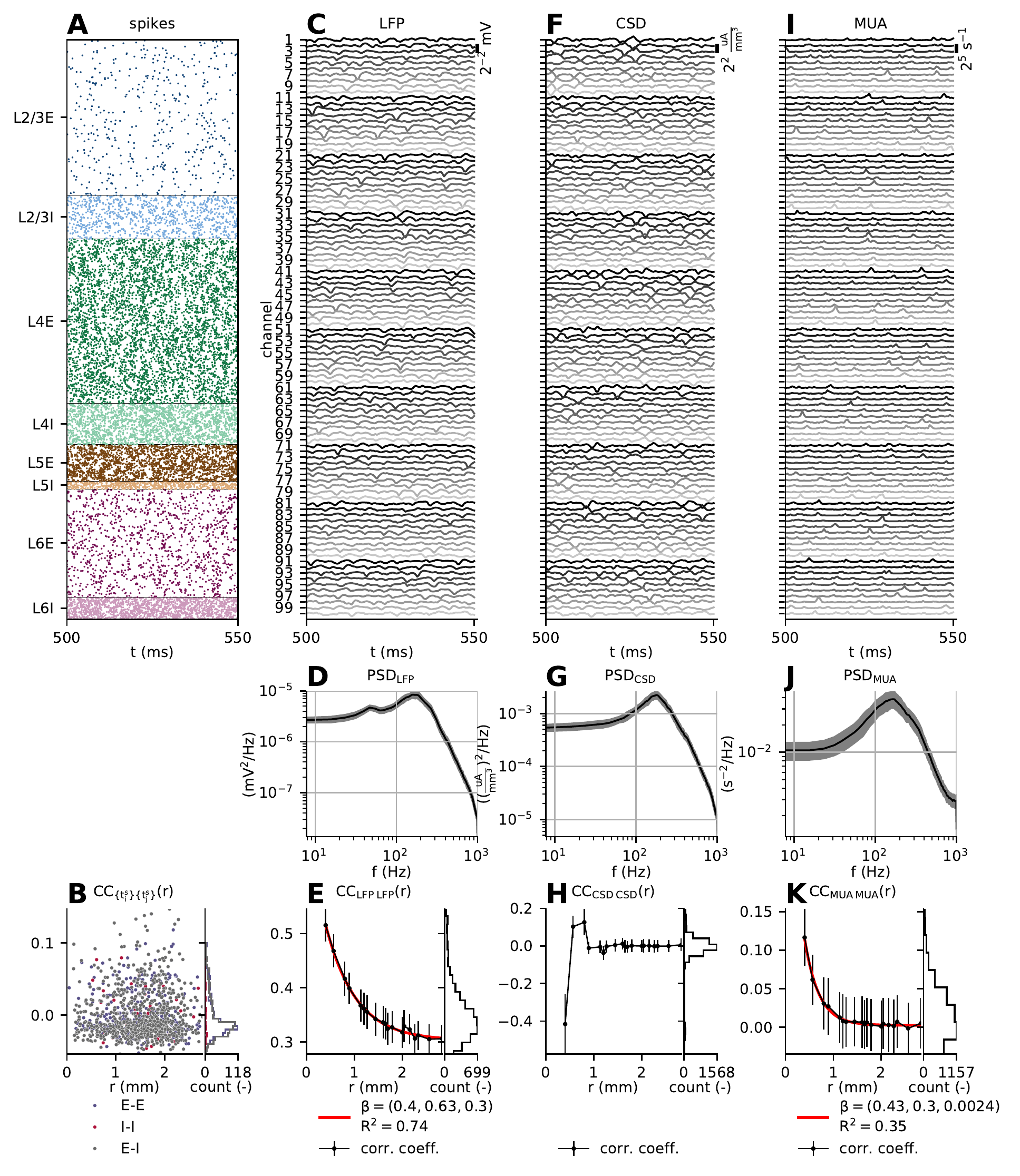}\caption{\textbf{Spikes, LFP, CSD, and MUA: Raw signals, power spectra, and
distance-dependent correlations in $\mathrm{L2/3}$.} \textbf{A~}Spike
raster (showing $\unit[10]{\%}$ of all neurons in $\unit[4\times4]{mm^{2}}$,
neurons are sorted as in \figref{rasters_ref_interm_ups}). \textbf{B}~Pairwise
spike-train correlations computed for pairs of excitatory (E-E, $n=40$),
inhibitory (I-I, $n=10$) and excitatory and inhibitory (E-I) $\mathrm{L2/3}$
neurons, sorted by inter-neuron distance $r$.\textbf{ C~}Local field
potentials (LFP) across the $10\times10$ electrode contact points
located at the center of layer $2/3$, each separated by $\unit[400]{\mu m}$
in the lateral directions. \textbf{D}~LFP power spectrum averaged
over channels (black line). The gray area denotes the average spectrum
plus/minus one standard deviation. \textbf{E}~Pearson correlation
coefficient between pairs of LFP signals as function of separation
between channels. The black line shows the mean at each unique separation,
whiskers denote one standard deviation. The red line shows the least-square
fit of an exponential function to all values. The coefficient of determination
($\mathrm{R}^{2}$) is given in the legend. \textbf{F}~Current-source
density (CSD) estimates from LFPs shown in panel C, calculated using
the kCSD method in 2D. \textbf{G}~CSD power spectrum (mean$\pm$one
standard deviation). \textbf{H}~Similar to panel E but for CSD signals,
minus fit to exponential function. \textbf{I}~Multi-unit activity
(MUA) approximated as the bin-wise spike rates of layer $2/3$ excitatory
and inhibitory point neurons, calculated using a spatial bin width
$\Delta h=\unit[400]{\mu m}$. \textbf{J}~MUA power spectrum (mean$\pm$one
standard deviation). \textbf{K}~Similar to panel E but for MUA signals.\label{fig:correlations_lfp_csd_mua_spikes}}
\end{figure}

For spontaneous spiking activity in the upscaled network (\figref{correlations_lfp_csd_mua_spikes}A),
we compute the LFP (panel C), reconstruct the underlying CSD from
the LFP (panel F), and compute the MUA (panel I) across the $100$
channel locations in layer $2/3$ illustrated in \figref{illustration_array_geometry}A.
The network parameters and corresponding network state are those resulting
from our upscaling procedure (see \subsecref{upscaling_procedure},
base parameters given in Tables \ref{tab:parameters_ref_and_ups}
and \ref{tab:parameters_ups}). Visual inspection of panel C reveals
that the LFP amplitude across channels is typically small ($\unit[\lesssim0.5]{mV}$)
as highlighted in \figref{illustration_array_geometry}B with occasional
transients which may be seen also on neighboring channels. These transient
events presumably result from spatially confined synchronization in
the network, but are not seen across every LFP channel as would be
the case with globally synchronous network events. The amplitudes
observed here are similar to those from the forward-model predictions
of LFPs from spontaneous activity in the original $\unit[1]{mm^{2}}$
network model \citep[Figure 8M]{Hagen16}, even if the total number
of neurons in the upscaled model is increased by a factor of $16$.
These similar amplitudes are partially explained by the suppression
of strong low-gamma oscillations in the upscaled network using modified
connection probabilities. An increase in network synchrony (that is,
increased correlations) can otherwise be expected to increase LFP
amplitudes overall due to an increased pairwise cross-correlation
between single-neuron contributions to the LFP \citep{Hagen16}. The
network upscaling procedure does not obliterate the high- and low-gamma
oscillations, which in the LFP spectra result in a large peak around
$\unit[200]{Hz}$ and a small peak around $\unit[50]{Hz}$. The network
receives background input with a flat power spectrum (driven by a
Poisson process with fixed rate) and has no internal sub-circuits
capable of generating rate fluctuations or slow oscillations. Hence,
the LFP in each channel contains little power towards small frequencies.
Another factor explaining the lack of low frequency power is active
decorrelation by inhibitory feedback, which is shown to suppress population-rate
fluctuations \citep{Tetzlaff12_e1002596,Helias14}. 

We next compute the Pearson product-moment correlation coefficient
between all possible pairs of LFP channels, and sort by inter-contact
distance (panel E). The mean and standard deviation for each discrete
contact separation are shown by the black line and corresponding error
bars. Due to the periodic boundary conditions of the network, the
longest possible inter-contact distance is $L/\sqrt{2}\approx\unit[2.8]{mm}$.
The mean values are well fit by a simple exponential function (red
line), with a spatial decay constant of $\text{\ensuremath{\unit[\sim0.63]{mm}}}$
and constant offset of $\sim0.3$. The histogram to the right is computed
for all observed correlation coefficients. The correlations in the
simulated LFP are lower compared to findings by \citet[Fig. 8]{Nauhaus09_70}
during spontaneous activity in anesthetized macaque (approximately
$0.95$ at $\unit[0.4]{mm}$ and $0.75$ at $\unit[2.4]{mm}$ electrode
separation, respectively) and cat (approximately $0.93$ at $\unit[0.4]{mm}$
and $0.83$ at $\unit[2.4]{mm}$ electrode separation, respectively).
With high-contrast drifting grating type stimuli, however, the correlations
between pairs of LFP signals are shown to decrease to values around
$0.5$ at an electrode separation of $\unit[2.4]{mm}$. Also \citet{Destexhe99_4595}
analyze spatial correlations in the LFP of cat suprasylvian cortex
during awake and different sleep states, and find mean correlations
of approximately $0.6$ at $\unit[2]{mm}$ contact separation in the
awake state. These LFP correlations computed from experimental data
are highly dependent on the choice of LFP reference which may introduce
a shared signal component (which increases correlations), while the
present model LFPs are computed with the assumption of an ideal reference
electrode at infinite distance from the sources. The point neuron
and corresponding LFP model also ignore rate fluctuations in their
background input (here represented as Poisson generators with fixed
rates) which is another source of spatial correlations. Global fluctuations
or shared input correlations in the background input can be expected
to increase pairwise LFP correlations \citep{Linden11_859,Leski13_e1003137,Hagen16}.

We next bring our attention to the estimated CSD signal in panel
F. By design the chosen CSD estimation method is expected to suppress
correlations among channels due to volume conduction by reconstructing
the sink/source pattern underlying the LFP \citep{Nicholson75_356,Mitzdorf85_37,Pettersen06_116,Pettersen08_291,Potworowski12_541}.
This can, for example, allow the identification of loci of strong
synaptic activity in experimental LFP data, which may be generated
locally or due to some external drive. A brief inspection of the CSD
traces computed from the model LFP reveals that \textquoteleft standout\textquoteright{}
LFP events (e.g., in channel $31$ at $\unit[525]{ms}$) result in
fluctuations in the corresponding CSD, but the traces appear overall
more variable than the LFP. Just as for the LFP, we show the power
spectra (panel G) and pairwise correlation coefficients with distance
(panel H). In contrast to the LFP spectra, the low-gamma peak around
$\unit[50]{Hz}$ is not present in the CSD spectra, but the high-frequency
peak remains. The overall positive correlations observed for the LFP
are largely canceled for the CSD. The CSD signals are typically anti-correlated
with mean around $-0.4$ at the shortest electrode separations ($\unit[0.4]{mm}$),
and then weakly correlated ($\sim0.1$) up to $\unit[1]{mm}$. This
CSD anti-correlation across proximal channels is expected, as a fraction
of capacitive and resistive (\textquoteleft leaky\textquoteright )
transmembrane return currents of synaptic input currents exits in
vicinity to the synapse site and at the soma. The return currents
are affected by intrinsic dendritic filtering \citep{Linden10} throughout
each individual LFP-generating neuron morphology. Our multicompartment
cells are effectively treated as closed electric circuits and the
basic principle of charge conservation must apply \citep[see, for example,][]{DeSchutter09_260}.
The correlations between channels are negligible beyond $\unit[1]{mm}$
electrode separation. This negligible correlation at greater distances
reflects in part that dendrites of each morphology (cf. \figref{morphologies})
are mostly confined within $\unit[\sim300]{\mu m}$ in the lateral
directions, and that local spontaneous network interactions for this
particular network parameterization do not readily propagate across
space. It is important to point out that the CSD estimate (cf. \subsecref{current_source_density})
is based on LFPs computed at a single depth only, and would change
if LFPs across all depths were taken into account.

As an approximation to the so-called multi-unit activity (MUA) signal,
we sum up spiking activity in layer $2/3$ in the vicinity of each
LFP contact point (cf. \subsecref{multi_unit_activity}), resulting
in the signals in \figref{correlations_lfp_csd_mua_spikes}I. Similar
to the computed LFP and CSD signals, we compute power spectra (panel
J) and distance-dependent pairwise correlations among MUA signals
(\figref{correlations_lfp_csd_mua_spikes}K). In contrast to pairwise
spike-train correlations (\figref{correlations_lfp_csd_mua_spikes}B),
a sharply decaying distance dependency is observed, which is well
fit by an exponential function with spatial decay constant of $\unit[\sim0.30]{mm}$
and vanishing offset from zero at greater distances. This sharp decay
contrasts with the longer spatial decay constant observed for the
LFP, and the anti-correlation between neighboring sites as observed
for the CSD does not occur. These differences reflect that the LFP
and CSD are measures resulting from synaptically driven transmembrane
currents, while the MUA is a measure of the network spiking activity
resulting from said synaptic input. Similar to the CSD spectra,
the low-gamma oscillation around $\unit[50]{Hz}$ is not seen, while
the high-gamma oscillation around $\unit[200]{Hz}$ is pronounced.

\subsubsection{Spatial coherence of local field potential is band-passed}

So far we have established that the model LFP is highly correlated
with distance in qualitative agreement with experimental findings,
while the corresponding CSD and MUA signals are hardly correlated
beyond electrode separations of $\unit[\sim1]{mm}$. We next extend
this analysis to the frequency domain by considering distance-dependent
coherences. This step is mainly motivated by two experimental observations:
LFP coherence across channels depends on inter-electrode distance
as described by \citet{Jia11_9390,Srinath14_741}, and a recent study
by \citet{Dubey16_1986} shows that the \textquoteleft spatial spread\textquoteright{}
of LFP has band-pass properties in the gamma range ($\unit[60-150]{Hz}$).
Another modeling study \citep{Leski13_e1003137} extends the study
of LFP \textquoteleft reach\textquoteright{} by \citet{Linden11_859}
to distance-dependent coherences, showing that dendritic filtering
\citep{Linden10} introduces a low-pass effect on the LFP reach of
uncorrelated synaptic input currents with an approximately white power
spectrum. In contrast to these latter modeling studies, our combined
point-neuron network and LFP-generating setup allows accounting for
weakly correlated spiking activity in the network, at realistic density
of neurons and connections.

\begin{figure}
\includegraphics[width=1\textwidth]{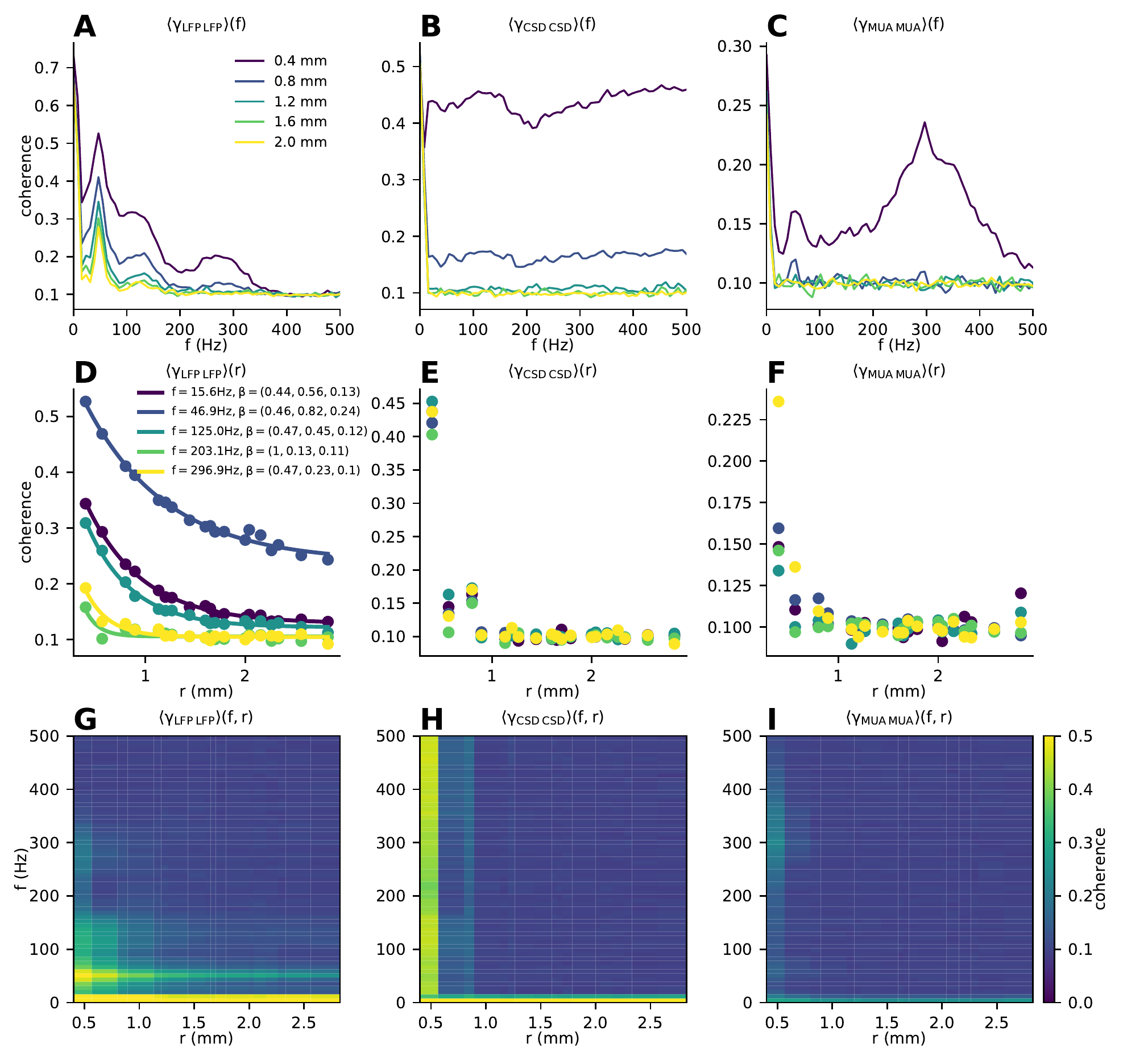}

\caption{\textbf{Distance dependency of LFP, CSD, and MUA coherences in $\mathrm{L2/3}$.}
\textbf{A}~Pairwise LFP coherences as function of frequency for different
distances (color-coded) between electrode contacts $r$, averaged
over pairs with identical electrode separation. \textbf{B}~Similar
to panel A but computed using the reconstructed CSD signal estimates
at each electrode. \textbf{C~}Similar to panel A and B but computed
using the MUA signal at each electrode. \textbf{D }Mean LFP coherences
as function of distance between electrode contacts for different frequencies
(color-coded) with exponential fit to mean values ($\mathrm{R^{2}=\{1,\,0.99,\,0.99,\,0.70,\,0.93\}}$
for $f=\{15.6,\,46.9,\,125.0,\,203.1,\,296.9\}\,\mathrm{Hz}$, respectively)\textbf{
E,F~}Mean CSD, and MUA coherences as function of distance between
electrode contacts for different frequencies (color-coded). \textbf{G\textendash I~}Color
image plot of mean LFP, CSD, and MUA coherences as function of frequency
and electrode separation. \label{fig:LFPcoherences} }
\end{figure}

From its spectra (\figref{correlations_lfp_csd_mua_spikes}D) we infer
that most of the variance in the spontaneous LFP data is due to a
high-frequency gamma oscillation above $\unit[200]{Hz}$ in the network
due to the ING mechanism present in each layer \citep{Bos16_1}. In
\figref{LFPcoherences}A we show the mean coherences $\langle\gamma_{\mathrm{LFP-LFP}}\rangle(f)$
between individual pairs of LFP signals from channels separated by
a distance $r=\unit[\left\{ 0.4,\,0.8,\,1.2,\,1.6,\,2.0\right\} ]{mm}$.
While the coherence is, as expected, highest for the lowest frequencies
($\unit[\lesssim10]{Hz}$) at all separations, it drops quickly for
frequencies $f\approx\unit[20]{Hz}$. For the shortest separation
($\unit[0.4]{mm}$), the coherence is around $0.35$ at this frequency,
and increases to $\sim0.5$ in the low-gamma range (around $\unit[50]{Hz}$).
Broader peaks in the coherence with magnitudes around $0.3$ and $2$
are also seen for $\unit[100-150]{Hz}$ and $\unit[250-300]{Hz}$,
respectively. Beyond this range, the coherence drops to around $0.1$.
The coherence across all frequencies is further reduced for increased
separations, but at $\unit[2]{mm}$ separation it still drops to the
same value of $\sim\!0.1$ at high frequencies. These model observations
resemble coherences computed for experimental LFP data during stimulus
conditions \citep[Figure 1A]{Srinath14_741}. There, a peak in low-gamma
coherence around $\unit[40]{Hz}$ is seen for distances up to $\unit[4]{mm}$
in two different subjects. Also an increase in coherence is seen for
frequencies around $\unit[80]{Hz}$. The baseline coherence (no visual
stimulus) shows no increase in the gamma range of frequencies, except
for sharp peaks seen at $\unit[100]{Hz}$ due to the CRT display frequency
and $\unit[120]{Hz}$ due to the second harmonic of noise \citep{Srinath14_741}.
This lack of gamma peaks of physiological origin differs from our
model predictions. We therefore conclude that the model LFP coherence
more closely resembles the stimulus-driven LFP, but we note also that
a baseline stationary thalamocortical activation level is assumed
in the reference network \citep{Potjans14_785}. This baseline activation
enters the $K_{X,\mathrm{ext}}^{\mathrm{u}}$ parameter for populations
$X\in\mathrm{\left\{ L4,L6\right\} \times\left\{ E,I\right\} }$ also
in the upscaled network. The corresponding mean-field theory \citep{Bos16_1}
identifies sub-circuits located in and across layers $2/3$ and $4$
as the origins of the low-gamma oscillations. Therefore, a reduced
external drive to layer $4$ (by turning off the baseline thalamic
activation altogether) should reduce the magnitude of this intrinsically
generated oscillation and consequently reduce the corresponding spatial
LFP coherence. An opposite effect on coherence can be expected by
increasing the thalamocortical drive in the model. At present we do
not pursue this possibility further. We also note that our LFP coherence
is smaller than the comparable experimental values \citep{Srinath14_741},
but see also \citet{Jia11_9390} and \citet{Dubey16_1986}. This smaller
coherence underlies the reduced correlation with distance noted above
which is likely due to the lack of temporally modulated input and
intrinsically generated low-frequency fluctuations, and the use of
an ideal reference. Some of these differences may also result from
the fact that these experimental studies rely on the multitaper method
\citep{Thomson82_1055} in order to compute coherences while we use
Welch's average periodogram (see \subsecref{statistical_measures}),
and that the experimental data have longer durations than our chosen
simulation period of $T_{\mathrm{sim}}=\unit[5]{s}$.

We next investigate the distance dependency of coherences for different
frequencies. \citet{Dubey16_1986} show an apparent band-pass effect
in the LFP, in that the phase coherence across sites is increased
and decays more slowly with distance in the gamma range compared to
higher and lower frequencies. In \figref{LFPcoherences}D we show
the LFP-LFP coherences as functions of distance for different frequencies
$f$, averaged over values computed for identical separation of channels.
We also show the corresponding least-square fit to exponential functions,
in order to investigate whether or not this model reproduces the experimentally
observed band-pass effect. Indeed, we find for $f\approx\unit[46.9]{Hz}$,
which is at the center of the low-gamma peak in panel A, an elevated
coherence with longer spatial decay constant ($\lambda=\unit[0.82]{mm}$)
than for frequencies where the overall coherence is reduced, such
as $f=\unit[15.6]{Hz}$ ($\lambda=\unit[0.56]{mm}$) and $f=\unit[203.1]{Hz}$
($\lambda=\unit[0.13]{mm}$). In the high-gamma band ($f=\unit[296.9]{Hz}$)
we again note a comparatively quick decay ($\unit[\lambda=0.23]{mm}$)
in coherence, which may reflect that the network interactions underlying
the generation of this oscillation frequency remain local. In panel
G we show the same data as displayed in panel A and D for all frequencies
up to $\unit[500]{Hz}$ and average for each distance up to $\unit[2.8]{mm}.$
As implied by the above findings, the low-gamma peak in the coherence
near $\unit[50]{Hz}$ is seen at all distances. 

In a similar manner we compute distance-dependent coherences for the
CSD (panels B,E,H) and MUA (panels C,F,I). The CSD shows only a weak
frequency dependence in its coherence at all tested distances (panel
B). The coherence is $\sim0.4$ for a contact separation of $\unit[0.4]{mm}$,
and drops to levels below $0.2$ at greater distances. The MUA coherence,
however, is increased for the shortest distances ($\unit[0.4]{mm}$)
around the high-frequency range of the high-gamma oscillation ($\unit[\sim\!300]{Hz}$)
as shown in panel C, but the coherence is at the baseline level at
all greater distances.

\section{Discussion\label{sec:discussion}}

The present work investigates a multi-layer point-neuron network
model covering $\unit[4\times4]{mm^{2}}$ of cortical surface at realistic
neuron and connection density, amounting to $\sim1.2\cdot10^{6}$
neurons and $\sim5.5\cdot10^{9}$ synapses. The model accounts for
spiking activity across excitatory and inhibitory neurons in layers
$2/3$, $4$, $5$, and $6$ and one external thalamocortical population,
as well as local field potentials (LFP). The $\unit[4\times4]{mm^{2}}$
area covered by the model is similar to the one covered by a $10\times10$
Utah multi-electrode array commonly used for electrophysiological
measurements in vivo in different cortical areas and species. The
model is a laterally extended version of the cortical microcircuit
under $\unit[1]{mm^{2}}$ of cortical surface by \citet{Potjans14_785},
but in contrast to this reference network the upscaled network accounts
for distance-dependent connection probabilities and delays. The biophysics-based
LFP predictions rely on the hybrid scheme for LFP predictions in point-neuron
networks by \citet{Hagen16}, which is here modified to account for
spatially structured networks. Earlier work has shown that correlations
are perturbed in downscaled networks \citep{Albada15}. The LFP reflects
the fluctuations caused by network correlations and depends also on
the spatial organization of networks \citep[see, for example,][]{Hagen16}.
Therefore, the development of biophysical network models that incorporate
the full density of connections as well as the spatial organization
of the observed system is crucial to aid the interpretation of the
corresponding experimental data.

Our upscaling procedure preserves the overall features of activity
in the reference network. This includes a stable network state with
asynchronous and irregular spiking activity for the different neuron
populations, distributed firing rates across neurons, spike trains
with variability in agreement with observed activity in sensory cortex,
weak pairwise spike-train correlations, and population firing rate
spectra with peaks in the low-gamma range ($\unit[40-80]{Hz}$) and
high-gamma range ($\unit[200-300]{Hz}$). Around this stable state,
we investigate the effect of varying key network parameters, namely
the weight of inhibitory connections and the external drive, as well
as the width of inhibitory connection profiles and the minimum delays.
We find that a strong external drive with reduced inhibitory feedback
results in high synchrony, that conduction delays strongly affect
the formation of temporal oscillations, and that wide inhibition results
in spatial instabilities. Furthermore, the model exhibits spatially
spreading activity evoked by thalamic pulses comparable to experiments
with a brief flash stimulus to a part of the visual field in terms
of the radial propagation speed of the evoked responses. Finally,
the model accounts for spatially correlated and coherent LFPs even
during spontaneous network activity when its pairwise spike-train
correlations are low on average. LFP coherences are distance-dependent
with a slower spatial decay around the frequency of the $\unit[50]{Hz}$
low-gamma oscillation compared to other frequencies, resulting in
an apparent band-pass filter effect on the LFP coherence.

\subsection{Comparison with other studies}

To our knowledge, this computational study is the first to simultaneously
account for both spiking activity and population activity measures
such as the LFP in a layered network model that covers several square
millimeters of cortical surface at the full density of neurons and
synaptic connections. Compared to experimentally reported cortical
neuron densities of $\sim10^{5}\text{ neurons/mm}^{2}$ \citep[see, for example,][]{Herculano-Houzel09,Ribeiro13_28},
other studies of laminar point-neuron networks with distance-dependent
connections \citep{Mehring03_395,Yger2011_229,Voges12,Rosenbaum14,Keane15_1591,Schnepel15_3818,Pyle17_18103,Rosenbaum16_107}
either rely on reducing the overall size of the network's geometry,
reduce the neuron densities per cortical area, consider only one layer
of excitatory and inhibitory neurons, or collapse all cortical layers
into one. \citet{Tomsett14_2333} also incorporate LFP predictions
from a recurrently connected network of $\sim10^{5}$ multicompartment
neurons, but consider only a thin cortical slice across layers similar
to in vitro experiments. While reduced cell and connection counts
speed up simulations, state-of-the art point-neuron simulation software
scales nearly linearly up to $\sim10^{9}$ neurons \citep{Nest2120,Jordan18_2}.
Hence, simulations of networks with $\sim10^{6}$ neurons such as
ours can be executed routinely on high-performance computing facilities. 

We here choose to start from a previously published model of the
cortical microcircuit by \citet{Potjans14_785}. In increasing the
model size, the choice of scaling procedure is critical. \citet{Albada15}
show that the reducibility (downscaling) of randomly connected asynchronous
networks is fundamentally limited if both spike rates and second-order
statistics (correlations) are to be preserved. Their proposed scaling
rules adjust the amplitudes of synaptic currents and mean and variance
of noisy background input to the decreasing numbers of synapses. However,
upscaling is different. In the microcircuit model each neuron receives
a realistic number of synapses, originating either from within the
circuit or attributed to the background. Increasing the network size
necessarily decreases the probability for two neurons to be connected.
The consideration of spatial organization, however, preserves a certain
level of local recurrence while the total network size is growing.
Consequently, our upscaling procedure works without the need to adjust
the amplitudes of synaptic currents of the reference network. The
distance-dependent connectivity results in modified in-degrees of
recurrent network connections and noisy background input such that
the mean input to each neuron is preserved, but not its variance.
As demonstrated here, the activity statistics of neurons in a $\unit[1]{mm^{2}}$
patch in the upscaled network is comparable to the statistics of the
reference network. This retrospectively validates the decisions made
in the construction of the microcircuit model by \citet{Potjans14_785}.

The modeled LFP has amplitudes in agreement with spontaneous LFP
amplitudes observed experimentally between $\unit[0.1-1]{mV}$ \citep[see, for example,][]{Maier10,Hagen15_182,Reyes-Puerta2016}.
The LFP spectra reveal a strong ongoing oscillation at high frequencies,
in the $\unit[200-300]{Hz}$ range, and around $\unit[50]{Hz}$. Spectra
of spontaneous potentials in visual cortex do not typically reveal
strong oscillations at these frequencies, but elevated LFP gamma power
in the $\unit[30-80]{Hz}$ range is frequently reported during stimuli
\citep{Jia11_9390,Ray2011,Berens08_1,Xing12b,Veit17_951,Katzner09_35,Jia13_17,Hadjipapas15_327}.
A functional role in computation and synchronization between areas
has therefore been hypothesized \citep{Ray2010,Jia13_762,Buzsaki12_203}.
The strong high-frequency oscillations here result from short interneuron
conduction delays \citep{Bos16_1}. Low frequencies are lacking in
our spontaneous LFP as our network receives external drive with a
stationary rate, does not intrinsically generate slow rate fluctuations,
and is subject to active decorrelation \citep{Tetzlaff12_e1002596},
as well as due to the assumption of an ideal reference at infinite
distance from the source. Nevertheless, the model produces highly
correlated LFPs with a distance dependence compatible with experimental
observations \citep{Destexhe99_4595,Nauhaus09_70}. The model also
reproduces elevated coherences in the low-gamma band as seen during
visual stimulation \citep{Jia11_9390,Srinath14_741}. The slower spatial
decay for frequencies around $\unit[50]{Hz}$ in the model is consistent
with a recent report of increased spatial LFP \textquoteleft reach\textquoteright{}
analogous to a spatial band-pass filter effect in the low-gamma band
\citep{Dubey16_1986}.

\subsection{Possible model refinements}

The upscaled model establishes local connections with a Gaussian
decay of connection probabilities up to a radius of $\unit[2]{mm}$.
However, pyramidal neurons can develop long horizontal axons spanning
several millimeters in addition to local axonal branching. In cat
and monkey visual cortex, these connections are typically clustered
or patchy and connect neurons with similar orientation tuning \citep{Livingstone84,Gilbert89,Bosking97,Tanigawa05,Buzas06_861,Binzegger07_12242}.
In contrast, the visual cortex of rodents exhibits a salt-and-pepper
layout without patchiness, but still some longer-distance connections
\citep{Ohki07_401,Laramee15_149}. Although less common, subsets of
inhibitory interneurons can also exhibit long-range connections \citep{McDonald93}.
\citet{Voges12} assess the influence of different types of remote
connections (none, random, or patchy) on the network activity of a
2D single-layer network, and conclude that the fraction of local versus
remote connections is crucial for the resulting network dynamics,
irrespective of the detailed spatial arrangement of remote connections.

For the type of model development conducted here, comprehensive datasets
with detailed (distance-dependent) connection probabilities are mostly
unavailable for all possible pairs of pre- and postsynaptic neuron
types and different cortical layers. Some exceptions exist \citep[for example,][]{Binzegger04},
but most connectivity studies focus on specific connections, and due
to differences in experimental methods, results may be difficult to
compare and reconcile; see, for example, \citet[Supplementary Material]{Schnepel15_3818}
on the limitations of their photostimulation technique and \citet{Stepanyants09_3555}
on truncated connections in brain slices. Neuron morphology appears
to provide a valid first approximation for the distance dependency
of connections \citep{Amirikian05,Brown09_1133,Hill12_E2885,Rees16},
but the overlap between dendrites and axons alone does not explain
connectivity patterns, due to target neuron-type specificity \citep{Potjans14_785},
specificity at the level of individual neurons \citep{Kasthuri2015_648},
and preferential locations of dendritic spines and synaptic boutons
on connected neurons \citep{Ohana12_e40601}. We make the conservative
choice to let the spatial widths of connections and shape of postsynaptic
potentials depend only on the presynaptic neuron type. Our hope is
that the algorithmic approach pursued within consortia such as the
Blue Brain Project \citep{Reimann15_1,Markram2015_456} and the Allen
Brain Institute \citep{Kandel13_659} will provide more accurate neuronal
connectomes of different brain regions across species in the future,
including their distance dependencies (as in, for instance, \citet{Reimann17}
for rat somatosensory cortex). 

Activity in finite-sized laminar networks is subject to effects that
depend on the choice of boundary conditions. Periodic boundary conditions
are frequently used in 1D networks (ring networks) \citep{Roxin05,Kriener14,Rosenbaum14}
and in 2D networks with torus connectivity \citep{Mehring03_395,Yger2011_229,Voges12,Rosenbaum14,Keane15_1591,Schnepel15_3818,Pyle17_18103,Rosenbaum16_107}
as also used here for the upscaled models. The model of a cortical
slice by \citet{Tomsett14_2333} incorporates connections only within
the confines of the modeled slice, but we consider networks that are
part of a larger system (the intact brain). An advantage of periodic
boundaries is the simplifying assumption that cortex is homogeneous
and isotropic, that is, the connectivity of a neuron is independent
of its location in the network. One disadvantage is that the maximal
distance for connections is only $L/2$ for a ring domain with circumference
$L$, or $L/\sqrt{2}$ for a square domain with side length $L$.
Here, we restrict connections to a radius $R=L/2$. Another disadvantage
is that propagating activity may travel across the boundary and directly
influence its own propagation, resulting in for example wave-front
annihilation \citep{Muller18_255}. An option to suppress such effects
would be to simulate a larger network and to sample only the activity
of neurons across a smaller domain. In Figures \ref{fig:rasters_ref_interm_ups}
and \ref{fig:statistics_ref_interm_ups}, we extract activity of neurons
within a center disc of $\unit[1]{mm^{2}}$. The network could be
further upscaled, for example to cover a full cortical area. The lateral
size of the unfolded cat striate cortex in one hemisphere is larger
by a factor of almost $25$ than the currently simulated upscaled
network of $\unit[16]{mm^{2}}$, estimated in the range of $\unit[310-400]{\mu m}$
\citep{Tusa78_213,VanEssen80_255,Olavarria85_191,Anderson88_2183}
). Striate cortex in macaque monkeys is even two to four times larger
than in cats \citep{VanEssen80_255}. Networks of a full cortical
area could also address the effects of borders to adjacent cortical
areas. Anatomical borders between distinct areas are shown to affect
wave propagation \citep{Xu07_119,Muller14_4675}.

Spontaneous activity in our models is driven by uncorrelated external
inputs with a fixed rate and Poisson statistics, to represent missing
connections from remote and neighboring cortices, subcortical structures,
and sensory inputs. Ongoing work aims to account for the structure
of one hemisphere of macaque vision-related cortex in a spiking model
\citep{Schuecker17,Schmidt18_1409}. Mutual interactions between recurrently
connected areas can be expected to profoundly affect their input statistics
in terms of rates, spectra, and correlations. Furthermore, we simulate
evoked potentials by short thalamic pulses of activity, but sensory
cortex receives continuously varying inputs. Ever more detailed models
of, for example, the response properties of relay cells in visual
thalamus are emerging \citep{Martinez-Canada18_e1005930}, representing
naturalistic image or movie stimuli to cortical models similar to
ours.

Activity statistics such as distributions of correlations depend
on simulation length \citep{Tetzlaff08_2133}. Here, we consider $\unit[5]{s}$
simulations, but experimental recordings are often longer (for example,
\citealp{Pan13_288,Chu2014_113,Chu2014_data}). Future work can address
how greater simulation durations affect the activity statistics, and
their convergence across time.

In terms of signal predictions, the tool LFPy (\href{http://LFPy.rtdf.io}{LFPy.rtdf.io})
embedded in the presently used hybrid scheme \citep{Hagen16}, facilitates
the calculation of current dipole moments of individual neurons and
associated contributions to electroencephalographic (EEG) signals
and magnetoencephalographic (MEG) signals as recorded on the surface
of the head \citep{Hamalainen93_413,Nunez06_2ed,Hagen18_biorxiv_281717}.
Forward-model predictions of macroscopic signals like EEG and MEG
are thus a tempting proposition, in particular under the consideration
of mutual interactions between areas. Among other applications, this
could provide an avenue towards a mechanistic model and understanding
of visually evoked potentials \citep{Sokol76_18}.

\subsection{Significance of work}

The present work represents a stepping-stone for understanding experimental
data obtained by multi-electrode arrays that cover several square
millimeters of cortical space. While the model description is highly
reduced, it simultaneously accounts for spiking activity and LFPs
and thereby enables a multi-scale comparison with corresponding experimental
data. At the same time, its simplicity makes mathematical analysis
in terms of mean-field and neural field theory viable \citep{Bos16_1,Senk18_arxiv_06046v1}.
Our hope is that the model facilitates a more principal understanding
of the dependence of spike correlations on distance, spatially coherent
and correlated LFPs, spike-LFP relationships, and emergent spatiotemporal
patterns such as waves. The article describes not only a particular
network model but a fully digitized ``integrative loop''. We therefore
envision the model as a starting point and building block for future
work iteratively modifying parameters and adding further constraints
to generate predictions for the activity of specific brain areas.


\begin{thebibliography}{}

\bibitem[\protect\citeauthoryear{Abeles}{1991}]{Abeles91}
Abeles M (1991)
\newblock {\em Corticonics: Neural Circuits of the Cerebral Cortex}
\newblock Cambridge University Press, Cambridge, 1st edition.

\bibitem[\protect\citeauthoryear{Amirikian}{2005}]{Amirikian05}
Amirikian B (2005)
\newblock A phenomenological theory of spatially structured local synaptic
  connectivity.
\newblock PLoS Comput. Biol.~1:\mbox{e11}.

\bibitem[\protect\citeauthoryear{Andersen  \bgroup et al.\egroup
  }{1978}]{Andersen78_11}
Andersen P, Silfvenius H, Sundberg SH, Sveen O, Wigstr\"{o}m H (1978)
\newblock Functional characteristics of unmyelinated fibres in the hippocampal
  cortex.
\newblock Brain Res.~144:\mbox{11--18}.

\bibitem[\protect\citeauthoryear{Anderson \bgroup et al.\egroup
  }{1988}]{Anderson88_2183}
Anderson PA, Olavarria J, Sluyters RCV (1988)
\newblock The overall pattern of ocular dominance bands in cat visual cortex.
\newblock J. Neurosci.~8:\mbox{2183--2200}.

\bibitem[\protect\citeauthoryear{Ascoli \bgroup et al.\egroup
  }{2007}]{Ascoli07_9247}
Ascoli GA, Donohue DE, Halavi M (2007)
\newblock {NeuroMorpho}.org: A central resource for neuronal morphologies.
\newblock J. Neurosci.~27:\mbox{9247--9251}.

\bibitem[\protect\citeauthoryear{Benucci \bgroup et al.\egroup
  }{2007}]{Benucci07_103}
Benucci A, Frazor RA, Carandini M (2007)
\newblock Standing waves and traveling waves distinguish two circuits in visual
  cortex.
\newblock Neuron~55:\mbox{103--117}.

\bibitem[\protect\citeauthoryear{Berens  \bgroup et al.\egroup
  }{2008}]{Berens08_1}
Berens P, Keliris GA, Ecker AS, Logothetis NK, Tolias AS (2008)
\newblock Comparing the feature selectivity of the gamma-band of the local
  field potential and the underlying spiking activity in primate visual cortex.
\newblock Front. Syst. Neurosci.~2:\mbox{1--9}.

\bibitem[\protect\citeauthoryear{Berg-Johnsen and
  Langmoen}{1992}]{Berg-Johnsen92_319}
Berg-Johnsen J, Langmoen IA (1992)
\newblock Temperature sensitivity of thin unmyelinated fibers in rat
  hippocampal cortex.
\newblock Brain Res.~576:\mbox{319--321}.

\bibitem[\protect\citeauthoryear{Binzegger \bgroup et al.\egroup
  }{2004}]{Binzegger04}
Binzegger T, Douglas RJ, Martin KAC (2004)
\newblock A quantitative map of the circuit of cat primary visual cortex.
\newblock J. Neurosci.~39:\mbox{8441--8453}.

\bibitem[\protect\citeauthoryear{Binzegger \bgroup et al.\egroup
  }{2007}]{Binzegger07_12242}
Binzegger T, Douglas RJ, Martin KAC (2007)
\newblock Stereotypical bouton clustering of individual neurons in cat primary
  visual cortex.
\newblock J. Neurosci.~27:\mbox{12242--12254}.

\bibitem[\protect\citeauthoryear{B\"{o}rgers and Kopell}{2003}]{Boergers03}
B\"{o}rgers C, Kopell N (2003)
\newblock Synchronization in networks of excitatory and inhibitory neurons with
  sparse, random connectivity.
\newblock Neural Comput.~15:\mbox{509--538}.

\bibitem[\protect\citeauthoryear{B\"{o}rgers and Kopell}{2005}]{Boergers05_557}
B\"{o}rgers C, Kopell N (2005)
\newblock Effects of noisy drive on rhythms in networks of excitatory and
  inhibitory neurons.
\newblock Neural Comput.~17:\mbox{557--608}.

\bibitem[\protect\citeauthoryear{Bos \bgroup et al.\egroup }{2016}]{Bos16_1}
Bos H, Diesmann M, Helias M (2016)
\newblock Identifying anatomical origins of coexisting oscillations in the
  cortical microcircuit.
\newblock PLoS Comput. Biol.~12:\mbox{e1005132}.

\bibitem[\protect\citeauthoryear{Bosking  \bgroup et al.\egroup
  }{1997}]{Bosking97}
Bosking WH, Zhang Y, Schofield B, Fitzpatrick D (1997)
\newblock Orientation selectivity and the arrangement of horizontal connections
  in tree shrew striate cortex.
\newblock J. Neurosci.~17:\mbox{2112--2127}.

\bibitem[\protect\citeauthoryear{Boucsein  \bgroup et al.\egroup
  }{2011}]{Boucsein11_1}
Boucsein C, Nawrot M, Schnepel P, Aertsen A (2011)
\newblock Beyond the cortical column: abundance and physiology of horizontal
  connections imply a strong role for inputs from the surround.
\newblock Front. Neurosci.~5:\mbox{32}.

\bibitem[\protect\citeauthoryear{Bressloff}{2012}]{Bressloff12}
Bressloff PC (2012)
\newblock Spatiotemporal dynamics of continuum neural fields.
\newblock Journal of Physics A: Mathematical and Theoretical~45:\mbox{033001}.

\bibitem[\protect\citeauthoryear{Bringuier  \bgroup et al.\egroup
  }{1999}]{Bringuier99_695}
Bringuier V, Chavane F, Glaeser L, Fr\'{e}gnac Y (1999)
\newblock Horizontal propagation of visual activity in the synaptic integration
  field of area 17 neurons.
\newblock Science~283:\mbox{695--699}.

\bibitem[\protect\citeauthoryear{Brown and Hestrin}{2009}]{Brown09_1133}
Brown SP, Hestrin S (2009)
\newblock Intracortical circuits of pyramidal neurons reflect their long-range
  axonal targets.
\newblock Nature~457:\mbox{1133--1136}.

\bibitem[\protect\citeauthoryear{Brunel}{2000}]{Brunel00_183}
Brunel N (2000)
\newblock Dynamics of sparsely connected networks of excitatory and inhibitory
  spiking neurons.
\newblock J. Comput. Neurosci.~8:\mbox{183--208}.

\bibitem[\protect\citeauthoryear{Brunel and Hakim}{1999}]{Brunel99}
Brunel N, Hakim V (1999)
\newblock Fast global oscillations in networks of integrate-and-fire neurons
  with low firing rates.
\newblock Neural Comput.~11:\mbox{1621--1671}.

\bibitem[\protect\citeauthoryear{Budd and Kisv\'{a}rday}{2001}]{Budd01}
Budd JM, Kisv\'{a}rday ZF (2001)
\newblock Local lateral connectivity of inhibitory clutch cells in layer 4 of
  cat visual cortex (area 17).
\newblock Exp. Brain Res.~140:\mbox{245--250}.

\bibitem[\protect\citeauthoryear{Buz\'{a}s  \bgroup et al.\egroup
  }{2006}]{Buzas06_861}
Buz\'{a}s P, Kov\'{a}cs K, Ferecsk\'{o} AS, Budd JML, Eysel UT, Kisv\'{a}rday
  ZF (2006)
\newblock Model-based analysis of excitatory lateral connections in the visual
  cortex.
\newblock J. Comp. Neurol.~499:\mbox{861--881}.

\bibitem[\protect\citeauthoryear{Buzs\'{a}ki \bgroup et al.\egroup
  }{2012}]{Buzsaki12_1322}
Buzs\'{a}ki G, Anastassiou CA, Koch C (2012)
\newblock The origin of extracellular fields and currents -- {EEG, ECoG, LFP
  and spikes}.
\newblock Nat. Rev. Neurosci.~13:\mbox{407--427}.

\bibitem[\protect\citeauthoryear{Buzs{\'a}ki and
  Mizuseki}{2014}]{buzsaki14_264}
Buzs{\'a}ki G, Mizuseki K (2014)
\newblock The log-dynamic brain: how skewed distributions affect network
  operations.
\newblock Nat. Rev. Neurosci.~15:\mbox{264--278}.

\bibitem[\protect\citeauthoryear{Buzs\'{a}ki and Wang}{2012}]{Buzsaki12_203}
Buzs\'{a}ki G, Wang XJ (2012)
\newblock Mechanisms of gamma oscillations.
\newblock Annu. Rev. Neurosci.~35:\mbox{203--225}.

\bibitem[\protect\citeauthoryear{Cain  \bgroup et al.\egroup
  }{2016}]{Cain16_e1005045}
Cain N, Iyer R, Koch C, Mihalas S (2016)
\newblock The computational properties of a simplified cortical column model.
\newblock PLoS Comput. Biol.~12:\mbox{e1005045}.

\bibitem[\protect\citeauthoryear{Camu\~{n}as Mesa and
  Quiroga}{2013}]{Camunas-Mesa2013}
Camu\~{n}as Mesa LA, Quiroga RQ (2013)
\newblock A detailed and fast model of extracellular recordings.
\newblock Neural Comput.~25:\mbox{1191--1212}.

\bibitem[\protect\citeauthoryear{Carnevale and Hines}{2006}]{Carnevale06}
Carnevale NT, Hines ML (2006)
\newblock {\em The {NEURON} Book}
\newblock Cambridge University Press, Cambridge.

\bibitem[\protect\citeauthoryear{Chow  \bgroup et al.\egroup }{1998}]{Chow1998}
Chow CC, White JA, Ritt J, Kopell N (1998)
\newblock Frequency {Control} in {Synchronized} {Networks} of {Inhibitory}
  {Neurons}.
\newblock J. Comput. Neurosci.~5:\mbox{407--420}.

\bibitem[\protect\citeauthoryear{Chu \bgroup et al.\egroup
  }{2014a}]{Chu2014_data}
Chu CCJ, Chien PF, Hung CP (2014a)
\newblock Multi-electrode recordings of ongoing activity and responses to
  parametric stimuli in macaque {V1}.
\newblock CRCNS.org~.

\bibitem[\protect\citeauthoryear{Chu \bgroup et al.\egroup
  }{2014b}]{Chu2014_113}
Chu CCJ, Chien PF, Hung CP (2014b)
\newblock Tuning dissimilarity explains short distance decline of spontaneous
  spike correlation in macaque {V1}.
\newblock Vision Res.~96:\mbox{113--132}.

\bibitem[\protect\citeauthoryear{Contreras \bgroup et al.\egroup
  }{1997}]{Contreras97_335}
Contreras D, Destexhe A, Steriade M (1997)
\newblock Intracellular and computational characterization of the intracortical
  inhibitory control of synchronized thalamic inputs in vivo.
\newblock J. Neurophysiol.~78:\mbox{335--350}.

\bibitem[\protect\citeauthoryear{Coombes}{2005}]{Coombes05}
Coombes S (2005)
\newblock Waves, bumps, and patterns in neural field theories.
\newblock Biol. Cybern.~93:\mbox{91--108}.

\bibitem[\protect\citeauthoryear{De~Schutter and
  Van~Geit}{2009}]{DeSchutter09_260}
De~Schutter E, Van~Geit W (2009)
\newblock Modeling complex neurons
\newblock In De~Schutter E, editor, {\em Computational Modeling Methods for
  Neuroscientists}, chapter~11, \mbox{pp. 260--283}. MIT Press, Cambridge, MA,
  1st edition.

\bibitem[\protect\citeauthoryear{Denker  \bgroup et al.\egroup
  }{2018}]{Denker18_1}
Denker M, Zehl L, Kilavik BE, Diesmann M, Brochier T, Riehle A, Gr\"{u}n S
  (2018)
\newblock {LFP} beta amplitude is linked to mesoscopic spatio-temporal phase
  patterns.
\newblock Scientific Reports~8:\mbox{1--21}.

\bibitem[\protect\citeauthoryear{Destexhe \bgroup et al.\egroup
  }{1999}]{Destexhe99_4595}
Destexhe A, Contreras D, Steriade M (1999)
\newblock Spatiotemporal analysis of local field potentials and unit discharges
  in cat cerebral cortex during natural wake and sleep states.
\newblock J. Neurosci.~19:\mbox{4595--4608}.

\bibitem[\protect\citeauthoryear{Doiron  \bgroup et al.\egroup
  }{2016}]{Doiron16_383}
Doiron B, Litwin-Kumar A, Rosenbaum R, Ocker GK, Josi\'{c} K (2016)
\newblock The mechanics of state-dependent neural correlations.
\newblock Nat. Neurosci.~19:\mbox{383--393}.

\bibitem[\protect\citeauthoryear{Douglas \bgroup et al.\egroup
  }{1989}]{Douglas89_480}
Douglas RJ, Martin KAC, Whitteridge D (1989)
\newblock A canonical microcircuit for neocortex.
\newblock Neural Comput.~1:\mbox{480--488}.

\bibitem[\protect\citeauthoryear{Dubey and Ray}{2016}]{Dubey16_1986}
Dubey A, Ray S (2016)
\newblock Spatial spread of local field potential is band-pass in the primary
  visual cortex.
\newblock J. Neurophysiol.~116:\mbox{1986--1999}.

\bibitem[\protect\citeauthoryear{Ecker  \bgroup et al.\egroup }{2010}]{Ecker10}
Ecker AS, Berens P, Keliris GA, Bethge M, Logothetis NK (2010)
\newblock Decorrelated neuronal firing in cortical microcircuits.
\newblock Science~327:\mbox{584--587}.

\bibitem[\protect\citeauthoryear{Einevoll  \bgroup et al.\egroup
  }{2013a}]{Einevoll13_770}
Einevoll GT, Kayser C, Logothetis NK, Panzeri S (2013a)
\newblock Modelling and analysis of local field potentials for studying the
  function of cortical circuits.
\newblock Nat. Rev. Neurosci.~14:\mbox{770--785}.

\bibitem[\protect\citeauthoryear{Einevoll  \bgroup et al.\egroup
  }{2013b}]{Einevoll13_37}
Einevoll GT, Lind\'{e}n H, Tetzlaff T, \L\c{e}ski S, Pettersen KH (2013b)
\newblock Local field potentials: {B}iophysical origin and analysis
\newblock In Quiroga RQ, Panzeri S, editors, {\em Principles of Neural Coding},
  \mbox{pp. 37--59}. CRC Press.

\bibitem[\protect\citeauthoryear{Einevoll  \bgroup et al.\egroup
  }{2007}]{Einevoll07_2174}
Einevoll GT, Pettersen KH, Devor A, Ulbert I, Halgren E, Dale AM (2007)
\newblock Laminar population analysis: Estimating firing rates and evoked
  synaptic activity from multielectrode recordings in rat barrel cortex.
\newblock J. Neurophysiol.~97:\mbox{2174--2190}.

\bibitem[\protect\citeauthoryear{Ermentrout}{1998}]{Ermentrout98b}
Ermentrout B (1998)
\newblock Neural networks as spatio-temporal pattern-forming systems.
\newblock Reports on Progress in Physics~61:\mbox{353--430}.

\bibitem[\protect\citeauthoryear{Gewaltig and
  Diesmann}{2007}]{Gewaltig_07_11204}
Gewaltig MO, Diesmann M (2007)
\newblock {NEST} ({NE}ural {S}imulation {T}ool).
\newblock Scholarpedia~2:\mbox{1430}.

\bibitem[\protect\citeauthoryear{Gilbert and Wiesel}{1989}]{Gilbert89}
Gilbert CD, Wiesel TN (1989)
\newblock Columnar specificity of intrinsic horizontal and corticocortical
  connections in cat visual cortex.
\newblock J. Neurosci.~9:\mbox{2432--2442}.

\bibitem[\protect\citeauthoryear{Grinvald  \bgroup et al.\egroup
  }{1994}]{Grinvald94_2545}
Grinvald A, Lieke EE, Frostig RD, Hildesheim R (1994)
\newblock Cortical point-spread function and long-range lateral interactions
  revealed by real-time optical imaging of macaque monkey primary visual
  cortex.
\newblock J. Neurosci.~14:\mbox{2545--2568}.

\bibitem[\protect\citeauthoryear{Hadjipapas  \bgroup et al.\egroup
  }{2015}]{Hadjipapas15_327}
Hadjipapas A, Lowet E, Roberts M, Peter A, Weerd PD (2015)
\newblock Parametric variation of gamma frequency and power with luminance
  contrast: {A} comparative study of human {MEG} and monkey {LFP} and spike
  responses.
\newblock NeuroImage~112:\mbox{327--340}.

\bibitem[\protect\citeauthoryear{Hagen  \bgroup et al.\egroup }{2016}]{Hagen16}
Hagen E, Dahmen D, Stavrinou ML, Lind\'{e}n H, Tetzlaff T, van Albada SJ,
  Gr\"{u}n S, Diesmann M, Einevoll GT (2016)
\newblock Hybrid scheme for modeling local field potentials from point-neuron
  networks.
\newblock Cereb. Cortex~26:\mbox{4461--4496}.

\bibitem[\protect\citeauthoryear{Hagen  \bgroup et al.\egroup
  }{2018}]{Hagen18_biorxiv_281717}
Hagen E, N{\ae}ss S, Ness TV, Einevoll GT (2018)
\newblock Multimodal modeling of neural network activity: computing {LFP},
  {ECoG}, {EEG} and {MEG} signals with {LFPy2.0}.
\newblock bioRxiv~.

\bibitem[\protect\citeauthoryear{Hagen  \bgroup et al.\egroup
  }{2015}]{Hagen15_182}
Hagen E, Ness TV, Khosrowshahi A, S{\o}rensen C, Fyhn M, Hafting T, Franke F,
  Einevoll GT (2015)
\newblock {ViSAPy}: A python tool for biophysics-based generation of virtual
  spiking activity for evaluation of spike-sorting algorithms.
\newblock J. Neurosci. Meth.~245:\mbox{182--204}.

\bibitem[\protect\citeauthoryear{Hagen  \bgroup et al.\egroup
  }{2016}]{Hagen16_cns_P167}
Hagen E, Senk J, van Albada SJ, Diesmann M (2016)
\newblock Local field potentials in a 4x4mm2 multi-layered network model
\newblock In {\em 25th Annual Computational Neuroscience Meeting: CNS-2016, BMC
  Neuroscience 2016, 17(Suppl 1):P167}.

\bibitem[\protect\citeauthoryear{Hahne  \bgroup et al.\egroup
  }{2017}]{Hahne17_34}
Hahne J, Dahmen D, Schuecker J, Frommer A, Bolten M, Helias M, Diesmann M
  (2017)
\newblock Integration of continuous-time dynamics in a spiking neural network
  simulator.
\newblock Front. Neuroinformatics~11:\mbox{34}.

\bibitem[\protect\citeauthoryear{H\"{a}m\"{a}l\"{a}inen  \bgroup et al.\egroup
  }{1993}]{Hamalainen93_413}
H\"{a}m\"{a}l\"{a}inen M, Haari R, Ilmoniemi RJ, Knuutila J, Lounasmaa OV
  (1993)
\newblock Magnetoencephalography --- theory, instrumentation, and application
  to noninvasive studies of the working human brain.
\newblock Rev. Mod. Phys.~65:\mbox{413--496}.

\bibitem[\protect\citeauthoryear{Hao \bgroup et al.\egroup }{2016}]{Hao16_104}
Hao Y, Riehle A, Brochier TG (2016)
\newblock Mapping horizontal spread of activity in monkey motor cortex using
  single pulse microstimulation.
\newblock Frontiers in neural circuits~10:\mbox{104}.

\bibitem[\protect\citeauthoryear{Hardingham and
  Larkman}{1998}]{Hardingham98_249}
Hardingham NR, Larkman AU (1998)
\newblock The reliability of excitatory synaptic transmission in slices of rat
  visual cortex in vitro is temperature dependent.
\newblock J. Physiol. (Lond.)~507:\mbox{249--256}.

\bibitem[\protect\citeauthoryear{Helias \bgroup et al.\egroup
  }{2014}]{Helias14}
Helias M, Tetzlaff T, Diesmann M (2014)
\newblock The correlation structure of local cortical networks intrinsically
  results from recurrent dynamics.
\newblock PLoS Comput. Biol.~10:\mbox{e1003428}.

\bibitem[\protect\citeauthoryear{Hellwig}{2000}]{Hellwig00_111}
Hellwig B (2000)
\newblock A quantitative analysis of the local connectivity between pyramidal
  neurons in layers 2/3 of the rat visual cortex.
\newblock Biol. Cybern.~2:\mbox{111--121}.

\bibitem[\protect\citeauthoryear{Herculano-Houzel}{2009}]{Herculano-Houzel09}
Herculano-Houzel S (2009)
\newblock The human brain in numbers: a linearly scaled-up primate brain.
\newblock Front. Hum. Neurosci.~3:\mbox{31}.

\bibitem[\protect\citeauthoryear{Hill  \bgroup et al.\egroup
  }{2012}]{Hill12_E2885}
Hill SL, Wang Y, Riachi I, Sch{\"u}rmann F, Markram H (2012)
\newblock Statistical connectivity provides a sufficient foundation for
  specific functional connectivity in neocortical neural microcircuits.
\newblock Proc. Natl. Acad. Sci. USA~109:\mbox{E2885--E2894}.

\bibitem[\protect\citeauthoryear{Hines \bgroup et al.\egroup
  }{2009}]{Hines09_1}
Hines ML, Davison AP, Muller E (2009)
\newblock {NEURON} and python.
\newblock Front. Neuroinformatics~3:\mbox{1}.

\bibitem[\protect\citeauthoryear{Hirsch and Gilbert}{1991}]{Hirsch91_1800}
Hirsch JA, Gilbert CD (1991)
\newblock Synaptic physiology of horizontal connections in the cat's visual
  cortex.
\newblock J. Neurosci.~11:\mbox{1800--1809}.

\bibitem[\protect\citeauthoryear{Holt and Koch}{1999}]{Holt99_169}
Holt GR, Koch C (1999)
\newblock Electrical interactions via the extracellular potential near cell
  bodies.
\newblock J. Comput. Neurosci.~6:\mbox{169--184}.

\bibitem[\protect\citeauthoryear{Ian~Nauhaus and
  Carandini}{2012}]{Nauhaus12_3088}
Ian~Nauhaus LB DLR, Carandini M (2012)
\newblock Robustness of traveling waves in ongoing activity of visual cortex.
\newblock J. Neurosci.~32:\mbox{3088--3094}.

\bibitem[\protect\citeauthoryear{Izhikevich and
  Edelman}{2008}]{Izhikevich08_3593}
Izhikevich EM, Edelman GM (2008)
\newblock Large-scale model of mammalian thalamocortical systems.
\newblock Proc. Natl. Acad. Sci. USA~105:\mbox{3593--3598}.

\bibitem[\protect\citeauthoryear{Jia \bgroup et al.\egroup }{2011}]{Jia11_9390}
Jia X, Smith MA, Kohn A (2011)
\newblock Stimulus selectivity and spatial coherence of gamma components of the
  local field potential.
\newblock J. Neurosci.~31:\mbox{9390--9403}.

\bibitem[\protect\citeauthoryear{Jia \bgroup et al.\egroup }{2013a}]{Jia13_762}
Jia X, Tanabe S, Kohn A (2013a)
\newblock Gamma and the coordination of spiking activity in early visual
  cortex.
\newblock Neuron~77:\mbox{762--774}.

\bibitem[\protect\citeauthoryear{Jia \bgroup et al.\egroup }{2013b}]{Jia13_17}
Jia X, Xing D, Kohn A (2013b)
\newblock No consistent relationship between gamma power and peak frequency in
  macaque primary visual cortex.
\newblock J. Neurosci.~33:\mbox{17--25}.

\bibitem[\protect\citeauthoryear{Jiang  \bgroup et al.\egroup
  }{2015}]{Jiang2015}
Jiang X, Shen S, Cadwell CR, Berens P, Sinz F, Ecker AS, Patel S, Tolias AS
  (2015)
\newblock Principles of connectivity among morphologically defined cell types
  in adult neocortex.
\newblock Science~350:\mbox{aac9462--aac9462}.

\bibitem[\protect\citeauthoryear{Jordan  \bgroup et al.\egroup
  }{2018}]{Jordan18_2}
Jordan J, Ippen T, Helias M, Kitayama I, Sato M, Igarashi J, Diesmann M, Kunkel
  S (2018)
\newblock Extremely scalable spiking neuronal network simulation code: From
  laptops to exascale computers.
\newblock Front. Neuroinformatics~12:\mbox{2}.

\bibitem[\protect\citeauthoryear{Kajikawa and Schroeder}{2011}]{Kajikawa11_847}
Kajikawa Y, Schroeder CE (2011)
\newblock How local is the local field potential?
\newblock Neuron~72:\mbox{847--858}.

\bibitem[\protect\citeauthoryear{Kandel  \bgroup et al.\egroup
  }{2013}]{Kandel13_659}
Kandel ER, Markram H, Matthews PM, Yuste R, Koch C (2013)
\newblock Neuroscience thinks big (and collaboratively).
\newblock Nat. Rev. Neurosci.~14:\mbox{659--664}.

\bibitem[\protect\citeauthoryear{Kang  \bgroup et al.\egroup
  }{1994}]{Kang94_280}
Kang Y, Kaneko T, Ohishi H, Endo K, Araki T (1994)
\newblock Spatiotemporally differential inhibition of pyramidal cells in the
  cat motor cortex.
\newblock J. Neurophysiol.~71:\mbox{280--293}.

\bibitem[\protect\citeauthoryear{Kasthuri  \bgroup et al.\egroup
  }{2015}]{Kasthuri2015_648}
Kasthuri N, Hayworth KJ, Berger DR, Schalek RL, Conchello JA, Knowles-Barley S,
  Lee D, V{\'a}zquez-Reina A, Kaynig V, Jones TR et~al. (2015)
\newblock Saturated reconstruction of a volume of
  neocortex~162:\mbox{648--661}.

\bibitem[\protect\citeauthoryear{Katz and Miledi}{1965}]{Katz65_656}
Katz B, Miledi R (1965)
\newblock The effect of temperature on the synaptic delay at the neuromuscular
  junction.
\newblock J. Physiol. (Lond.)~181:\mbox{656--670}.

\bibitem[\protect\citeauthoryear{K\"{a}tzel  \bgroup et al.\egroup
  }{2011}]{Katzel2011_100}
K\"{a}tzel D, Zemelman BV, Buetfering C, W\"{o}lfel M, Miesenb\"{o}ck G (2011)
\newblock The columnar and laminar organization of inhibitory connections to
  neocortical excitatory cells.
\newblock Nat. Neurosci.~14:\mbox{100--107}.

\bibitem[\protect\citeauthoryear{Katzner  \bgroup et al.\egroup
  }{2009}]{Katzner09_35}
Katzner S, Nauhaus I, Benucci A, Bonin V, Ringach DL, Carandini M (2009)
\newblock Local origin of field potentials in visual cortex.
\newblock Neuron~61:\mbox{35--41}.

\bibitem[\protect\citeauthoryear{Keane and Gong}{2015}]{Keane15_1591}
Keane A, Gong P (2015)
\newblock Propagating waves can explain irregular neural dynamics.
\newblock J. Neurosci.~35:\mbox{1591--1605}.

\bibitem[\protect\citeauthoryear{Kisv\'{a}rday and Eysel}{1992}]{Kisvarday92}
Kisv\'{a}rday ZF, Eysel UT (1992)
\newblock Cellular organization of reciprocal patchy networks in layer {III} of
  cat visual cortex (area 17).
\newblock Neuroscience~46:\mbox{275--286}.

\bibitem[\protect\citeauthoryear{Klein  \bgroup et al.\egroup
  }{2016}]{Klein16_143}
Klein C, Evrard HC, Shapcott KA, Haverkamp S, Logothetis NK, Schmid MC (2016)
\newblock Cell-targeted optogenetics and electrical microstimulation reveal the
  primate koniocellular projection to supra-granular visual cortex.
\newblock Neuron~90:\mbox{143--151}.

\bibitem[\protect\citeauthoryear{Kriener  \bgroup et al.\egroup
  }{2009}]{Kriener09_177}
Kriener B, Helias M, Aertsen A, Rotter S (2009)
\newblock Correlations in spiking neuronal networks with distance dependent
  connections.
\newblock J. Comput. Neurosci.~27:\mbox{177--200}.

\bibitem[\protect\citeauthoryear{Kriener  \bgroup et al.\egroup
  }{2014}]{Kriener14}
Kriener B, Helias M, Rotter S, Diesmann M, Einevoll GT (2014)
\newblock How pattern formation in ring networks of excitatory and inhibitory
  spiking neurons depends on the input current regime.
\newblock Front. Comput. Neurosci.~7:\mbox{1--21}.

\bibitem[\protect\citeauthoryear{Kunkel  \bgroup et al.\egroup
  }{2017}]{Nest2120}
Kunkel S, Morrison A, Weidel P, Eppler JM, Sinha A, Schenck W, Schmidt M,
  Vennemo SB, Jordan J, Peyser A, Plotnikov D, Graber S, Fardet T, Terhorst D,
  M{\o}rk H, Trensch G, Seeholzer A, Deepu R, Hahne J, Blundell I, Ippen T,
  Schuecker J, Bos H, Diaz S, Hagen E, Mahmoudian S, Bachmann C, Lepper{\o}d
  ME, Breitwieser O, Golosio B, Rothe H, Setareh H, Djurfeldt M, Schumann T,
  Shusharin A, Garrido J, Muller EB, Rao A, Vieites JH, Plesser HE (2017)
\newblock Nest 2.12.0.

\bibitem[\protect\citeauthoryear{Laram\'{e}e and Boire}{2015}]{Laramee15_149}
Laram\'{e}e ME, Boire D (2015)
\newblock Visual cortical areas of the mouse: comparison of parcellation and
  network structure with primates~8:\mbox{149}.

\bibitem[\protect\citeauthoryear{Larkum \bgroup et al.\egroup
  }{2001}]{Larkum01_447}
Larkum ME, Zhu JJ, Sakmann B (2001)
\newblock Dendritic mechanisms underlying the coupling of the dendritic with
  the axonal action potential initiation zone of adult rat layer 5 pyramidal
  neurons.
\newblock J. Physiol. (Lond.)~533:\mbox{447--466}.

\bibitem[\protect\citeauthoryear{{\L}{\k{e}}ski  \bgroup et al.\egroup
  }{2013}]{Leski13_e1003137}
{\L}{\k{e}}ski S, Lind\'{e}n H, Tetzlaff T, Pettersen KH, Einevoll GT (2013)
\newblock Frequency dependence of signal power and spatial reach of the local
  field potential.
\newblock PLoS Comput. Biol.~9:\mbox{e1003137}.

\bibitem[\protect\citeauthoryear{{\L}{\k{e}}ski  \bgroup et al.\egroup
  }{2011}]{Leski11_401}
{\L}{\k{e}}ski S, Pettersen KH, Tunstall B, Einevoll GT, Gigg J, W{\'o}jcik DK
  (2011)
\newblock Inverse current source density method in two dimensions: Inferring
  neural activation from multielectrode recordings.
\newblock Neuroinformatics~9:\mbox{401--425}.

\bibitem[\protect\citeauthoryear{Leung}{1982}]{Leung1982}
Leung LS (1982)
\newblock Nonlinear feedback model of neuronal populations in hippocampal {CAl}
  region.
\newblock J. Neurophysiol.~47:\mbox{845--868}.

\bibitem[\protect\citeauthoryear{Levy and Reyes}{2012}]{Levy2012_5609}
Levy RB, Reyes AD (2012)
\newblock Spatial profile of excitatory and inhibitory synaptic connectivity in
  mouse primary auditory cortex.
\newblock J. Neurosci.~32:\mbox{5609--5619}.

\bibitem[\protect\citeauthoryear{Lind\'{e}n  \bgroup et al.\egroup
  }{2014}]{Linden14}
Lind\'{e}n H, Hagen E, \L{\k{e}}ski S, Norheim ES, Pettersen KH, Einevoll GT
  (2014)
\newblock {LFPy}: a tool for biophysical simulation of extracellular potentials
  generated by detailed model neurons.
\newblock Front. Neuroinformatics~7:\mbox{41}.

\bibitem[\protect\citeauthoryear{Lind{\'e}n \bgroup et al.\egroup
  }{2010}]{Linden10}
Lind{\'e}n H, Pettersen KH, Einevoll GT (2010)
\newblock Intrinsic dendritic filtering gives low-pass power spectra of local
  field potentials.
\newblock J. Comput. Neurosci.~29:\mbox{423--444}.

\bibitem[\protect\citeauthoryear{Lind\'{e}n  \bgroup et al.\egroup
  }{2011}]{Linden11_859}
Lind\'{e}n H, Tetzlaff T, Potjans TC, Pettersen KH, Gr\"{u}n S, Diesmann M,
  Einevoll GT (2011)
\newblock Modeling the spatial reach of the {LFP}.
\newblock Neuron~72:\mbox{859--872}.

\bibitem[\protect\citeauthoryear{Livingstone and Hubel}{1984}]{Livingstone84}
Livingstone MS, Hubel DH (1984)
\newblock Specificity of intrinsic connections in primate primary visual
  cortex.
\newblock J. Neurosci.~4:\mbox{2830--2835}.

\bibitem[\protect\citeauthoryear{Lohmann and R\"{o}rig}{1994}]{Lohmann94}
Lohmann H, R\"{o}rig B (1994)
\newblock Long-range horizontal connections between supragranular pyramidal
  cells in the extrastiate visual cortex of the rat.
\newblock J. Comp. Neurol.~344:\mbox{543--558}.

\bibitem[\protect\citeauthoryear{Maier  \bgroup et al.\egroup }{2010}]{Maier10}
Maier A, Adams GK, Aura C, Leopold DA (2010)
\newblock Distinct superficial and deep laminar domains of activity in the
  visual cortex during rest and stimulation.
\newblock Front. Syst. Neurosci.~4:\mbox{31}.

\bibitem[\protect\citeauthoryear{Mainen and Sejnowski}{1996}]{Mainen96}
Mainen ZF, Sejnowski TJ (1996)
\newblock Influence of dendritic structure on firing pattern in model
  neocortical neurons.
\newblock Nature~382:\mbox{363--366}.

\bibitem[\protect\citeauthoryear{Markram  \bgroup et al.\egroup
  }{2015}]{Markram2015_456}
Markram H, Muller E, Ramaswamy S, Reimann MW, Abdellah M, Sanchez CA, Ailamaki
  A, Alonso-Nanclares L, Antille N, Arsever S, Kahou GAA, Berger TK, Bilgili A,
  Buncic N, Chalimourda A, Chindemi G, Courcol JD, Delalondre F, Delattre V,
  Druckmann S, Dumusc R, Dynes J, Eilemann S, Gal E, Gevaert ME, Ghobril JP,
  Gidon A, Graham JW, Gupta A, Haenel V, Hay E, Heinis T, Hernando JB, Hines M,
  Kanari L, Keller D, Kenyon J, Khazen G, Kim Y, King JG, Kisvarday Z, Kumbhar
  P, Lasserre S, B{\'{e}} JVL, Magalh{\~{a}}es BR, Merch{\'{a}}n-P{\'{e}}rez A,
  Meystre J, Morrice BR, Muller J, Mu{\~{n}}oz-C{\'{e}}spedes A, Muralidhar S,
  Muthurasa K, Nachbaur D, Newton TH, Nolte M, Ovcharenko A, Palacios J, Pastor
  L, Perin R, Ranjan R, Riachi I, Rodr{\'{\i}}guez JR, Riquelme JL, R\"{o}ssert
  C, Sfyrakis K, Shi Y, Shillcock JC, Silberberg G, Silva R, Tauheed F,
  Telefont M, Toledo-Rodriguez M, Tr\"{a}nkler T, Geit WV, D{\'{\i}}az JV,
  Walker R, Wang Y, Zaninetta SM, DeFelipe J, Hill SL, Segev I, Sch\"{u}rmann F
  (2015)
\newblock Reconstruction and simulation of neocortical microcircuitry.
\newblock Cell~163:\mbox{456--492}.

\bibitem[\protect\citeauthoryear{Mart{\'{\i}}nez-Ca{\~{n}}ada  \bgroup et
  al.\egroup }{2018}]{Martinez-Canada18_e1005930}
Mart{\'{\i}}nez-Ca{\~{n}}ada P, Mobarhan MH, Halnes G, Fyhn M, Morillas C,
  Pelayo F, Einevoll GT (2018)
\newblock Biophysical network modeling of the {dLGN} circuit: Effects of
  cortical feedback on spatial response properties of relay cells.
\newblock PLoS Comput. Biol.~14:\mbox{e1005930}.

\bibitem[\protect\citeauthoryear{Maynard \bgroup et al.\egroup
  }{1997}]{Maynard97}
Maynard EM, Nordhausen CT, Normann RA (1997)
\newblock The {U}tah intracortical electrode array: {A} recording structure for
  potential brain-computer interfaces.
\newblock EEG Clin. Neurophysiol.~102:\mbox{228--239}.

\bibitem[\protect\citeauthoryear{McDonald and Burkhalter}{1993}]{McDonald93}
McDonald CT, Burkhalter A (1993)
\newblock Organisation of long-range inhibitory connections within rat visual
  cortex.
\newblock J. Neurosci.~13:\mbox{768--781}.

\bibitem[\protect\citeauthoryear{Mehring  \bgroup et al.\egroup
  }{2003}]{Mehring03_395}
Mehring C, Hehl U, Kubo M, Diesmann M, Aertsen A (2003)
\newblock Activity dynamics and propagation of synchronous spiking in locally
  connected random networks.
\newblock Biol. Cybern.~88:\mbox{395--408}.

\bibitem[\protect\citeauthoryear{Mitzdorf}{1985}]{Mitzdorf85_37}
Mitzdorf U (1985)
\newblock Current source-density method and application in cat cerebral cortex:
  Investigation of evoked potentials and {EEG} phenomena.
\newblock Physiol. Rev.~65:\mbox{37--100}.

\bibitem[\protect\citeauthoryear{Mochizuki  \bgroup et al.\egroup
  }{2016}]{Mochizuki16_5736}
Mochizuki Y, Onaga T, Shimazaki H, Shimokawa T, Tsubo Y, Kimura R, Saiki A,
  Sakai Y, Isomura Y, Fujisawa S, Shibata KI, Hirai D, Furuta T, Kaneko T,
  Takahashi S, Nakazono T, Ishino S, Sakurai Y, Kitsukawa T, Lee JW, Lee H,
  Jung MW, Babul C, Maldonado PE, Takahashi K, Arce-McShane FI, Ross CF, Sessle
  BJ, Hatsopoulos NG, Brochier T, Riehle A, Chorley P, Gr\"{u}n S, Nishijo H,
  Ichihara-Takeda S, Funahashi S, Shima K, Mushiake H, Yamane Y, Tamura H,
  Fujita I, Inaba N, Kawano K, Kurkin S, Fukushima K, Kurata K, Taira M,
  Tsutsui KI, Ogawa T, Komatsu H, Koida K, Toyama K, Richmond BJ, Shinomoto S
  (2016)
\newblock {Similarity in neuronal firing regimes across mammalian species}.
\newblock J. Neurosci.~36:\mbox{5736--5747}.

\bibitem[\protect\citeauthoryear{Muller  \bgroup et al.\egroup
  }{2018}]{Muller18_255}
Muller L, Chavane F, Reynolds J, Sejnowski TJ (2018)
\newblock Cortical travelling waves: mechanisms and computational principles.
\newblock Nat. Rev. Neurosci.~19:\mbox{255--268}.

\bibitem[\protect\citeauthoryear{Muller and Destexhe}{2012}]{Muller12_222}
Muller L, Destexhe A (2012)
\newblock Propagating waves in thalamus, cortex and the thalamocortical system:
  Experiments and models.
\newblock J. Physiol. (Paris)~106:\mbox{222--238}.

\bibitem[\protect\citeauthoryear{Muller  \bgroup et al.\egroup
  }{2014}]{Muller14_4675}
Muller L, Reynaud A, Chavane F, Destexhe A (2014)
\newblock The stimulus-evoked population response in visual cortex of awake
  monkey is a propagating wave.
\newblock Nature Communications~5.

\bibitem[\protect\citeauthoryear{Murakoshi \bgroup et al.\egroup
  }{1993}]{Murakoshi93_211}
Murakoshi T, Guo JZ, Ichinose T (1993)
\newblock Electrophysiological identification of horizontal synaptic
  connections in rat visual cortex in vitro.
\newblock Neuroscience Letters~163:\mbox{211--214}.

\bibitem[\protect\citeauthoryear{Nauhaus  \bgroup et al.\egroup
  }{2009}]{Nauhaus09_70}
Nauhaus I, Busse L, Carandini M, Ringach DL (2009)
\newblock Stimulus contrast modulates functional connectivity in visual cortex.
\newblock Nat. Neurosci.~12:\mbox{70--76}.

\bibitem[\protect\citeauthoryear{Nelson and Pouget}{2010}]{Nelson10_2315}
Nelson MJ, Pouget P (2010)
\newblock Do electrode properties create a problem in interpreting local field
  potential recordings?
\newblock J. Neurophysiol.~103:\mbox{2315--2317}.

\bibitem[\protect\citeauthoryear{Nelson  \bgroup et al.\egroup
  }{2008}]{Nelson08_141}
Nelson MJ, Pouget P, Nilsen EA, Patten CD, Schall JD (2008)
\newblock Review of signal distortion through metal microelectrode recording
  circuits and filters~169:\mbox{141--157}.

\bibitem[\protect\citeauthoryear{Ness  \bgroup et al.\egroup }{2015}]{Ness2015}
Ness TV, Chintaluri HC, Potworowski J, \L{\k{e}}ski S, G\l{\k{a}}bska H,
  W\'{o}jcik DK, Einevoll GT (2015)
\newblock Modelling and analysis of electrical potentials recorded in
  multielectrode arrays ({MEAs}).
\newblock Neuroinformatics~13:\mbox{403--426}.

\bibitem[\protect\citeauthoryear{Nicholson and Freeman}{1975}]{Nicholson75_356}
Nicholson C, Freeman JA (1975)
\newblock Theory of current source-density analysis and determination of
  conductivity tensor for anuran cerebellum.
\newblock J. Neurophysiol.~2:\mbox{356--368}.

\bibitem[\protect\citeauthoryear{Nordlie \bgroup et al.\egroup
  }{2009}]{Nordlie-2009_e1000456}
Nordlie E, Gewaltig MO, Plesser HE (2009)
\newblock Towards reproducible descriptions of neuronal network models.
\newblock PLoS Comput. Biol.~5:\mbox{e1000456}.

\bibitem[\protect\citeauthoryear{Nunez and Srinivasan}{2006}]{Nunez06_2ed}
Nunez PL, Srinivasan R (2006)
\newblock {\em Electric Fields of the Brain, The Neurophysics of {EEG}}
\newblock Oxford University Press, Inc., 2nd edition.

\bibitem[\protect\citeauthoryear{Ohana \bgroup et al.\egroup
  }{2012}]{Ohana12_e40601}
Ohana O, Portner H, Martin KAC (2012)
\newblock Fast recruitment of recurrent inhibition in the cat visual cortex.
\newblock {PLoS} {ONE}~7:\mbox{e40601}.

\bibitem[\protect\citeauthoryear{Ohki and Reid}{2007}]{Ohki07_401}
Ohki K, Reid RC (2007)
\newblock Specificity and randomness in the visual cortex.
\newblock Curr. Opin. Neurobiol.~17:\mbox{401--407}.

\bibitem[\protect\citeauthoryear{Olavarria and
  Sluyters}{1985}]{Olavarria85_191}
Olavarria J, Sluyters RCV (1985)
\newblock Unfolding and flattening the cortex of gyrencephalic brains.
\newblock J. Neurosci. Methods~15:\mbox{191--202}.

\bibitem[\protect\citeauthoryear{Packer and Yuste}{2011}]{Packer2011_13260}
Packer AM, Yuste R (2011)
\newblock Dense, unspecific connectivity of neocortical parvalbumin-positive
  interneurons: A canonical microcircuit for inhibition?
\newblock J. Neurosci.~31:\mbox{13260--13271}.

\bibitem[\protect\citeauthoryear{Pan  \bgroup et al.\egroup }{2013}]{Pan13_288}
Pan WJ, Thompson GJ, Magnuson ME, Jaeger D, Keilholz S (2013)
\newblock Infraslow {LFP} correlates to resting-state {fMRI} {BOLD} signals.
\newblock {NeuroImage}~74:\mbox{288--297}.

\bibitem[\protect\citeauthoryear{Perin \bgroup et al.\egroup }{2011}]{Perin11}
Perin R, Berger TK, Markram H (2011)
\newblock A synaptic organizing principle for cortical neuronal groups.
\newblock Proc. Natl. Acad. Sci. USA~108:\mbox{5419--5424}.

\bibitem[\protect\citeauthoryear{Perkel \bgroup et al.\egroup
  }{1967}]{Perkel67a}
Perkel DH, Gerstein GL, Moore GP (1967)
\newblock Neuronal spike trains and stochastic point processes. {I}.~{T}he
  single spike train.
\newblock Biophys. J.~7:\mbox{391--418}.

\bibitem[\protect\citeauthoryear{Pettersen  \bgroup et al.\egroup
  }{2006}]{Pettersen06_116}
Pettersen KH, Devor A, Ulbert I, Dale AM, Einevoll GT (2006)
\newblock Current-source density estimation based on inversion of electrostatic
  forward solution: Effects of finite extent of neuronal activity and
  conductivity discontinuities.
\newblock J. Neurosci. Methods~154:\mbox{116--133}.

\bibitem[\protect\citeauthoryear{Pettersen \bgroup et al.\egroup
  }{2008}]{Pettersen08_291}
Pettersen KH, Hagen E, Einevoll GT (2008)
\newblock Estimation of population firing rates and current source densities
  from laminar electrode recordings.
\newblock J. Comput. Neurosci.~24:\mbox{291--313}.

\bibitem[\protect\citeauthoryear{Peyrache  \bgroup et al.\egroup
  }{2012}]{Peyrache12_1731}
Peyrache A, Dehghani N, Eskandar EN, Madsen JR, Anderson WS, Donoghue JA,
  Hochberg LR, Halgren E, Cash SS, Destexhe A (2012)
\newblock Spatiotemporal dynamics of neocortical excitation and inhibition
  during human sleep.
\newblock Proc. Natl. Acad. Sci. USA~109:\mbox{1731--1736}.

\bibitem[\protect\citeauthoryear{Potjans and Diesmann}{2014}]{Potjans14_785}
Potjans TC, Diesmann M (2014)
\newblock The cell-type specific cortical microcircuit: Relating structure and
  activity in a full-scale spiking network model.
\newblock Cereb. Cortex~24:\mbox{785--806}.

\bibitem[\protect\citeauthoryear{Potworowski  \bgroup et al.\egroup
  }{2012}]{Potworowski12_541}
Potworowski J, Jakuczun W, {\L}{\k{e}}ski S, W{\'{o}}jcik D (2012)
\newblock Kernel current source density method.
\newblock NeuralComput~24:\mbox{541--575}.

\bibitem[\protect\citeauthoryear{Pyle and Rosenbaum}{2017}]{Pyle17_18103}
Pyle R, Rosenbaum R (2017)
\newblock Spatiotemporal dynamics and reliable computations in recurrent
  spiking neural networks~118.

\bibitem[\protect\citeauthoryear{Quiroga}{2007}]{Quiroga07_3583}
Quiroga RQ (2007)
\newblock Spike sorting.
\newblock Scholarpedia~2:\mbox{3583}.

\bibitem[\protect\citeauthoryear{Ray and Maunsell}{2010}]{Ray2010}
Ray S, Maunsell JHR (2010)
\newblock Differences in gamma frequencies across visual cortex restrict their
  possible use in computation.
\newblock Neuron~67:\mbox{885--896}.

\bibitem[\protect\citeauthoryear{Ray and Maunsell}{2011}]{Ray2011}
Ray S, Maunsell JHR (2011)
\newblock Different {Origins} of {Gamma} {Rhythm} and {High}-{Gamma} {Activity}
  in {Macaque} {Visual} {Cortex}.
\newblock PLoS Comput. Biol.~9:\mbox{e1000610}.

\bibitem[\protect\citeauthoryear{Rees \bgroup et al.\egroup }{2016}]{Rees16}
Rees CL, Moradi K, Ascoli GA (2016)
\newblock Weighing the evidence in {P}eters' rule: Does neuronal morphology
  predict connectivity?
\newblock Trends Neurosci.~40:\mbox{63--71}.

\bibitem[\protect\citeauthoryear{Reimann  \bgroup et al.\egroup
  }{2017}]{Reimann17}
Reimann MW, Horlemann AL, Ramaswamy S, Muller EB, Markram H (2017)
\newblock Morphological diversity strongly constrains synaptic connectivity and
  plasticity.
\newblock Cereb. Cortex~27:\mbox{4570--4585}.

\bibitem[\protect\citeauthoryear{Reimann  \bgroup et al.\egroup
  }{2015}]{Reimann15_1}
Reimann MW, King JG, Muller EB, Ramaswamy S, Markram H (2015)
\newblock An algorithm to predict the connectome of neural microcircuits.
\newblock Front. Comput. Neurosci.~9:\mbox{120}.

\bibitem[\protect\citeauthoryear{Renart  \bgroup et al.\egroup
  }{2010}]{Renart10_587}
Renart A, {De La Rocha} J, Bartho P, Hollender L, Parga N, Reyes A, Harris KD
  (2010)
\newblock The asynchronous state in cortical circuits.
\newblock Science~327:\mbox{587--590}.

\bibitem[\protect\citeauthoryear{Reyes-Puerta  \bgroup et al.\egroup
  }{2016}]{Reyes-Puerta2016}
Reyes-Puerta V, Yang JW, Siwek ME, Kilb W, Sun JJ, Luhmann HJ (2016)
\newblock Propagation of spontaneous slow-wave activity across columns and
  layers of the adult rat barrel cortex in vivo.
\newblock Brain Structure Function~221:\mbox{4429--4449}.

\bibitem[\protect\citeauthoryear{Ribeiro  \bgroup et al.\egroup
  }{2013}]{Ribeiro13_28}
Ribeiro PFM, Ventura-Antunes L, Gabi M, Mota B, Grinberg LT, Farfel JM,
  Ferretti-Rebustini REL, Leite REP, Filho WJ, Herculano-Houzel S (2013)
\newblock The human cerebral cortex is neither one nor many: neuronal
  distribution reveals two quantitatively different zones in the gray matter,
  three in the white matter, and explains local variations in cortical
  folding~7:\mbox{28}.

\bibitem[\protect\citeauthoryear{Riehle  \bgroup et al.\egroup
  }{2013}]{Riehle13_48}
Riehle A, Wirtssohn S, Gr\"{u}n S, Brochier T (2013)
\newblock Mapping the spatio-temporal structure of motor cortical lfp and
  spiking activities during reach-to-grasp movements.
\newblock Frontiers in Neural Circuits~7:\mbox{48}.

\bibitem[\protect\citeauthoryear{Robinson}{1968}]{Robinson1968}
Robinson DA (1968)
\newblock The electrical properties of metal microelectrodes.
\newblock Proceedings of the {IEEE}~56:\mbox{1065--1071}.

\bibitem[\protect\citeauthoryear{Rosenbaum and Doiron}{2014}]{Rosenbaum14}
Rosenbaum R, Doiron B (2014)
\newblock Balanced networks of spiking neurons with spatially dependent
  recurrent connections.
\newblock Physical Review X~4:\mbox{021039}.

\bibitem[\protect\citeauthoryear{Rosenbaum  \bgroup et al.\egroup
  }{2017}]{Rosenbaum16_107}
Rosenbaum R, Smith MA, Kohn A, Rubin JE, Doiron B (2017)
\newblock The spatial structure of correlated neuronal variability.
\newblock Nat. Neurosci.~20:\mbox{107--114}.

\bibitem[\protect\citeauthoryear{Rotter and Diesmann}{1999}]{Rotter99a}
Rotter S, Diesmann M (1999)
\newblock Exact digital simulation of time-invariant linear systems with
  applications to neuronal modeling.
\newblock Biol. Cybern.~81:\mbox{381--402}.

\bibitem[\protect\citeauthoryear{Roxin \bgroup et al.\egroup }{2005}]{Roxin05}
Roxin A, Brunel N, Hansel D (2005)
\newblock The role of delays in shaping spatio-temporal dynamics of neuronal
  activity in large networks.
\newblock Phys. Rev. Lett.~94:\mbox{238103}.

\bibitem[\protect\citeauthoryear{Rubino \bgroup et al.\egroup
  }{2006}]{Rubino-2006_1549}
Rubino D, Robbins KA, Hatsopoulos NG (2006)
\newblock Propagating waves mediate information transfer in the motor cortex.
\newblock Nat. Neurosci.~9:\mbox{1549--1557}.

\bibitem[\protect\citeauthoryear{Sabatini and Regehr}{1996}]{Sabatini96_170}
Sabatini BL, Regehr WG (1996)
\newblock Timing of neurotransmission at fast synapses in the mammalian brain.
\newblock Nature~384:\mbox{170--172}.

\bibitem[\protect\citeauthoryear{Salin and Prince}{1996}]{Salin96_1589}
Salin PA, Prince DA (1996)
\newblock Electrophysiological mapping of {GABAA} receptor-mediated inhibition
  in adult rat somatosensory cortex.
\newblock J. Neurophysiol.~75:\mbox{1589--1600}.

\bibitem[\protect\citeauthoryear{Sato \bgroup et al.\egroup
  }{2012}]{Sato12_218}
Sato TK, Nauhaus I, Carandini M (2012)
\newblock Traveling waves in visual cortex.
\newblock Neuron~75:\mbox{218--229}.

\bibitem[\protect\citeauthoryear{Schmidt  \bgroup et al.\egroup
  }{2018}]{Schmidt18_1409}
Schmidt M, Bakker R, Hilgetag CC, Diesmann M, van Albada SJ (2018)
\newblock Multi-scale account of the network structure of macaque visual
  cortex.
\newblock Brain Structure \& Function~223:\mbox{1409--1435}.

\bibitem[\protect\citeauthoryear{Schnepel  \bgroup et al.\egroup
  }{2015}]{Schnepel15_3818}
Schnepel P, Kumar A, Zohar M, Aertsen A, Boucsein C (2015)
\newblock Physiology and impact of horizontal connections in rat neocortex.
\newblock Cereb. Cortex~25:\mbox{3818--3835}.

\bibitem[\protect\citeauthoryear{Schuecker  \bgroup et al.\egroup
  }{2017}]{Schuecker17}
Schuecker J, Schmidt M, van Albada SJ, Diesmann M, Helias M (2017)
\newblock Fundamental activity constraints lead to specific interpretations of
  the connectome.
\newblock PLoS Comput. Biol.~13:\mbox{e1005179}.

\bibitem[\protect\citeauthoryear{Schwalger \bgroup et al.\egroup
  }{2017}]{Schwalger17_e1005507}
Schwalger T, Deger M, Gerstner W (2017)
\newblock Towards a theory of cortical columns: From spiking neurons to
  interacting neural populations of finite size.
\newblock PLoS Comput. Biol.~13:\mbox{e1005507}.

\bibitem[\protect\citeauthoryear{Senk  \bgroup et al.\egroup }{2015}]{Senk_15}
Senk J, Hagen E, van Albada SJ, Diesmann M (2015)
\newblock From randomly connected to spatially organized multi-layered cortical
  network models
\newblock In {\em 11th G{\"o}ttingen Meeting of the German Neuroscience
  Society}.

\bibitem[\protect\citeauthoryear{Senk  \bgroup et al.\egroup
  }{2018}]{Senk18_arxiv_06046v1}
Senk J, Korvasov\'{a} K, Schuecker J, Hagen E, Tetzlaff T, Diesmann M, Helias M
  (2018)
\newblock Conditions for traveling waves in spiking neural networks.
\newblock arXiv preprint arXiv:1801.06046v1~.

\bibitem[\protect\citeauthoryear{Senk  \bgroup et al.\egroup
  }{2017}]{Senk17_243}
Senk J, Yegenoglu A, Amblet O, Brukau Y, Davison A, Lester DR, L\"{u}hrs A,
  Quaglio P, Rostami V, Rowley A, Schuller B, Stokes AB, van Albada SJ,
  Zielasko D, Diesmann M, Weyers B, Denker M, Gr\"{u}n S (2017)
\newblock A collaborative simulation-analysis workflow for computational
  neuroscience using {HPC}
\newblock In {Di Napoli} E, Hermanns MA, Iliev H, Lintermann A, Peyser A,
  editors, {\em High-Performance Scientific Computing. {JHPCS} 2016. Lecture
  Notes in Computer Science, vol 10164.}, \mbox{pp. 243--256}.

\bibitem[\protect\citeauthoryear{Sheng}{1985}]{Sheng85}
Sheng TK (1985)
\newblock The distance between two random points in plane regions.
\newblock Adv. Appl. Prob.~17:\mbox{748--773}.

\bibitem[\protect\citeauthoryear{Shinomoto \bgroup et al.\egroup
  }{2003}]{Shinomoto03_2823}
Shinomoto S, Shima K, Tanji J (2003)
\newblock Differences in spiking patterns among cortical neurons.
\newblock Neural Comput.~15:\mbox{2823--2842}.

\bibitem[\protect\citeauthoryear{Smith  \bgroup et al.\egroup
  }{2012}]{Smith2012}
Smith MA, Jia X, Zandvakili A, Kohn A (2012)
\newblock Laminar dependence of neuronal correlations in visual cortex.
\newblock J. Neurophysiol.~\mbox{pp. 940--947}.

\bibitem[\protect\citeauthoryear{Smith and Kohn}{2008}]{Smith08_12591}
Smith MA, Kohn A (2008)
\newblock Spatial and temporal scales of neuronal correlation in primary visual
  cortex.
\newblock J. Neurosci.~28:\mbox{12591--12603}.

\bibitem[\protect\citeauthoryear{Softky and Koch}{1993}]{Softky93}
Softky WR, Koch C (1993)
\newblock The highly irregular firing of cortical cells is inconsistent with
  temporal integration of random {EPSP}s.
\newblock J. Neurosci.~13:\mbox{334--350}.

\bibitem[\protect\citeauthoryear{Sokol}{1976}]{Sokol76_18}
Sokol S (1976)
\newblock Visually evoked potentials: Theory, techniques and clinical
  applications.
\newblock Survey of Ophthalmology~21:\mbox{18--44}.

\bibitem[\protect\citeauthoryear{Srinath and Ray}{2014}]{Srinath14_741}
Srinath R, Ray S (2014)
\newblock Effect of amplitude correlations on coherence in the local field
  potential.
\newblock J. Neurophysiol.~112:\mbox{741--751}.

\bibitem[\protect\citeauthoryear{Stepanyants  \bgroup et al.\egroup
  }{2008}]{Stepanyants2008_13}
Stepanyants A, Hirsch JA, Martinez LM, Kisv\'{a}rday ZF, Ferecsk\'{o} AS,
  Chklovskii DB (2008)
\newblock Local potential connectivity in cat primary visual cortex.
\newblock Cereb. Cortex~18:\mbox{13--28}.

\bibitem[\protect\citeauthoryear{Stepanyants  \bgroup et al.\egroup
  }{2009}]{Stepanyants09_3555}
Stepanyants A, Martinez LM, Ferecsk\'{o} AS, Kisv\'{a}rday ZF (2009)
\newblock The fractions of short- and long-range connections in the visual
  cortex.
\newblock Proc. Natl. Acad. Sci. USA~106:\mbox{3555--3560}.

\bibitem[\protect\citeauthoryear{Swadlow \bgroup et al.\egroup
  }{2002}]{Swadlow02_7766}
Swadlow HA, Gusev AG, Bezdudnaya T (2002)
\newblock Activation of a cortical column by a thalamocortical impulse.
\newblock J. Neurosci.~22:\mbox{7766--7773}.

\bibitem[\protect\citeauthoryear{Takahashi  \bgroup et al.\egroup
  }{2015}]{Takahashi15}
Takahashi K, Kim S, Coleman TP, Brown KA, Suminski AJ, Best MD, Hatsopoulos NG
  (2015)
\newblock Large-scale spatiotemporal spike patterning consistent with wave
  propagation in motor cortex.
\newblock Nature Communications~6.

\bibitem[\protect\citeauthoryear{Tanigawa \bgroup et al.\egroup
  }{2005}]{Tanigawa05}
Tanigawa H, Wang Q, Fujita I (2005)
\newblock Organization of horizontal axons in the inferior temporal cortex and
  primary visual cortex of the macaque monkey.
\newblock Cereb. Cortex~15:\mbox{1887--1899}.

\bibitem[\protect\citeauthoryear{Tetzlaff  \bgroup et al.\egroup
  }{2012}]{Tetzlaff12_e1002596}
Tetzlaff T, Helias M, Einevoll GT, Diesmann M (2012)
\newblock Decorrelation of neural-network activity by inhibitory feedback.
\newblock PLoS Comput. Biol.~8:\mbox{e1002596}.

\bibitem[\protect\citeauthoryear{Tetzlaff  \bgroup et al.\egroup
  }{2008}]{Tetzlaff08_2133}
Tetzlaff T, Rotter S, Stark E, Abeles M, Aertsen A, Diesmann M (2008)
\newblock Dependence of neuronal correlations on filter characteristics and
  marginal spike-train statistics.
\newblock Neural Comput.~20:\mbox{{2133--2184}}.

\bibitem[\protect\citeauthoryear{Thomson  \bgroup et al.\egroup
  }{2002}]{Thomson02_936}
Thomson AM, West DC, Wang Y, Bannister AP (2002)
\newblock Synaptic connections and small circuits involving excitatory and
  inhibitory neurons in layer 2-5 of adult rat and cat neocortex: Triple
  intracellular recordings and biocytin labelling in vitro.
\newblock Cereb. Cortex~12:\mbox{936--953}.

\bibitem[\protect\citeauthoryear{Thomson}{1982}]{Thomson82_1055}
Thomson DJ (1982)
\newblock Spectrum estimation and harmonic analysis.
\newblock Proc. IEEE~70:\mbox{1055--1096}.

\bibitem[\protect\citeauthoryear{Tomsett  \bgroup et al.\egroup
  }{2014}]{Tomsett14_2333}
Tomsett RJ, Ainsworth M, Thiele A, Sanayei M, Chen X, Gieselmann MA,
  Whittington MA, Cunningham MO, Kaiser M (2014)
\newblock {V}irtual {E}lectrode {R}ecording {T}ool for {EX}tracellular
  potentials ({VERTEX}): comparing multi-electrode recordings from simulated
  and biological mammalian cortical tissue.
\newblock Brain Structure and Function~220:\mbox{2333--2353}.

\bibitem[\protect\citeauthoryear{Tusa \bgroup et al.\egroup
  }{1978}]{Tusa78_213}
Tusa RJ, Palmer LA, Rosenquist AC (1978)
\newblock The retinotopic organization of area 17 (striate cortex) in the cat.
\newblock J. Comp. Neurol.~177:\mbox{213--235}.

\bibitem[\protect\citeauthoryear{van Albada  \bgroup et al.\egroup
  }{2018}]{VanAlbada18_291}
van Albada SJ, Rowley AG, Senk J, Hopkins M, Schmidt M, Stokes AB, Lester DR,
  Diesmann M, Furber SB (2018)
\newblock Performance comparison of the digital neuromorphic hardware
  {SpiNNaker} and the neural network simulation software {NEST} for a
  full-scale cortical microcircuit model.
\newblock Front. Neurosci.~12:\mbox{291}.

\bibitem[\protect\citeauthoryear{van Albada \bgroup et al.\egroup
  }{2015}]{Albada15}
van Albada SJ, Helias M, Diesmann M (2015)
\newblock Scalability of asynchronous networks is limited by one-to-one mapping
  between effective connectivity and correlations.
\newblock PLoS Comput. Biol.~11:\mbox{e1004490}.

\bibitem[\protect\citeauthoryear{van Essen and Maunsell}{1980}]{VanEssen80_255}
van Essen DC, Maunsell JHR (1980)
\newblock Two-dimensional maps of the cerebral cortex.
\newblock J. Comp. Neurol.~191:\mbox{255--281}.

\bibitem[\protect\citeauthoryear{van Kerkoerle  \bgroup et al.\egroup
  }{2014}]{vanKerkoerle14}
van Kerkoerle T, Self MW, Dagnino B, Gariel-Mathis MA, Poort J, van~der Togt C,
  Roelfsema PR (2014)
\newblock Alpha and gamma oscillations characterize feedback and feedforward
  processing in monkey visual cortex.
\newblock Proc. Natl. Acad. Sci. USA~111:\mbox{14332--14341}.

\bibitem[\protect\citeauthoryear{Veit  \bgroup et al.\egroup
  }{2017}]{Veit17_951}
Veit J, Hakim R, Jadi MP, Sejnowski TJ, Adesnik H (2017)
\newblock Cortical gamma band synchronization through somatostatin
  interneurons.
\newblock Nat. Neurosci.~20:\mbox{951--959}.

\bibitem[\protect\citeauthoryear{Voges and Perrinet}{2010}]{Voges10_51}
Voges N, Perrinet L (2010)
\newblock Phase space analysis of networks based on biologically realistic
  parameters.
\newblock J. Physiol. (Paris)~104:\mbox{51--60}.

\bibitem[\protect\citeauthoryear{Voges and Perrinet}{2012}]{Voges12}
Voges N, Perrinet L (2012)
\newblock Complex dynamics in recurrent cortical networks based on spatially
  realistic connectivities.
\newblock Front. Comput. Neurosci.~6:\mbox{41}.

\bibitem[\protect\citeauthoryear{Voges  \bgroup et al.\egroup
  }{2010}]{Voges10_277}
Voges N, Sch\"{u}z A, Aertsen A, Rotter S (2010)
\newblock A modeler's view on the spatial structure of intrinsic horizontal
  connectivity in the neocortex.
\newblock Prog. Neurobiol.~92:\mbox{277--292}.

\bibitem[\protect\citeauthoryear{Wagatsuma  \bgroup et al.\egroup
  }{2011}]{Wagatsuma11_00031}
Wagatsuma N, Potjans TC, Diesmann M, Fukai T (2011)
\newblock Layer-dependent attentional processing by top-down signals in a
  visual cortical microcircuit model.
\newblock Front. Comput. Neurosci.~5:\mbox{31}.

\bibitem[\protect\citeauthoryear{Wang and Buzs\'{a}ki}{1996}]{Wang1996}
Wang XJ, Buzs\'{a}ki G (1996)
\newblock Gamma {Oscillation} by {Synaptic} {Inhibition} in a {Hippocampal}
  {Interneuronal} {Network} {Model}.
\newblock J. Neurosci.~16:\mbox{6402--6413}.

\bibitem[\protect\citeauthoryear{Welch}{1967}]{Welch1967}
Welch PD (1967)
\newblock The use of fast fourier transform for the estimation of power
  spectra: A method based on time averaging over short, modified periodograms.
\newblock IEEE Transactions on Audio Electroacoustics~15:\mbox{70--73}.

\bibitem[\protect\citeauthoryear{Whittington  \bgroup et al.\egroup
  }{2000}]{Whittington2000}
Whittington MA, Traub RD, Kopell N, Ermentrout B, Buhl EH (2000)
\newblock Inhibition-based rhythms: experimental and mathematical observations
  on network dynamics~38:\mbox{315--336}.

\bibitem[\protect\citeauthoryear{Whittington \bgroup et al.\egroup
  }{1995}]{Whittington1995}
Whittington MA, Traub RD, Jefferys JGR (1995)
\newblock Synchronized oscillations in interneuron networks driven by
  metabotropic glutamate receptor activation.
\newblock Nature~373:\mbox{612--615}.

\bibitem[\protect\citeauthoryear{Wu \bgroup et al.\egroup }{2008}]{Wu08_487}
Wu JY, Huang X, Zhang C (2008)
\newblock Propagating waves of activity in the neocortex: What they are, what
  they do.
\newblock The Neuroscientist~14:\mbox{487--502}.

\bibitem[\protect\citeauthoryear{Xing  \bgroup et al.\egroup }{2012}]{Xing12b}
Xing D, Yeh CI, Burns S, Shapley RM (2012)
\newblock Laminar analysis of visually evoked activity in the primary visual
  cortex.
\newblock Proc. Natl. Acad. Sci. USA~109:\mbox{13871--13876}.

\bibitem[\protect\citeauthoryear{Xu  \bgroup et al.\egroup }{2007}]{Xu07_119}
Xu W, Huang X, Takagaki K, young Wu J (2007)
\newblock Compression and reflection of visually evoked cortical waves.
\newblock Neuron~55:\mbox{119--129}.

\bibitem[\protect\citeauthoryear{Yger  \bgroup et al.\egroup
  }{2011}]{Yger2011_229}
Yger P, {El Boustani} S, Destexhe A, Fr\'{e}gnac Y (2011)
\newblock Topologically invariant macroscopic statistics in balanced networks
  of conductance-based integrate-and-fire neurons.
\newblock J. Comput. Neurosci.~31:\mbox{229--245}.

\bibitem[\protect\citeauthoryear{Zanos  \bgroup et al.\egroup
  }{2015}]{Zanos15_615}
Zanos TP, Mineault PJ, Nasiotis KT, Guitton D, Pack CC (2015)
\newblock A sensorimotor role for traveling waves in primate visual cortex.
\newblock Neuron~85:\mbox{615--627}.

\end{thebibliography}
\end{document}